\documentclass[11pt,a4paper]{article}

\pdfoutput=1
\usepackage{jheppub}
\usepackage{color}
\usepackage{appendix}
\usepackage{graphicx}
\usepackage{wrapfig,enumerate,slashed}
\usepackage{footmisc}
\usepackage{amsmath}
\usepackage{graphicx}
\usepackage{color}
\usepackage{upgreek}
\usepackage{comment}
\usepackage{hyperref}
\usepackage{epstopdf}
\usepackage{caption}
\usepackage{subcaption}
\usepackage{braket}
\usepackage{bm}

\DeclareMathAlphabet{\mathbold}{U}{zeur}{b}{n}

\usepackage[dvipsnames]{xcolor}
\usepackage[labelfont=bf,font={sl,small}]{caption}

\renewcommand\({\left(}
\renewcommand\){\right)}
\renewcommand\[{\left[}
\renewcommand\]{\right]}

\newcommand{\csar}{|\{a\}\rangle}
\newcommand{\csal}{\langle\{a\}|}
\newcommand{\csbr}{|\{b\}\rangle}
\newcommand{\csbl}{\langle\{b\}|}
\newcommand{\vac}{|0\rangle}

\newcommand{\eq}{\begin{equation}}
\newcommand{\eqq}{\end{equation}}
\newcommand{\an}{\hat{a}}
\newcommand{\ad}{\hat{a}^\dagger}
\newcommand{\aln}{\hat{\alpha}}
\newcommand{\ald}{\hat{\alpha}^\dagger}

\newcommand{\ank}[1]{\hat{a}_{\mathrm{\textbf{#1}}}}
\newcommand{\adk}[1]{\hat{a}^\dagger_{\mathrm{\textbf{#1}}}}
\newcommand{\bec}[1]{\mathrm{\textbf{#1}}}
\newcommand{\omk}{\omega_\bec{k}}
\newcommand{\dk}{d \bec{k}}
\newcommand{\dep}{d \bec{p}}
\newcommand{\dx}{d \bec{x}}
\newcommand{\tc}{t_\mathbb{C}}

\newcommand{\refsec}{section~}

\def\beq{\begin{equation}}
\def\eeq{\end{equation}}
\def\[{\begin{equation}}
\def\]{\end{equation}}

\begin{document}
\numberwithin{equation}{section}

\title{Review of the semiclassical formalism for multiparticle production at high energies
}

\author{Valentin V. Khoze}
\author{and Joey Reiness}

\affiliation{Institute for Particle Physics Phenomenology, Department of Physics, Durham University,\\ 
Durham, DH1 3LE, UK}


\abstract{
These notes provide a comprehensive review of the semiclassical approach for calculating multiparticle production rates for
initial states with few particles at very high energies. In this work we concentrate on a scalar field theory with a mass gap.
Specifically, we look at a weakly-coupled theory in the high-energy limit, where the number of particles in the final state scales with energy, 
$n\sim E\to \infty$, and the coupling $\lambda\to 0$ with $n \lambda$ held fixed. In this regime, the semiclasical approach allows us to 
calculate multiparticle rates non-perturbatively.}

\date{}

\preprint{IPPP/18/46}
\maketitle


\section{Introduction}
\label{sec:intro}

There is a renewed interest among particle theorists in re-examining our understanding of basic
predictions of quantum field theory in the regime where production of a very large number of elementary
massive bosons becomes energetically possible. 
Specifically, in 
quantum field theoretical models with microscopic massive scalar fields at weak coupling, $\lambda \ll 1$,
the regime of interest is characterised by 
$few \to many$ particle production processes,
\[
 \quad X\, \to\,  n\times \phi\,,
\label{eq:Xplosion}
\]
at ultra-high centre of mass
energies $\sqrt{s} \gg m$. 
In these reactions, $X$ is the initial state with a small particle number, generally 1 or 2,  
and the final state is a multiparticle state with $n \propto \sqrt{s}/ m \gg 1$ Higgs-like neutral massive scalar particles.
For the initial states $X$ being the 2-particle states, the processes \eqref{eq:Xplosion} correspond to particle
collisions at very high $\sqrt{s}$ centre of mass energies. 

If $X$ is a single particle state  $|1^*\rangle$ with the virtuality $p^2=s$, \eqref{eq:Xplosion} describes its
decay into $n$-particle final states.
The authors of~\cite{Khoze:2017tjt} conjectured that the partial width of $X$ to decay into $n$ relatively soft elementary 
Higgs-like scalars can become exponentially large above a certain energy scale $s \gtrsim E_*^2$.  
This scenario is called Higgsplosion~\cite{Khoze:2017tjt}. It allows all super-heavy or highly-virtual states to be destroyed via 
rapid decays into multiple Higgs bosons. 

The aim of this paper is to provide a comprehensive review of the semiclassical calculation
of ${\rm few}\to n$-particle processes in the limit of ultra-high particle multiplicity, $n$.
The underlying semiclassical formalism was originally developed by Son in Ref.~\cite{Son:1995wz}, 
and generalised to the $\lambda n \gg 1$ regime in~\cite{Khoze:2018kkz,Khoze:2017ifq}.
We will give a detailed justification of the formalism and its derivation, and show its application to non-perturbative 
calculations in the weakly-coupled high-multiplicity regime.

Scattering processes at very high energies with $n\gg 1$ particles in the final state were studied 
in depth in the early literature~\cite{Cornwall:1990hh,Goldberg:1990qk,Brown:1992ay,Argyres:1992np,
Voloshin:1992rr,Voloshin:1992nu,Smith:1992rq,Argyres:1993wz,Libanov:1994ug,Libanov:1995gh,Voloshin:1994yp}, and more recently in 
\cite{Khoze:2014zha,Khoze:2014kka,Degrande:2016oan}.
These papers largely relied on perturbation theory which is robust in the regime of relatively low multiplicities, $n \ll 1/\lambda$.
However, in the regime of interest for Higgsplosion,  
$n \gtrsim 1/\lambda \gg 1$, perturbative results for $n$-particle amplitudes and rates can no longer be trusted. Perturbation theory becomes 
efectively strongly coupled in terms of the expansion parameter $\lambda n \gtrsim 1$. 
This calls for a robust non-perturbative formalism. 
Semiclassical methods~\cite{Gorsky:1993ix,Son:1995wz, Libanov:1997nt} 
provide a way to achieve this in the large $\lambda n$ regime~\cite{Khoze:2018kkz,Khoze:2017ifq}.
It is for this reason that the semiclassical method is at the centre of much of these notes.

\medskip

We consider a real scalar field $\phi(x)$ in $(d+1)$-dimensional spacetime, with the Lagrangian,
	\begin{equation}
	\label{eq:lag0}
	\mathcal{L}=\dfrac{1}{2}(\partial_\mu\phi)^2-\dfrac{1}{2}m^2\phi^2- \mathcal{L}_{\rm int}(\phi),
	\end{equation}
where $\mathcal{L}_{\rm int}$ is the interaction term. The two simplest examples are the $\phi^4$ model in the unbroken phase,
with $\mathcal{L}_{\rm int} =(\lambda/{4})\, \phi^4$, and the model with the spontaneously broken $Z_2$ symmetry,
\[
{\cal L} \,= \, 
\frac{1}{2}\, \partial^\mu h \, \partial_\mu h\, -\,  \frac{\lambda}{4} \left( h^2 - v^2\right)^2
\,. \label{eq:L}
\]
The classical equation for the model \eqref{eq:L} is the familiar Euler-Lagrange equation,
\[
\partial^\mu \partial_\mu h \,+\, \frac{\lambda}{4} \,h\, (h^2-v^2)\,=\, 0\,.
\label{eq:ELssb}
\]
As in Refs.~\cite{Khoze:2017ifq,Khoze:2017tjt}, we are ultimately interested in the scalar sector of the Standard Model, 
for which we use a simplified description in terms of the model \eqref{eq:L}.

\medskip

We will concentrate on the simplest realisation of Higgsplosion where $X$ is a single-particle state $|1^*\rangle$. 
In high-energy $2\to n$ scattering processes, the highly-virtual state $1^*$ would correspond to the $s$-channel 
resonance created by two incoming colliding particles. For example in the gluon fusion process,
$gg \to h^* \to n\times h$, the highly-virtual Higgs boson $h^*$ is created by the two initial gluons before decaying into $n$ Higgs bosons in
the final state. In this example, the $1^* \to n$ decay rate of interest corresponds to the  $h^* \to n\times h$ part of the
process. We will not discuss the complete $2\to n$ scattering 
in this paper.\footnote{In particular,
we will not attempt to apply the semiclassical approximation for the initial states that are not point-like, for example contributions
to scattering
processes dominated by 
exchanges in the $t$-channel. This is beyond the scope of this work.}
This paper focuses on explaining how the semiclassical calculation of the $n$-particle decay rates works.
We consider the method itself and its applications, rather than its potential phenomenological implications. 
The calculation we present is aimed to develop a theoretical foundation 
for the phenomenon of Higgsplosion~\cite{Khoze:2017tjt}.

If Higgsplosion can be realised in the Standard Model, its consequences for particle theory would be quite remarkable.
Higgsplosion would result in an exponential suppression of quantum fluctuations beyond the Higgsplosion energy scale and have observable
consequences at future high-energy colliders and in 
cosmology, some of which were discussed in~\cite{Khoze:2017lft,Jaeckel:2014lya,Gainer:2017jkp,Khoze:2017uga,Khoze:2018bwa}.
However, of course, the formalism we review is general and
not limited to Higgsplosion nor its applications.

\medskip

This work broadly consists of two halves: the first half provides context, and reviews the complex tools needed for much of the non-perturbative calculation presented in the second half. 
We begin by recalling the known results for multiparticle scattering rates via tree-level perturbation theory in \refsec\ref{sec:class}. 
In sections \ref{sec:CSQM} and \ref{sec:CSFQFT} we move on to summarising the basics of coherent states in quantum mechanics and quantum field theory respectively. The coherent state formalism in quantum field theory (for reviews and some applications 
see~\cite{Faddeev:1980be,Tinyakov:1992dr,Khlebnikov:1990ue,Libanov:1997nt}) 
 forms much of the foundation for the semiclassical method in question. Its summary helps provide context, familiarity and referencable results for the method's derivation, which is presented in \refsec\ref{sec:son}. 

With the necessary tools reviewed, we begin to calculate the rate for the $1^*\to n$ process semiclassically in \refsec\ref{sec:main}. 
The resulting set-up is ideal for using the thin-wall approach, which we develop is sections~\ref{sec:son_loops}
and~\ref{sec:thin_w}. In particular, in section~\ref{sec:son_loops} we recover tree-level results discussed in section \ref{sec:class} 
along with the prescription for computing the quantum corrections. These quantum contributions to the multiparticle rate are computed 
in section~\ref{sec:thin_w} using thin-walled singular classical solutions. 
In section~\refsec\ref{sec:Landau} we compare 
the semiclassical method in quantum field theory that we rely upon and use in this work
to Landau and Lifshitz' WKB calculation of quasi-classical matrix elements in quantum mechanics~\cite{Landau2},
and discuss the similarities and differences between these two methods. 
In section~\ref{sec:lowd} we consider multiparticle processes in a lower than 4 number of spacetime dimensions 
and provide a successful test for the semiclassical results.
Finally, we present our conclusions in section~\ref{sec:concl}.

\bigskip
\section{First glance at classical solutions for tree-level amplitudes}
\label{sec:class}
\medskip

In later sections of this paper we will compute the amplitudes and corresponding probabilistic rates for processes 
involving multiparticle 
final states in the large $\lambda n$ limit non-perturbatively using a semiclassical approach with no reference to perturbation theory 
and without artificially separating the result into tree-level and `quantum corrections' contributions. 
Their entire combined contribution should emerge from the unified semiclassical algorithm. 
Before beginning our review of the semiclassical formalism \cite{Son:1995wz,Libanov:1997nt,Khoze:2018kkz}, 
it is worth setting the scene for its application in this computation.
In this introductory section our aim is to recall the known properties 
of the tree-level amplitudes and their relation with certain classical solutions. We will also discuss the ways to analytically continue 
such classical solutions by complexifying the time variable in section~\ref{sec:class2}

\bigskip
\subsection{Classical solutions for tree-level amplitudes}
\label{sec:class1}
\medskip

Let us begin with tree-level $n$-point scattering amplitudes 
computed on the $n$-particle mass thresholds. This is the kinematic regime where all $n$
final state particles are produced at rest. These amplitudes for all $n$ are conveniently assembled 
into a single object -- the amplitude generating function -- which at tree-level is described by 
a particular solution of the Euler-Lagrange equations.
The classical solution, which provides the generating function of tree-level amplitudes on multi-particle
mass thresholds in the model \eqref{eq:L}, is given by~\cite{Brown:1992ay},
\[
h_{0} (z_0;t) \,=\, v\, \left(\frac{1+z_0\, e^{im t}/(2v)}{1-z_0\, e^{im t}/(2v)}\right)\,, \quad m=\sqrt{2\lambda} v\,,
\label{clas_sol}
\]
where $z_0$ is an auxiliary variable. It is easy to check by direct substitution that the
expression in \eqref{clas_sol} satisfies the the time-dependent ODE,
\[
\partial_t^2 h\,+\, \frac{\lambda}{4} \,h\, (h^2-v^2)\,=\, 0\,,
\label{eq:ELtssb}
\] 
for any value of the parameter $z_0$. 
Since the expression for $h_{0} (z_0;t)$ is uniform in space, it automatically satisfies
the full Euler-Lagrange equation \eqref{eq:ELssb}.
In fact, the configuration \eqref{clas_sol} is the unique non-trivial solution of  \eqref{eq:ELssb} with only
outgoing waves.

It then follows 
that all $1^*\to n$ tree-level scattering amplitudes on the $n$-particle mass thresholds are given by the differentiation 
of $h_{0} (z_0;t)$ with respect to $z_0$,
\[
{\cal A}_{1\to n}\,=\, \langle n|S \phi(0)| 0\rangle \,=\,\, \left.\left(\frac{\partial}{\partial z_0}\right)^n h_{0} \,\right|_{z_0=0}.
 \label{eq:A5}
\]
The classical solution in \eqref{clas_sol} is uniquely specified by requiring that it is a holomorphic function of the complex variable 
$z(t) = z_0\, e^{im t}$,
\[
h_{0} (z) \,=\,  v\,+\, 2v\,\sum_{n=1}^{\infty} \left(\frac{z}{2v}\right)^n
\,, \quad z=z(t) = z_0\, e^{im t}\,,
\label{gen-funh2}
\]
so that the amplitudes in \eqref{eq:A5} are given by the coefficients of the Taylor expansion in \eqref{gen-funh2} 
with a factor of $n!$ from differentiating $n$ times over $z$,
\[
{\cal A}_{1\to n}\,=\, 
\left.\left(\frac{\partial}{\partial z}\right)^n h_{0} (z) \,\right|_{z=0} \,=\, n!\, \left(\frac{1}{2v}\right)^{{n-1}}\,=\, 
n!\, \left(\frac{\lambda}{2m^2}\right)^{{\frac{n-1}{2}}}
\,.
\label{eq:ampln2}
\]
The connection between the classical solution $h_{0} (z_0;t)$ in \eqref{clas_sol} and the $1\to n$ tree-level amplitudes 
in \eqref{eq:ampln2} is nontrivial, but it can be verified that \eqref{eq:ampln2} is the correct answer following 
 the elegant formalism pioneered by Brown in Ref.~\cite{Brown:1992ay}, for a recent review see section 2 of
 Ref.~\cite{Khoze:2014zha}.
The approach of~\cite{Brown:1992ay} focuses on solving classical equations of motion and bypasses the summation 
over individual Feynman diagrams. 
In the following sections we will see how these (and also more general solutions describing full quantum processes) emerge
from the semiclassical approach of \cite{Son:1995wz} which we shall follow.
For now we just note the most interesting for us feature of the tree-level expressions expressions in \eqref{eq:ampln2} --  the  factorial growth of $n$-particle amplitudes,  ${\cal A}_{n} \sim \lambda^{n/2} n!$.

 \begin{figure*}[t]
\begin{center}
\includegraphics[width=0.8\textwidth]{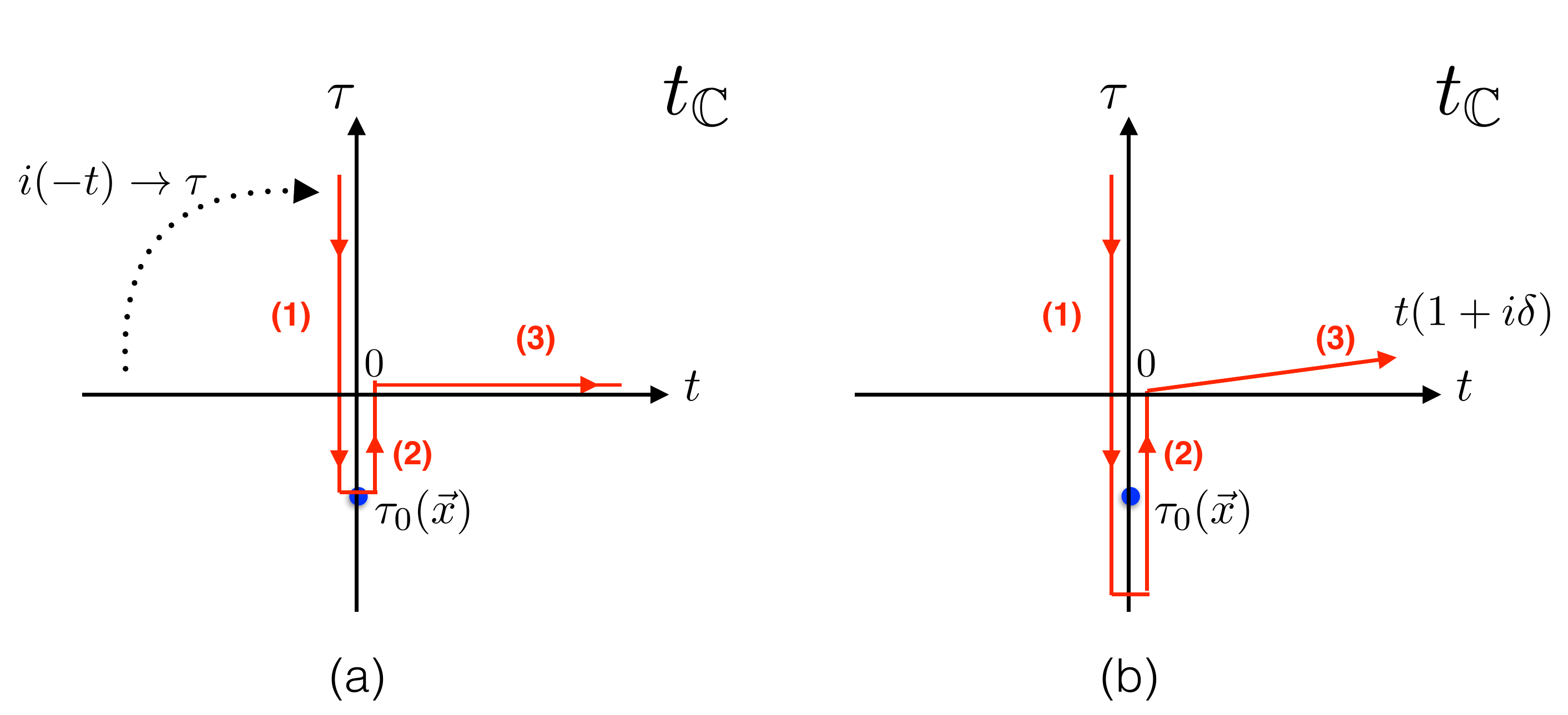}
\end{center}
\vskip-.5cm
\caption{Time evolution contour on the complex time plane $t_{\mathbb{C}}$.  {\bf Plot (a)} shows the contour obtained after deforming the 
the evolution along the real time axis $-\infty<t<+\infty$ where the early-time ray $-\infty <t<0$ is rotated by $\pi/2$ into the ray along the vertical axis,
$\infty>\tau>\tau_0(\bec{x})$ and ending at the singularity surface of the solution $\tau_0(\bec{x})$.
{\bf Plot (b)} shows a refinement of this contour: (1) rather than touching the singularity, the contour surrounds it; (2) at late times, the contour approaches $t\to +\infty$ along the ray with an infinitesimally small positive angle $\delta$ to the real time axis.}
\label{fig:contour}
\end{figure*}

\medskip 

Next, we would like to  draw the reader's attention to the fact that the classical solution \eqref{gen-funh2} is complex-valued, in spite of the fact that we are working with the
real-valued scalar field theory model \eqref{eq:L}. The classical solution $h_0$ that generates tree-level amplitudes via \eqref{eq:ampln2}
does not have to be real, in fact it is manifestly complex 
and this is a consequence of the fact that this solution will emerge
as an extremum of the action in the path integral using the steepest descent method. 
In this case, the integration contours in path integrals are deformed to enable them to pass through extrema (or encircle singularities) 
that are generically complex-valued.

It makes sense however to consider whether one can avoid explicitly deforming the fields as functional integration variables 
in the functional integral and instead analytically continue the time variable. 
Such an approach can simplify the calculation if it allows the saddle-point field configuration to be real-valued, even if only 
for part of its time evolution path.
Thus let us consider field configurations that depend on 
the complexified time $t_\mathbb{C}$. We promote the real time variable $t$ into the variable $t_\mathbb{C}$ that takes 
values on the complex time plane,
\[
t\,  \longrightarrow\, t_{\mathbb{C}} = t+i\tau\,,
\label{eq:tC}
\]
where $t$ and $\tau$ are real-valued. We will use the deformation the time-evolution contour from the real time axis 
$-\infty <t<+\infty$
to the contour in the complex $t_{\mathbb{C}}$ plane (depicted in Fig.~\ref{fig:contour})
in such a way that the initial time, $t=-\infty$, maps to the imaginary time, ${\rm Im}\, t_{\mathbb{C}} \,=\,\tau=+\infty$. 
This corresponds to the 
$(- t) \times e^{i\pi} = \tau$ rotation, 
\begin{eqnarray}
{\rm at \, early \, times,\,} -\infty<t<0: &&\quad  t \to i\tau.
\label{eq:ourAC}
\end{eqnarray}
We also note that $\tau$ corresponds to minus the Euclidean time $t_{E}$ defined by the standard Wick rotation $t\to -i t_{E}$.
The rationale for choosing this slightly bizarre looking `down-up-right' analytic continuation on the complex plane of $t_{\mathbb{C}}$
-- i.e. the contour shown in Fig.~\ref{fig:contour} --  will be discussed at the end of this section. First we would like to
explain the analytic structure of the field configurations relevant to us with a simple example of the classical solution \eqref{clas_sol}.

Expressed as a function of the complexified time variable, $t_{\mathbb{C}}$, the classical solution \eqref{clas_sol} reads,
\[
h_{0} (t_{\mathbb{C}}) \,=\, v\, \left(\frac{ 1\,+\,e^{im (t_{\mathbb{C}}-i\tau_\infty)}}
{1\,-\, e^{im (t_{\mathbb{C}}-i\tau_\infty)}}\right)\,,
\label{clas_sol2}
\]
where  $\tau_\infty$ is a constant,
\[ 
\tau_\infty := \frac{1}{m} \log \left(\frac{z_0}{2v}\right)\,,
\]
and parameterises the location (or the centre)
of the solution in imaginary time. If the time-evolution contour of the solution in the $t_{\mathbb{C}}$ plane is along 
the imaginary time axis with real time $t=0$, the field configuration \eqref{clas_sol2} becomes real-valued,
\[
h_{0} (t_{\mathbb{C}} =i \tau) \,=\, v\, \left(\frac{ 1\,+\,e^{-m (\tau-\tau_\infty)}}
{1\,-\, e^{-m (\tau-\tau_\infty)}}\right)\,,
\label{clas_sol22}
\]
and singular at $\tau=\tau_\infty$. 

For future reference it will be useful to define the profile function of $\tau$ 
\[
h_{0E}(\tau) \,=\,  v\, \left(\frac{ 1\,+\,e^{-m \tau}}
{1\,-\, e^{-m \tau}}\right)\,,
\label{eq:h0Edef}
\]
so that Eq.~\eqref{clas_sol22} becomes
$h_{0} (t_{\mathbb{C}} =i \tau) \,=\, h_{0E}(\tau -\tau_\infty)$.
By construction, $h_{0E}(\tau)$ is a real-valued function of its argument, is ${\bf x}$-independent,
 and is a solution
of the Euclidean-time analogue of the equation of motion \eqref{eq:ELtssb},
\[
-\partial_\tau^2 h_{0E}(\tau)\,+\, \frac{\lambda}{4} \,h_{0E}(\tau)\, (h_{0E}^2(\tau)-v^2)\,=\, 0\,.
\label{eq:ELtssbE}
\]

 \begin{figure*}[t]
\begin{center}
\includegraphics[width=0.55\textwidth]{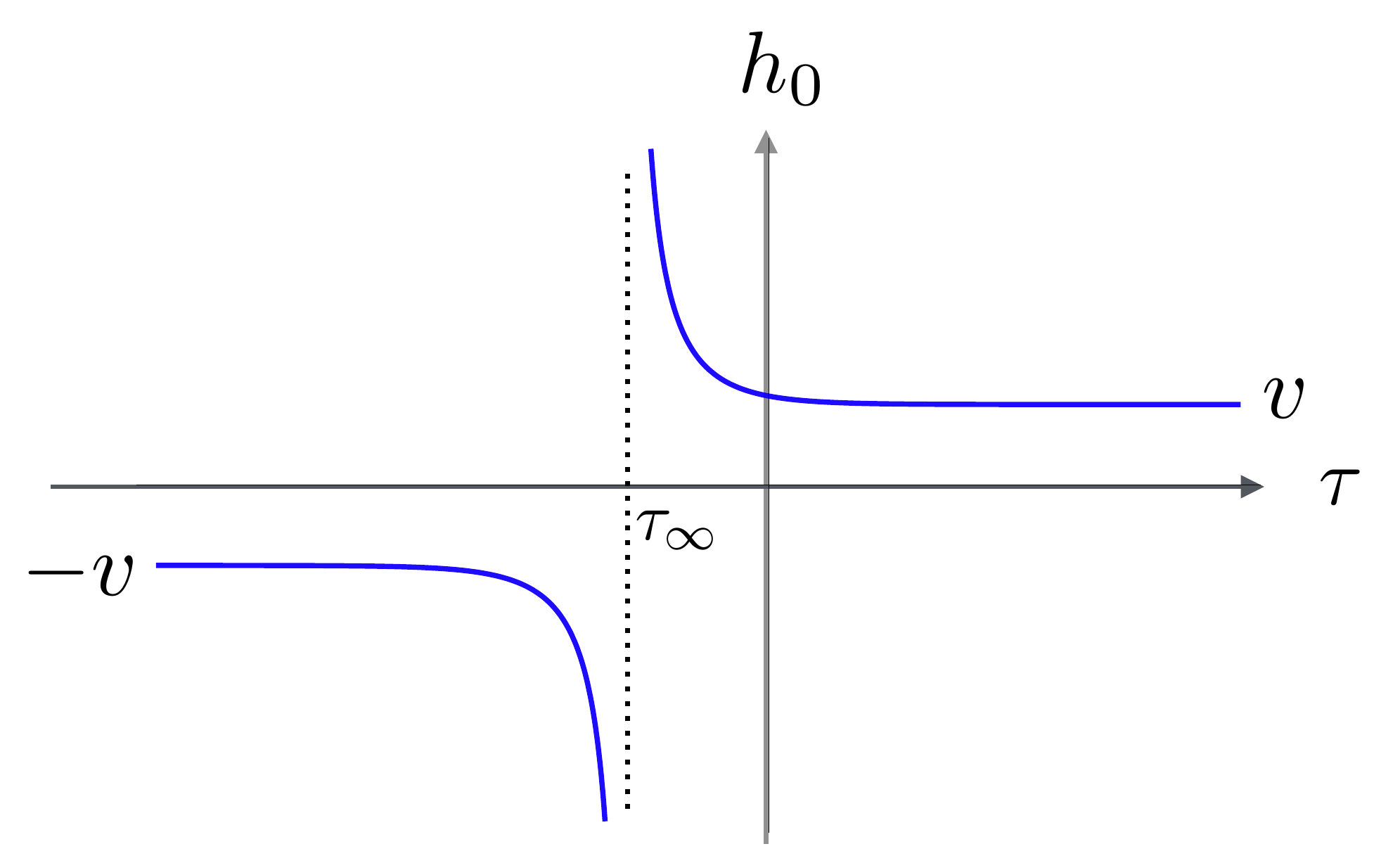}
\end{center}
\vskip-.5cm
\caption{Singular classical solution \eqref{clas_sol22}. This configuration defines a uniform in space flat singular domain wall located at imaginary time $\tau=\tau_\infty$.}
\label{fig:kink}
\end{figure*}

The expression on the right-hand side of \eqref{clas_sol22} has an obvious interpretation 
in terms of a singular domain wall, located at $\tau=\tau_\infty$, that separates two domains of the field $h(\tau, \bec{x})$ as shown in Fig.~\ref{fig:kink}.
The domain on the right of the wall 
$\tau \gg \tau_\infty$ has $h = +v$,
and the domain on the left of the wall, $\tau \ll \tau_\infty $, is characterised by $h = -v$.
The field configuration is singular at the position of the wall, $\tau=\tau_\infty$, for all values of $\bec{x}$, i.e. the singularity surface is 
flat (or uniform in space). The thickness of the wall is set by the inverse mass $1/m$.

\bigskip
\subsection{More on the analytic continuation in time}
\label{sec:class2}
\medskip

In the previous section we 
reviewed two important general features of the classical solution \eqref{gen-funh2} describing simple tree-level scattering amplitudes:

\newpage

\begin{enumerate}
\item the classical solution is complex-valued in real time;
\item it has a singularity on the complex plane located at a point $(t,\tau) = (0,\tau_{\infty})$,
where $\tau_{\infty}$ is a free parameter (a collective coordinate).
\end{enumerate}
We have also noted that the analytic continuation of $h_0$ to the imaginary time, $t_{\mathbb{C}} =i \tau$, 
gives a manifestly real-valued scalar field configuration in \eqref{clas_sol22} or equivalently \eqref{eq:h0Edef}.
As a result, the classical solution is real-valued along the two vertical parts of the red contour in Fig.~\ref{fig:contour}.
This fact turns out to be a general feature of all saddle-point solutions that will be relevant for our scattering problem, and
is a consequence of the initial-time boundary condition, which will be derived in Eq.~\eqref{eq:al2},
\[
\lim_{t\to - \infty}\,h(x)  \,=\, v\,+\, 
\int \frac{d^d k}{(2\pi)^{d/2}} \frac{1}{\sqrt{2\omega_{\bf k}}}\,\, a^*_{\bf k}\, e^{ik_\mu x^\mu} \,.
\label{eq:al2_I}
\]
Notice that the $\sim e^{i \omega_{\bf k}t}$ terms appearing on the right hand side are not accompanied by the opposite-sign
 frequencies -- the latter are not allowed in this
expression. Hence, when analytically continued to the imaginary time $t=i\tau$, the above equation gives,
\[
\lim_{\tau\to + \infty}\,h(x)  \,=\, v\,+\, 
\sim  a^*_{\bf k}\, e^{- \omega_{\bf k} \tau} \,,
\label{eq:al2_II}
\]
which amounts to a real-valued field configuration that is well-behaved at large $\tau$. 
Time evolving this initial condition along the first (downwards) part of the 
contour in Fig.~\ref{fig:contour} results in a real-valued classical solution along the Euclidean time axis $\tau \le \infty$.

The obvious question is then why shouldn't we just remain in the Euclidean time and define the entire contour 
as $\int_{+\infty}^{-\infty} d\tau$ instead of the `up-down-right' zig-zag contour in Fig.~\ref{fig:contour}.
The reason is that the final-time boundary conditions are also specified for our problem. As we will show in
Eq.~\eqref{eq:al3}, in general (i.e. for the saddle-point solution giving a dominant contribution to the functional integral representation for the scattering rate, in the regime where $\lambda n$ is not small)  these boundary conditions state:
\begin{eqnarray}
\lim_{t\to + \infty}\,h(x) &=& v\,+\, 
\int \frac{d^d k}{(2\pi)^{d/2}} \frac{1}{\sqrt{2\omega_{\bf k}}}\left( c_{\bf k}\, e^{-ik_\mu x^\mu}\,+\, b^*_{\bf k}\, e^{ik_\mu x^\mu}\right)\,.
\label{eq:al3_I}
\end{eqnarray}
The coefficients $c_{\bf k}$ and $b^*_{\bf k}$ for both positive and negative frequencies are non-vanishing in the general case, which
is incompatible with
any naive continuation of the complete solution to $\tau \to -\infty$, as it will diverge,\footnote{In this respect
the simple configuration $h_{0E}(\tau)$ in \eqref{eq:h0Edef} is an exception of the general rule
\eqref{eq:al3_I}, as it can be written in the form $\lim_{\tau\to \pm \infty}\, h_{0E}(\tau) \,=\, \pm v \,+\, {\cal O} (e^{-m |\tau|})$
and it appears that only the decaying exponents are present at large positive or large negative values of $\tau$.
However, this is an accidental simplification specific to this particularly simple solution describing tree-level amplitudes 
(i.e. $\lambda n \ll 1$ limit) on the mass thresholds. }
\[
h \to v \,+\, {\cal O} (e^{-m \tau}) \,+\, {\cal O} (e^{m \tau})\,.
\label{eq:al3_II}
\]
In general,
$c_{\bf k} \neq b_{\bf k}$, giving a genuinely complex-valued field configuration in Minkowski time.

To implement \eqref{eq:al3_I} it is thus unavoidable that the final part of the contour should be along
the real time axis -- the time variable cannot run to infinite values in any other direction if we are to avoid exponentially-divergent  
field configurations. 
On the first two vertical parts of the contour in Fig.~\ref{fig:contour}, the solution and its Euclidean action are real-valued quantities.
However, on the real-time part of the contour, the classical solution is a complex field.

Finally, the contour should also encounter the singularity of the solution, as shown in Fig.~\ref{fig:contour} a or b.
This ensures that the contour cannot be continuously
 deformed and shifted away to infinite values of $|t_{\mathbb{C}}|$ in the upper-right quadrant of the complex plane. 
 If this were possible, we could keep the contour at infinite values of $\tau$, which would contradict the boundary condition
 \eqref{eq:al3_II}. We will see later on that encountering the singularity of the solution on its time evolution trajectory 
 is precisely what allows for the jump in the energy $E$ carried by the solution. The energy changes from $E=0$ carried by the
configuration \eqref{eq:al2_I} at early times 
 to $E>0$ computed at late times from \eqref{eq:al3_I}. 
 
 A brief summary of the analytic continuation discussion above consists of the following steps.  
 The starting point for the formalism is the 
 functional integral over real-valued fields in Minkowski spacetime. The saddle-point (a.k.a. steepest descent) field configurations 
always  turn out to be complex-valued functions in real time. This fact by itself does not at all contradict the original requirement that
 the fields we are integrating over in a real scalar field theory, are real by definition. The steepest descent saddle-point is 
 simply not on the real-field-valued functional contour and in order to pass through it, one simply deforms the functional integration 
 field variables into complex fields. The main point instead is whether this approach can be simplified by also analytically continuing 
 the time variable for the fields. The answer is `yes', but only in so far as the standard Wick rotation $t\to i\tau$ is allowed.
 We showed that this can be achieved on the first two (vertical) parts of the time-evolution contour in Fig.~\ref{fig:contour}. 
 The steepest descent field configurations on these parts of the contour are real and well-defined in the $\tau\to \infty$ limit.
 We have thus avoided complex-valued fields and actions on the saddle point -- but only for part of the time-evolution contour.
 We further explained that the last part of the contour cannot be rotated into the imaginary time direction and must remain real
 or at least run parallel to the real-time axis (we neglect the infinitesimal angle $\delta$ in this discussion). This is forced 
 on us by the final-time boundary condition \eqref{eq:al3_I} for the saddle point field. On this final right-most part of the contour 
 in Fig.~\ref{fig:contour}, the saddle-point field configuration cannot be made real and remains complex. 
 This does not in any way present an obstacle or an ambiguity for using the steepest descent method.
 It is more of a technical point to keep in mind: on this final segment of the time evolution, the fields $h(x)$  in the functional integration measure ${\mathcal D} h(x)$ 
should be analytically continued to
 allow the integration contour to pass through complex-valued saddle-point field configurations. 
 Incidentally, it will turn out in our calculation that 
 classical action contributions on this part of the time-evolution contour will simply amount to certain boundary terms 
 that will be easy to account for.
 This summary concludes our discussion of the analytic continuation.
 We will return to its implementation in section~\ref{sec:5.3}.
Until then we will be following instead the original first-principles formulation in real Minkowski time.

\medskip
We now move on to reviewing the semiclassical formalism, starting with a brief discussion of coherent states in quantum mechanics, which form the foundation for the coherent state representation used heavily in later sections.

\bigskip
\section{Coherent states in quantum mechanics}
\label{sec:CSQM}
\medskip

Much of this section is basic quantum mechanics, but we review it nonetheless to ensure that our conventions are clear from the beginning. Furthermore, many of the more complex and notation-heavy equations presented in \refsec\ref{sec:CSFQFT} can be understood as analagous to the more simple relations discussed here. Thus the formulae below provide a useful reference for the more advanced calculations to come. We begin with a brief summary of coherent states in quantum mechanics, using the familiar canonical example of the quantum harmonic oscillator.

\subsection{Review of the quantum harmonic oscillator}
\label{sec:QHO}

Consider the 1D quantum harmonic oscillator, with Hamiltonian, $\hat{H}_0$,
\begin{equation}
\hat{H}_0 = \dfrac{\hat{p}^2}{2}+\dfrac{1}{2}\omega^2\hat{q}^2,
\end{equation}
where $\hat{p}$ and $\hat{q}$ are the momentum and position operators respectively, satisfying the usial commutation relation, $[\hat{q},\hat{p}]=i$. The angular frequency of the oscillator is denoted by $\omega$. Note that we set $\hbar=1$ and choose a unit mass $m=1$ in this quantum mechanical example.

In quantum mechanics, one seeks the energy spectrum of this system. This is usually done using the so-called raising and lowering
 operators, $\aln$ and $\ald$,\footnote{In this section we use Greek letters $\aln$ and $\ald$  to denote the 
 lowering/raising operators. The complex-number-valued eigenvalue of $\aln$ is denoted by Latin letter $a$, and its complex conjugate is  $a^*$. When dealing with the QFT generalisation starting from section~\ref{sec:CSFQFT}, we will use a more compact notation with Latin letters denoting both the operator-valued expressions and their eignevalues.}
\begin{equation}
\aln = \sqrt{\omega/2}\,(\hat{q}+i\hat{p}/\omega) 
\qquad
\ald = \sqrt{\omega/2}\,(\hat{q}-i\hat{p}/\omega),
\end{equation}
which satisfy the commutation relation,
\[
[\aln,\ald]=1,
\]
and enable the Hamiltonian to be rewritten as,
\begin{equation}
\hat{H}_0 = \omega(\ald\aln+1/2)=\omega(\hat{n}+1/2).
\end{equation}
We find that the stationary states are eigenstates, $|n\rangle$, of the operator $\hat{n}=\ald\aln$, which is commonly referred to as the 
occupation number operator, with integer eigenvalues $n\ge 0$. Given the Schr$\ddot{o}$dinger equation, we see that unique energy levels are uniformly separated by intevals $\Delta E=E_n-E_{n-1}= \omega$, with a ground state energy $E_0 = \omega/2$:
\begin{equation}
\hat{H}_0|n\rangle=E_n|n\rangle \qquad
E_n = \omega(n+1/2)
\qquad n\in\mathbb{Z}\ge0.
\end{equation}
A simple consequence of the commutation relations for $\ald$ and $\aln$ is that,
\begin{equation}
\ald|n\rangle = \sqrt{n+1}|n+1\rangle \qquad
\aln|n\rangle = \sqrt{n}|n-1\rangle,
\end{equation}
and so $\ald$ increases the energy of a state by $\omega$ where $\aln$ decreases it in equal measure. Given the ``vacuum'' state, $|0\rangle$, for which $\aln|0\rangle=0$, one can generate the full spectrum using the raising operator,
\begin{equation}
|n\rangle = \dfrac{(\ald)^n}{\sqrt{n!}}|0\rangle.
\end{equation}

\subsection{Coherent states as eigenstates of the lowering operator}
\label{sec:CS}
With the energy spectrum and associated states found, we now want to find eigenstates of the lowering operator $\aln$. These eigenstates are known as coherent states~\cite{Glauber:1963fi,Glauber:1963tx,Zhang:1990fy}.

We note that the states $|n\rangle$ form a complete set (since they are the eignestates of a Hermitian operator $\hat{H}_0$) and thus any state $|\psi\rangle$ can be written as,
\begin{equation}
|\psi\rangle = \sum_{n=0}^{\infty}\psi_n|n\rangle=\sum_{n=0}^{\infty}\psi_n \dfrac{(\ald)^n}{\sqrt{n!}}|0\rangle\,, \qquad
\mathrm{with}\qquad
\psi_n\in\mathbb{C}.
\end{equation}
From this we can see that an eigenstate of the raising operator is not possible as the lowest $n$ component in the decomposition will not be present after acting with $\ald$. However, one can find eigenstates of the lowering operator,
\begin{equation}
\aln|\psi\rangle = \sum_{n=0}^{\infty}\psi_n \aln|n\rangle=
\sum_{n=0}^{\infty}\psi_{n+1}\sqrt{n+1}\,|n\rangle
:=\, a\,|\psi\rangle\,, \qquad
\mathrm{for}\quad
\psi_{n+1}\sqrt{n+1}=\,a\, \psi_n,
\end{equation}
where $a$ is the eigenvalue of $\aln$. In other words, by iterating the second equation above, we find,
\begin{equation}
\psi_n \,=\, a^n\dfrac{\psi_0}{\sqrt{n!}}\quad \to \quad
|\psi\rangle \,=\, \psi_0 \sum_{n=0}^\infty \dfrac{(a \ald)^n}{n!}|0\rangle \,=\,
\psi_0\, e^{a\ald}|0\rangle.
\end{equation}
These eigenstates of the lowering operator are known as \emph{coherent states} and will prove to be a powerful tool in the functional integral QFT framework. We treat $\psi_0$ as an optional normalisation, which we set to $1$ in accordance with coherent state convention, despite the state's subsequently non-unit norm. In preparation for the calculations to follow, where many independent sets of coherent states can appear in a single expression is it useful to clearly define a coherent state in terms of one Latin letter $a$,
\begin{equation}
|a\rangle = e^{a\ald}|0\rangle \qquad \aln|a\rangle  = a|a\rangle
\qquad
\langle a| = \langle 0| e^{a^*\aln}\qquad \langle a|\ald = a^*\langle a|.
\end{equation}
Note that stars simply indicate a complex conjugate. It is important to distinguish between:
\begin{itemize}
	\item the quantum state $|a\rangle$ which lives in a Hilbert space,
	\item the raising and lowering operators $\ald$ and $\aln$, which have hats and act on states in this space,
	\item the complex number $a$, which is the eigenvalue associated with the action of operator $\aln$ on state $|a\rangle$, and
	$a^*$ which is the complex conjugate of the eigenvalue $a$.
\end{itemize}
With this in mind, one can introduce any number of coherent states $|b\rangle,|c\rangle,...$, with eigenvalues $\{b,c,...\}$ under the \emph{same} single set of operators: $\ald$ and $\aln$. For example,
\begin{equation}
|b\rangle = e^{b\ald}\vac \qquad
\aln|b\rangle = b|b\rangle .
\end{equation}
Hence we established a one-to-one correspondence between a complex number $z \in \mathbb {C}$ and a coherent state
$|z\rangle$, defined via,
\[
|z\rangle = e^{z \ald}\vac \,.
\]
The set of coherent states $\{ |a\rangle\}$ obtained by the complex number $z$ spanning the entire complex plane is known to be
an over-complete set. Mathematically, this is the statement that,
\[
1 = \int \dfrac{dz^*dz}{2\pi i}\,e^{-z^*z}|z\rangle\langle z| \,,
\label{eq:overcomplz}	
\]
where the 2-real-dimensional integral is over the complex plane $z \in \mathbb {C}$ . The \emph{over}-completeness
of the set manifests itself in the presence of the exponential factor $e^{-z^*z}$ on the right hand side.
We will derive Eq.~\eqref{eq:overcomplz} in the following section, see Eq.~\eqref{eq:overcompl} below.

The transition to QFT in section~\ref{sec:CSFQFT}
will be achieved by generalising the simple 1-dimensional QM example considered so far, to an infinite number of dimensions (i.e. infinite number of coupled harmonic oscillators). Hence we will need an infinite set of creation and annihilation operators 
$\aln_{\bf k}$ and $\ald_{\bf k}$, and correspondingly a set of coherent states $\{ |a\rangle\}_{\bf k}$ parameterised by 
complex-valued \emph{functions}
$a({\bf k}) \equiv a_{\bf k}$ where
${\bf k}$ is the momentum variable.

\subsection{Properties of coherent states in quantum mechanics}
\label{sec:CSprop}
We now discuss some of the useful properties of coherent states, which form the basis of many more complex derivations. As the
eigenstates of the lowering operator, one might ask how 
 coherent states are changed by application of the raising operator. It follows directly from the definition, $|a\rangle = e^{a\ald}|0\rangle$, that
 \begin{equation}
\ald|a\rangle = \dfrac{\partial}{\partial a}|a\rangle.
\end{equation}
Next we need the inner product of two coherent states,
\begin{equation}
\braket{b|a}=\braket{0|e^{b^*\aln}e^{a\ald}|0} = e^{b^*a}.
\label{eq:nonortn}
\end{equation}
The expression on the right-hand side was obtained  using the Baker-Campbell-Hausdorff (BCH) relation,
\begin{equation}
e^{\hat{A}}e^{\hat{B}}=e^{[\hat{A},\hat{B}]}e^{\hat{B}}e^{\hat{A}},
\end{equation}
which is valid so long as the commutator $[\hat{A},\hat{B}]\in\mathbb{C}$

Since $a$ is just a complex number, for a given set of raising and lowering operators there is an infinite set $\{|a\rangle\}$ of coherent states: one for every point in the complex plane. Accounting for their non-unit norm, the analogue of the completeness relation for coherent states is,
	\begin{equation}
	1 = \int \dfrac{da^*da}{2\pi i}\,e^{-a^*a}|a\rangle\langle a| \,:=\, \int d(a^*,a)\,e^{-a^*a}|a\rangle\langle a|.
\label{eq:overcompl}	
	\end{equation}
The identity \eqref{eq:overcompl} is often called the over-completeness relation due to the non-trivial exponential factor in the integral, with the
basis of coherent states described as over-complete. Also note that \eqref{eq:nonortn} implies that coherent states are not orthonormal either.

Using the relations \eqref{eq:nonortn}-\eqref{eq:overcompl}, one can write quantum-mechanical objects in a coherent state representation, where they appear as functions of one or more of the complex coherent state variables. We define the coherent state representation  of the state $|\psi\rangle$ as $\braket{a|\psi}=\psi(a^*)$, from which we find that the inner product of two states,
	\begin{equation}
	\braket{\psi_A|\psi_B} = \int d(a^*,a)\, e^{-a^*a}\psi^*_A(a)\psi_B(a^*),
	\end{equation}
	where $\psi_A^*(a)$ is just $[\psi_A(a^*)]^*$. Similarly, we define the matrix element of an operator $\hat{A}$ between two coherent states as $\braket{b|\hat{A}|a}=A(b^*,a)$. The action of such an operator on an arbitrary state $|\psi\rangle$ can be written as,
	\begin{equation}
	(\hat{A}\psi)(b^*) = \int d(a^*,a)\, e^{-a^*a}A(b^*,a)\psi(a^*).
	\end{equation} 
	Furthermore, one can write the matrix element for the product of two operators as,
	\begin{equation}
	\label{eq:cscombo}
	(AB)(b^*,a) = \braket{b|\hat{A}\hat{B}|a}=\int d(c^*,c)\, e^{-c^*c}A(b^*,c)B(c^*,a).
	\end{equation}
	The above logic is used extensively in the rest of this section and \refsec\ref{sec:CSFQFT}.
	
	A quantity which will prove to be useful is the coherent state representation of a position eigenstate, $\braket{q|a}$. We rewrite the raising operator in the coherent state exponent in terms of the original position and momentum operators, $\hat{q}$ and $\hat{p}$,
	\begin{equation}
	\braket{q|a}=\braket{q|\,e^{a\hat{a}^\dagger}|0}=\braket{q|\,e^{a\sqrt{\omega/2}\left(\hat{q}-\dfrac{i\hat{p}}{\omega}\right)}|0}\,=\,
	e^{a\sqrt{\omega/2}\left(q-\dfrac{1}{\omega}\dfrac{d}{dq}\right)}\braket{q|0},
	\end{equation}
	where the operators are now in their position-space representations: $q$ and $-id/dq$. It is well-known in quantum mechanics that the vacuum state is a Gaussian distribution centred at the origin of the potential well, $q=0$,
	\begin{equation}
	\braket{q|0}=\psi_0(q) = Ne^{-(q/q_0)^2/2}=Ne^{-\omega q^2/2},
	\end{equation}
	where $N$ is a normalisation constant and $q_0=\sqrt{1/\omega}$. We now make the substitution $y=q/q_0$ and use another BCH-like relation,
	\begin{equation}
	e^{\hat{A}+\hat{B}} = e^{-[\hat{A},\hat{B}]/2}e^{\hat{A}}e^{\hat{B}},
	\end{equation}
	to show that,
	\begin{equation}
	\braket{q|a}=N\exp\left( -\dfrac{1}{2}a^2 -\dfrac{1}{2}\omega q^2 +\sqrt{2\omega}aq \right).
	\label{eq:QMov}
	\end{equation}
	In quantum field theory, where the lowering operator is instead understood as an annihilation operator for quanta of the field, the analogous operator to position $\hat{q}$ will be the real scalar quantum field itself: $\hat{\phi}$.
	
	Finally, consider the action of a time evolution operator, $\hat{U}_0
	(t)=e^{-i\hat{H}_0 t}$, on a coherent state,\footnote{The subscripts on $\hat{U}_0$ and $\hat{H}_0$ are used to remind that the Hamiltonian is that
	of the simple harmonic oscillator, it will play the role of the free part of the Hamiltonian in interacting models, in particular the QFT settings considered in the following section.}
	\begin{equation}
	\label{eq:cstime}
	\begin{split}
	\hat{U}_0(t)\ket{a}\,=&\, e^{-i\hat{H}_0t}e^{a\ald}\vac\,=\, e^{-i\hat{H}_0t} \sum_{n=0}^{\infty}a^n\dfrac{(\ald)^n}{n!}\vac\\
	=&\, \sum_{n=0}^{\infty}a^n \, e^{-i\hat{H}_0t} \ket{n} \frac{1}{\sqrt{n!}}\,=\,  \sum_{n=0}^{\infty}(ae^{-i\omega t})^n \ket{n} \frac{1}{\sqrt{n!}}\\
	=&\,  \sum_{n=0}^{\infty}(ae^{-i\omega t})^n\dfrac{(\ald)^n}{n!}\vac\,=\,  \ket{ae^{-i\omega t}}.
	\end{split}
	\end{equation}
	We see that time evolution operators simply shift the phase of the coherent state variable associated with the coherent state. This property will be useful in computing the scattering $S$-matrix operator $\hat{S}$ in \refsec\ref{sec:skern}.

\bigskip
\section{Coherent state formalism in QFT and the S-matrix}
\label{sec:CSFQFT}
\medskip
	
In this section we develop the coherent state formalism for the functional integral representation of the S-matrix in quantum field theory, which is 
an important ingredient in the formulation of the semi-classical method for computing multiparticle production rates. We begin with the nuances associated with the move from quantum mechanics (QM) to quantum field theory (QFT), where the concept of coherent states is somewhat more abstract. We then explore their use in the calculation of amplitudes via path integrals. Sections \ref{sec:qftjump} and \ref{sec:skern} outline the coherent-states-based approach for writing matrix elements in QFT that was originally presented in 
Refs.~\cite{Khlebnikov:1990ue,Tinyakov:1992dr,Libanov:1997nt}, and in a slightly different formulation, called the holomorphic representation, in the textbook \cite{Faddeev:1980be}.
In section~\ref{sec:son} we will use the results derived here for matrix elements to efficiently implement the phase space integration and thus write down formulae for probabilistic rates for multiparticle production following the semiclassical formalism of Ref.~\cite{Son:1995wz}.

\subsection{QFT in $d+1$ dimensions as the infinite-dimensional QM system}
\label{sec:qftjump}

	In \refsec\ref{sec:CSQM} we discussed coherent states as eigenstates of the lowering operator for the quantum harmonic oscillator (QHO). Now we instead discuss the real scalar quantum field in $d+1$ dimensions, which has many mathematical parallels to the harmonic oscillator.
	
	A free real scalar field  is described by the Klein-Gordon Lagrangian, which can be manipulated into a Klein-Gordon Hamiltonian $H_0$,	\begin{equation}
H_0=\int d^d x\, \mathcal{H}_0\,, \quad {\rm where} \quad	\mathcal{H}_0 = \dfrac{1}{2}\uppi^2 + \dfrac{1}{2}|\nabla\phi|^2+\dfrac{m^2}{2}\phi^2.
	\end{equation}
Here $\phi(x)$ is the scalar field, $\uppi(x)$ is the momentum conjugate to the field, and $\nabla$ denotes a spatial derivative in $d$ dimensions. The conjugate momentum field $\uppi=\partial_t \phi= \dot{\phi}$ is just a change in variable associated with the Legendre transformation linking the Lagrangian and Hamiltonian formalisms. The final term is a mass term with mass set to unity. When transitioning from classical field theory to quantum field theory, the fields $\phi$ and $\uppi$ gain operator status. We will always be in the quantum regime and so their hats are omitted. Finally, note that $x$ now represents a $(1,d)$-vector $x=(t,\bec{x})$. 
	
	The above Hamiltonian already looks similar to that of the QHO but with the so-called generalised coordinate now being a field $\phi$ rather than a position $q$. This is made more obvious if we integrate the spatial derivative by parts,
	\begin{equation}
	\mathcal{H}_0 = \dfrac{\uppi^2}{2}+\dfrac{1}{2}\phi(m^2-\nabla^2)\phi,
\label{eq:Hfree1}		
	\end{equation}
	and so our frequency $\omega^2=(m^2-\nabla^2)$ is no longer a constant parameter. In a Fourier expansion, the Laplacian, $-\nabla^2$, will bring out a factor of the square $d$-dimensional momentum~$\bec{k}^2$. We thus expect a dispersion relation: $\omega_\bec{k}^2 = m^2+\bec{k}^2$. For every $d$-momentum, $\bec{k}$, there is an associated harmonic oscillator with frequency, $\omega_\bec{k}$, which we can solve by introducing raising and lowering operators, $\adk{k}$ and $\ank{k}$, as shown in \refsec\ref{sec:CSQM},\footnote{As already mentioned, to avoid overly comlicated notation in QFT, we will now use Latin letters for both, the creation/annihilation operators and for their eignevalues, see Eqs~\eqref{eq:freenr} and \eqref{eq:qfteigen}.}
	\begin{equation}
	\hat{H}_0 = \int  d^d k \, \omk\, \adk{k}\,\ank{k} \,+\, \mathcal{V}.
\label{eq:Hfree2}	
	\end{equation}
Here, $\mathcal{V}$ is analogous to the ground-state energy in the QHO and can be thought of as the energy of the vacuum. We ignore this term in the rest of this work by assuming the normal ordering prescription $:H_0:$ as is standard.
	 
	 The usual interpretation of the free quantum field is that it consists of an infinite number of harmonic oscillators. The raising and lowering operators can now be reinterpreted as creation and annihilation operators: the operator $\adk{k}$ creates a quantum of the field with 3-momentum $\bec{k}$, whereas operator $\ank{k}$ annihilates it. Considering the parallels with the QHO it should not be surprising that they obey the commutation relation, 
	\[
	[\ank{k},\adk{p}]=(2\pi)^{d/2}\, \delta(\bec{k}-\bec{p}).
\label{eq:ccr}	
	\]
	
	Inverting the definitions of the creation and annihilation operators and accounting for their momentum-space representation, one obtains the definition of the scalar field operator in terms of Fourier modes,
	\begin{equation}
	\hat\phi(x) = \int \dfrac{d^d k}{(2\pi)^{d/2}}\dfrac{1}{\sqrt{2\omk}}(\ank{k}\,e^{-ikx}\,+\,\adk{k}\,e^{+ikx}),
\label{eq:freenr}	
	\end{equation}
	where $kx=\,k_0t-\bec{k}\cdot\bec{x}$. An on-shell field satisfies the Klein-Gordon equation, implying $k_0=\omk=\sqrt{m^2+\bec{k}^2}$. The coefficients of the modes are the creation and annihilation operators in the quantum theory.\footnote{We use a non-relativistic normalisation for the integration measure in \eqref{eq:freenr}. In the relativistic normalisation, 
$\hat\phi(x) = \int \dfrac{d^dk}{(2\pi)^{d/2}}\dfrac{1}{2\omk}(\ank{k}e^{-ikx}+\adk{k}e^{+ikx})$ and one rescales $\ank{k}$ and $\adk{p}$ such that $[\ank{k},\adk{p}]=(2\pi)^{d /2}{2\omk}\delta(\bec{k}-\bec{p})$.}

In analogy with QM, a coherent state in QFT is a common eigenstate of \emph{all} annihilation operators, with eigenvalues dependent on the momentum $\bec{k}$ of the annihilation operator. We follow the notational rationale put forward in \refsec\ref{sec:CSQM}, by labelling the coherent state $|\{a\}\rangle$, and denoting its eigenvalue under operator $\ank{k}$ as $a_\bec{k}$,
	\begin{equation}
	\ank{k}\csar=a_\bec{k}\csar \qquad \forall \bec{k}.
	\label{eq:qfteigen}
	\end{equation}
	When converting from QM to QFT we have to take into account that we have moved from one oscillator to an infinite set, indexed by a free $d$-dimensional momentum $\bec{k}$. The curly braces in the state label, $\csar$, serve as a constant reminder. Therefore, in terms of the vacuum, our coherent state can be written as,
\begin{equation}
	\csar = e^{\int \dk\, a_\bec{k}\ad_\bec{k}}\vac.     
\label{eq:defcs}
\end{equation}
To avoid notational clutter, we use $\dk$ to represent the $d$-dimensional momentum integration measure, 
	\[
	\dk\, :=\, d^d k. \label{dkdefint}
	\] 
The Fourier transformation of the field operator is defined via,
	\begin{equation}
	{\tilde{\hat{\phi}}}(\bec{k}):={\tilde{\hat{\phi}}}(t,\bec{k}) := \int \dfrac{d^dx}{(2\pi)^{d/2}} \, e^{-i\bec{k}\cdot\bec{x}}\, \hat{\phi}(t,\bec{x}),
	\label{eq:FTdef}
	\end{equation}
so that the Fourier transform of the free field \eqref{eq:freenr} becomes simply the linear combination of the annihilation and creation operators with the positive and
negative frequencies,
\[
\tilde{\hat{\phi}}(\bec{k})  \,=\, 
\dfrac{1}{\sqrt{2\omk}}(\an_{\bec{k}}\,e^{-i\omega_{\bec{k}}t}\,+\,\ad_{-\bec{k}}\,e^{i\omega_{\bec{k}}t}).
\label{eq:freeFTop}	
\]	
In exact analogy to the operator-valued expressions for the Fourier-transforms in \eqref{eq:FTdef}-\eqref{eq:freeFTop}
we can also define the Fourier transforms of the $c$-valued scalar field $\phi(x)$,
\[
\tilde{\phi}(\bec{k})  \,=\, 
\dfrac{1}{\sqrt{2\omk}}(a_{\bec{k}}\,e^{-i\omega_{\bec{k}}t}\,+\,a^*_{-\bec{k}}\,e^{i\omega_{\bec{k}}t}).
\label{eq:freeFT}	
\]
where in this case the annihilation and creation
operators are substituted by the complex-valued eigenfunction $a_{\bf k}$ and it complex conjugate $a^*_{\bf k}$
as per \eqref{eq:qfteigen}.	
Note that $[\tilde{\phi}(\bec{k})]^*=\tilde{\phi}(-\bec{k})$ because $\phi(t,\bec{x})\in\mathbb{R}$.

	As in QM, we can find an inner product of two coherent states,
\begin{equation}
	\braket{\{b\}|\{a\}}\, =\, e^{\int \dk \,b^*_\bec{k}a_\bec{k}},
\label{eq:abprod}	
\end{equation}
	the over-completeness relation reads,
\begin{equation}
	\label{eq:csid}
	1 = \int d(\{a^*\},\{a\})e^{-\int \dk\, a^*_\bec{k}a_\bec{k}}\csar\csal,
\end{equation}
and for the inner product between the eigenstate of the field $\tilde\phi_{\bec{k}}$ and the coherent state we have,
	\begin{equation}
	\label{eq:projqft}
	\braket{\phi|\{a\}}=N\exp\left(-\dfrac{1}{2}\int \dk\, a_\bec{k}a_{-\bec{k}} - \dfrac{1}{2}\int \dk\, \omk \tilde{\phi}(\bec{k})\tilde{\phi}(-\bec{k})+\int \dk\, \sqrt{2\omk}a_\bec{k}\phi(\bec{k})\right).
	\end{equation}
Here $N$ is some constant normalisation factor irrelevant to our purposes.
The expression \eqref{eq:projqft} is the generalisation of the quantum mechanical overlap formula \eqref{eq:QMov} to the QFT case at hand which corresponds to an infinite number of QM oscillator degrees of freedom.
We note that the states $\bra{\phi}$ and $\ket{\{a\}}$ in the expression \eqref{eq:projqft} are defined as the eigenstates of the operators 
$\hat{\phi}$ and $\hat{a}$ respectively; with both operators taken at the same time $t$, which we take to be $t=0$.\footnote{The operators
$\hat{\phi}$ and $\hat{a}$ are defined in a theory with the Hamiltonian $H_0$, and it is straightforward to time-evolve them from $t=0$ to any $t$
with $e^{\pm i H_0 t}$. This will be done in \eqref{eq:BTs} in the next section, but in \eqref{eq:projqft} we use $t=0$.}
Hence the Fourier components of the field
$\tilde{\phi}(\bec{k})$ are given by the spatial Fourier transform \eqref{eq:FTdef} of $\phi(t,\bec{x})$ at $t=0$.
For completeness of our presentation,
the formula \eqref{eq:projqft} is derived in Appendix \ref{app:proj}.

	\subsection{Application to path integrals and amplitude calculation}
	\label{sec:skern}
	
We now consider an interacting quantum field theory in $d+1$ dimensions with the Hamiltonian $H$.
The object central to scattering theory is the S-matrix.  Given an initial state, $\ket{\phi_i(t_i)}$, the S-matrix 
defines the probability amplitude 
of arriving at a final state, $\ket{\phi_f(t_f)}$.

In the interaction picture, where we split the Hamiltonian into the free part $H_0$, and the interacting part $V$, 
\[
\hat{H}= \hat{H}_0+\hat{V},
\] 
the S-matrix $S_{fi}$ is defined as,
\begin{equation}
S_{fi} =\braket{\phi_f|\hat{S}|\phi_i}= \lim_{t_f,t_i\to\pm\infty}\braket{\phi_f|\, e^{i\hat{H}_0t_f}\hat{U}(t_f,t_i)e^{-i\hat{H}_0t_i}|\phi_i},
\label{eq:sdef}
\end{equation}
where $\ket{\phi_f}$ and $\ket{\phi_i}$ are free states, i.e. eigenstates of the free Hamiltonian $\hat{H}_0$, prepared at the times
$t=t_f$ and $t=t_i$ respectively.
The S-matrix operator, appearing on the right-hand side of  \eqref{eq:sdef},
\[
\hat{S} = \lim_{t_f,t_i\to\pm\infty}\, e^{i\hat{H}_0t_f}\hat{U}(t_f,t_i)e^{-i\hat{H}_0t_i},
\label{eq:evolint}
\]
implements the time-evolution of the interaction-picture-state $\ket{\phi_i}$ from $t_i$ to $t_f$ where it is contracted with the final state
$\ket{\phi_f}$. The operator $U(t_f,t_i)$ in \eqref{eq:evolint} is the time-evolution operator for the Heisenberg fields,
\begin{equation}
\hat{U}(t_f,t_i)\,=\,\mathcal{T}\exp{\big(  -i\int_{t_i}^{t_f} \hat{H} dt \big)},
\end{equation}
with $\mathcal{T}$ denoting a time-ordered product. 
Given that the fields in the interaction picture are free fields, 
one has,
\[
\ket{\phi(t')} \,=\, e^{-i \hat{H}_0 (t'-t)} \ket{\phi(t)},
\]
which explains the $e^{ i \hat{H}_0 t_{f}}$ and $e^{ -i \hat{H}_0 t_{i}}$ factors in \eqref{eq:evolint}.

In the infinite future and past, the initial and final particles are sufficiently separated 
in the $d$-dimensional space so as not to experience interactions (apart from the effects accounted for by UV renormalisation of fields and parameters of the theory). Thus, by taking the limit 
$\lim_{t_f,t_i\to\pm\infty}$, 
the free Hamiltonian eigenstates in \eqref{eq:sdef} are a good approximation to the actual initial and final states.

Of course, for a non-interacting theory, $\hat{S}$ is simply the identity operator. More generally, $\hat{S}=1+i\hat{T}$ and we define the matrix element, $\mathcal{M}$, by the relation,
\begin{equation}
\braket{\phi_f|\hat{S}-1|\phi_i}=\braket{\phi_f|i\hat{T}|\phi_i} = (2\pi)^4\delta^{(d+1)}\left( \sum k_f-\sum k_i \right)i\mathcal{M},
\end{equation}
where the delta function simply enforces momentum conservation.
	
\medskip

We now want to express the S-matrix \eqref{eq:evolint} in the basis of coherent states. This is
the kernel of the $S$-matrix,
	\begin{equation}
	S(b^*,a):=\lim_{t_f,t_i\to\pm\infty}\braket{\{b\}|e^{i\hat{H}_0t_f}\hat{U}(t_f,t_i)e^{-i\hat{H}_0t_i}|\{a\}}.
	\end{equation}
Equation \eqref{eq:cstime} implies that the free evolution operators simply shift the phase of the coherent states, giving,
	\begin{equation}
	S(b^*,a):=\lim_{t_f,t_i\to\pm\infty}\braket{\{be^{-i\omega t_f}\}|\hat{U}(t_f,t_i)|\{ae^{-i\omega t_i}\}}.
\label{eq:Skernel}	
	\end{equation}
	Note that $|\{ae^{-i\omega t_i}\}\rangle$ refers to a coherent state much like $|\{a\}\rangle$ but with $a_\bec{k}\to a_\bec{k}e^{-i\omk t_i}$ for all $\bec{k}$.
	
The derivation of the $S$-matrix kernel will closely follow that presented in 
 \cite{Khlebnikov:1990ue,Tinyakov:1992dr}.
 Using the completeness relation,
\begin{equation}
	1=\int d\phi_f\ket{\phi_f}\bra{\phi_f},
\end{equation}
	and similarly for $\phi_i$, we can re-write \eqref{eq:Skernel} as,
\begin{equation}
	S(b^*,a)=\lim_{t_f,t_i\to\pm\infty}\int d\phi_fd\phi_i\braket{\{be^{-i\omega t_f}\}|\phi_f}\braket{\phi_f|\hat{U}(t_f,t_i)|\phi_i}\braket{\phi_i|\{ae^{-i\omega t_i}\}}.
\end{equation}
	We recognise $\braket{\phi_f|\hat{U}(t_f,t_i)|\phi_i}$ as the Feynman path integral,
	\begin{equation}
	\braket{\phi_f|\hat{U}(t_f,t_i)|\phi_i}=\int\mathcal{D}\phi \, e^{iS[\phi]_{t_i}^{t_f}},
	\end{equation}
over the fields satisfing the boundary conditions,
\begin{equation}
	\label{eq:funcbcs}
	\phi(t_i)=\phi_i,\quad
	\phi(t_f)=\phi_f, \quad
\end{equation}
where $S[\phi]_{t_i}^{t_f}$ is the action, 
\begin{equation}
	S[\phi]_{t_i}^{t_f}=\int_{t_i}^{t_f} dt\int d^d x \, \mathcal{L}(\phi).
\end{equation} 
Inserting the projections of the initial and final states in the coherent state basis \eqref{eq:projqft}, we arrive at the following result,
\begin{equation}
	S(b^*,a) =\, \lim\limits_{t_f,t_i\to\pm\infty}\int d\phi_f\, d\phi_i \, e^{B_i(\phi_i;a)+B_f(\phi_f;b^*)}\int\mathcal{D}\phi \, e^{iS[\phi]_{t_i}^{t_f}}.
\end{equation}
Here the boundary terms, $B_i(\phi_i;a)=\braket{\phi_i|\{ae^{-i\omega t_i}\}}$ and 
$B_f(\phi_f;b^*)=\braket{\{be^{-i\omega t_f}\}|\phi_f}$, are given by ({\it cf.}~\eqref{eq:projqft}),
\begin{equation}
	\label{eq:BTs}
	\begin{split}
	B_i(\phi_i;a)=-\dfrac{1}{2}\int \dk\, a_\bec{k} a_{-\bec{k}}e^{-2i\omega_k t_i}-\dfrac{1}{2}\int \dk\, \omk \tilde{\phi}_i(\bec{k})\tilde{\phi}_i(-\bec{k})+\int \dk \sqrt{2\omk}a_\bec{k}\tilde{\phi}_i(\bec{k})e^{-i\omk t_i},\\
	B_f(\phi_f;b^*)=-\dfrac{1}{2}\int \dk\, b^*_\bec{k} b^*_{-\bec{k}}e^{2i\omega_\bec{k} t_f}-\dfrac{1}{2}\int \dk\, \omega_\bec{k} \tilde{\phi}_f(\bec{k})\tilde{\phi}_f(-\bec{k})+\int \dk\sqrt{2\omega_\bec{k}}b^*_\bec{k}\tilde{\phi}_f(-\bec{k})e^{i\omk t_f}.
	\end{split}
\end{equation}
In these expressions, $\tilde{\phi}_i(\bec{k})$ and $\tilde{\phi}_f(\bec{k})$ are the $d$-dimensional Fourier transforms of the boundary fields,
$\phi_i(\bec{x}) = \phi(t_i, \bec{x})$ and $\phi_f(\bec{x}) = \phi(t_f, \bec{x})$, so that,
\begin{equation}
\begin{split}
	\tilde{\phi}_i(\bec{k}) =& \int \dx\, e^{-i\bec{k}\cdot\bec{x}}\, \phi(t_i,\bec{x}), \\
	\tilde{\phi}_f(\bec{k}) =& \int \dx \, e^{-i\bec{k}\cdot\bec{x}}\, \phi(t_f,\bec{x}).
	\label{eq:FTdef2}
\end{split}	
\end{equation}
where in analogy with \eqref{dkdefint} the $d$-dimensional coordinate integration measure is defined via,
\[
\dx := d^dx  
\,. \label{dxdefint}
\]
Thus in comparison to the simple overlaps in \eqref{eq:projqft} at $t=0$, the boundary terms in \eqref{eq:BTs} contain the dependence on
$t_i$ or $t_f$ via the phase factors accompanying the $a$ and $b^*$ in \eqref{eq:BTs}, as well as in the definitions of the boundary fields \eqref{eq:FTdef2}.

Before concluding this section, we mention a particularly useful property of the coherent state basis for scattering theory, that allows one to circumvent the LSZ reduction formulae.
The kernel, $A(b^*,a)=\braket{b|\hat{A}|a}$, of any operator $\hat{A}$ in the coherent state representation is the generating functional for the same operator in the Fock space,	
\begin{equation}
	\braket{q_1...q_m|\hat{A}|p_1...p_n}\,=\,
	\dfrac{\partial}{\partial b^*_{\bec{q}_1}}...\dfrac{\partial}{\partial b^*_{\bec{q}_m}}\, 
	\dfrac{\partial}{\partial a_{\bec{p}_1}}...\dfrac{\partial}{\partial a_{\bec{p}_n}}\,A(b^*,a)|_{a=b^*=0},
\end{equation}
where $\ket{p_1...p_n}$ is an $n$-particle state with particle $d$-momenta $\bec{p}_i$, $i=1,\ldots,n$.
This formula follows immediately from the definition of the coherent sate \eqref{eq:defcs}, since
\begin{equation}
	\dfrac{\partial}{\partial a_{\bec{p}_1}}...\dfrac{\partial}{\partial a_{\bec{p}_n}}\,\,
	e^{\int \dk\, a_\bec{k}\ad_\bec{k}}\vac |_{a=0}\,= \, \ket{p_1...p_n}.
\label{eq:diffcs}
\end{equation}
Applied to the $S$-matrix operator we find,
	\begin{equation}
	\braket{q_1...q_m|S|p_1...p_n}\,=\, \dfrac{\partial}{\partial b^*_{\bec{q}_1}}...\dfrac{\partial}{\partial b^*_{\bec{q}_m}}\, 
	\dfrac{\partial}{\partial a_{\bec{p}_1}}...\dfrac{\partial}{\partial a_{\bec{p}_n}}\, S(b^*,a)|_{a=b^*=0}.
	\end{equation}
	The left-hand side is just the $S$-matrix element for the $n\to m$ process. Hence, just differentiating with respect to coherent state variables, we can calculate any scattering amplitudes directly from the kernel of the S-matrix. Thus the coherent state representation allows 
one to bypass the LSZ reduction formulae, by simply differentiating the path integral for the kernel of the S-matrix. 
This coherent state formulation is of course equivalent to the LSZ procedure,\footnote{In our derivation we have neglected the 
$Z$ factors arising from the wave-function renormalisation. Of course they can be painstakingly restored, but this will not be required for our applications of the semiclassical approach.} but gives a more direct route for semiclassical applications, given the exponential nature of $S(b^*,a)$.
	
\bigskip
\section{The semiclassical method for multi-particle production}
\label{sec:son}
\medskip
	
In this section we review the semiclassical method of Son \cite{Son:1995wz} for calculating probabilistic rates or crosssections 
for processes \eqref{eq:Xplosion}. There are two types of initial states $X$ that are of particular interest,
\begin{eqnarray}
{\rm Scattering\, process}: \quad && |X(\sqrt{s})\rangle = |2\rangle \,\,\,\to \,\, |n\rangle \quad \Rightarrow \quad {\rm cross\,\, section} \,\, \sigma_n(\sqrt{s})\,,
\label{eq:2pH}
\\
{\rm Resonance\, decay}: \quad &&|X(\sqrt{s})\rangle = |1^*\rangle \,\to\,\, |n\rangle \quad \Rightarrow \quad {\rm partial\, width} \,\, \Gamma_n(s)\,.
\label{eq:1pH}
\end{eqnarray}
For the 2-particle initial state, the $n$-particle production process \eqref{eq:2pH} is characterised by the cross section $\sigma_n(\sqrt{s})$; for the single-particle state of virtuality $p^2=s$ in \eqref{eq:1pH}, the relevant quantity is the partial decay width $\Gamma_n (s)$.
	Final states contain a large number $n \gtrsim 1/({\rm coupling\, constant}) \gg 1$ of elementary Higgs-like scalar particles 
	of mass $m$.

\medskip
	
	As we already mentioned in the Introduction, this paper concentrates primarily on the process of type \eqref{eq:1pH} to simplify the presentation. Formally, both processes in 
	\eqref{eq:2pH}-\eqref{eq:1pH}
	can be treated
	simultaneously in the semiclassical approach of Son \cite{Son:1995wz}, where the initial state $X$ is approximated by a local operator                
	${\cal O}(x)$ acting on the vacuum state.
In $|2\rangle \,\to \, |n\rangle$ scattering with $n$ large, the original 2 particles exchange large momentum and thus 
come within a short distance of one another. This justifies a description with 
a local operator source.\footnote{The effect of smearing of the local operator
would be important in the description of $2\to n$ processes in order account for the effect of a finite impact parameter 
between the two incoming particles in the collision (see also the footnote$^1$) and to maintain unitarity in the asymptotic
high-energy regime, $\sqrt{s}\to \infty$, with fixed coupling constant $\lambda={\rm fixed}\ll 1$.}
	
	We will use the notation 
	$\ket{1^*}\to \ket{n}$ for the process \eqref{eq:1pH}, where $\ket{1^*}$ denotes a highly-virtual particle that, for example, 
	can be produced as an intermediate state in a high-energy collision, and $\ket{n}$ denotes an $n$-particle final state. 
	We are interested in the regime of high-multiplicity $(n\gg 1)$ in a weakly coupled theory $(\lambda \ll 1)$  with 
	$\lambda n$~held at a fixed value that we ultimately take to be large.

Our discussion in this section follows the construction in \cite{Son:1995wz} and also borrows from 
Refs.~\cite{Khlebnikov:1990ue,Rubakov:1991fb,Rubakov:1992az,Libanov:1997nt}.

\medskip	
\subsection{Setting up the problem}
\label{sec:settingup}
\medskip
	
Consider a real scalar field $\phi(x)$ in $(d+1)$-dimensional spacetime, with the Lagrangian,
	\begin{equation}
	\label{eq:lag}
	\mathcal{L}(\phi)=\dfrac{1}{2}(\partial_\mu\phi)^2-\dfrac{1}{2}m^2\phi^2- \mathcal{L}_{\rm int}(\phi),
	\end{equation}
where $\mathcal{L}_{\rm int}$ is the interaction term. The two simplest examples are the $\phi^4$ model in the unbroken phase,
with $\mathcal{L}_{\rm int} =\dfrac{\lambda}{4}\phi^4$, and the theory \eqref{eq:L} with the spontaneously broken $Z_2$ symmetry,
\[
{\cal L} \,= \, 
\frac{1}{2}\, \partial^\mu h \, \partial_\mu h\, -\,  \frac{\lambda}{4} \left( h^2 - v^2\right)^2
\,. \label{eq:L2}
\]
The theory \eqref{eq:L2} has a non-zero vacuum expectation value $\langle h\rangle=v$ and we introduce the shifted field
of  mass $m = \sqrt{2\lambda}\,v$,
\[
\phi(x)=h(x)-v \,, \qquad m = \sqrt{2\lambda}\,v\,.
\label{eq:shift}
\]
Our considerations in this section are general and the expressions that follow, unless stated otherwise, will be written in terms of the
manifestly VEV-less field $\phi$ with the Lagrangian \eqref{eq:lag}. If the VEV is non-zero, as in the model \eqref{eq:L2}, the $\phi$ field is defined by subtracting the VEV
from the original field via \eqref{eq:shift}.
	
Our main goal is to derive the probability rate or the `crosssection' for the process where a single highly virtual off-shell particle
produced as an intermediate state in a high-energy collision, or alternatively a few energetic 
on-shell particles in the initial state $\ket{\phi_i}$, produce an $n$-particle final state with $n\gg 1$. Most importantly,
this probability rate should be written in a form suitable for a semiclassical treatment. In other words, the functional integral representation 
for the multiparticle rate should be calculable by some appropriate incarnation of the steepest descent method.

We begin by specifying the initial state. Instead of using the coherent state $\ket{\{a\}}$ as we have done in the previous section, we now
assume that the initial state is prepared by acting with a certain local operator $\hat{\cal O}(x)$ on the vacuum,
\begin{equation}
	\ket{\phi_i}\,=\, \hat{\cal O}(x)\ket{0}.
\label{eq:Xdef}	
\end{equation}
We will see that the operator $\hat{\cal O}(x)$ will act as a local injection of energy (or more precisely the virtuality characterising the off-shell state
$\ket{\phi_i}$)
into the vacuum state $\ket{0}$ at the spacetime point $x$. 
From now on, and without loss of generality, we will place the operator insertion point $x$ at the origin, $x=0$.

	In a general local QFT, any field ${\cal O}(x)$ that is sharply defined at a point $x$ is in fact an operator-valued distribution.
	In order to define an operator one has to smear the field with a test function
	that belongs to an appropriate set of well-behaved smooth functions with finite 
	support in spacetime~\cite{Jaffe:1967nb}.
	This implies that  ${\cal O}(x)$ in \eqref{eq:opdef2} should be averaged with a test function $g(x)$.
	The operator localized in the vicinity of a point $x$ is,
	\[
	{\cal O}_g (x)\,=\, \int d^4 x' \, g(x'-x)\, {\cal O}(x')\,,
	\]
	and the prescription \eqref{eq:Xdef} for defining the initial state should be refined~\cite{Khoze:2018qhz} using,
	\[
	\ket{\phi_i}\,=\,  {\cal O}_g(0) \ket{0}\,=\, \int d^4 x' \, g(x')\, {\cal O}(x') \, \ket{0}
	\,.
	\label{eq:opdef11}	
	\]
	This gives a well-defined state in the Hilbert space. For the rest of this section we will ignore
	the averaging of the operators with the test functions. Their effect can be recovered from the distribution-valued 
	rate ${\cal R}_n (\sqrt{s}=E)$ that we will concentrate on from now on and refer the reader to~\cite{Khoze:2018qhz}
	for more detail on the topic of the operator smearing.

\medskip

For a given final state, $\ket{\phi_f}$, one can isolate the parts with the desired energy and multiplicity using projection operators 
$\hat{P}_E$ and $\hat{P}_n$
on states with the fixed energy $E$ and particle number $n$. 
The probability rate ${\cal R}_n (E)$ for a transition between the initial state and the final state with the energy $E$ and particle number $n$ is given by
the square of the matrix element of the S-matrix with the projection operators $\hat{P}_E$ and $\hat{P}_n$,
\[
\braket{\phi_f|\hat{P}_E\hat{P}_n\hat{S}|\phi_i}\,=\,
\braket{\phi_f|\hat{P}_E\hat{P}_n\hat{S}\,\hat{\cal O}|0},
\label{eq:mPPSO}
\]
integrated over the final states phase space, $\int d\phi_f \ket{\phi_f}\bra{\phi_f}$, to give
\begin{equation}
	\begin{split}
{\cal R}_n(E) 
	=& \int d\phi_f \braket{0|\hat{\cal O}^\dagger\hat{S}^\dagger\hat{P}_E\hat{P}_n|\phi_f}\braket{\phi_f|\hat{P}_E\hat{P}_n\hat{S}\hat{\cal O}|0}\\
	 &\, = \braket{0|\hat{\cal O}^\dagger\hat{S}^\dagger\hat{P}_E\hat{P}_n\hat{S}\hat{\cal O}|0}.
	\end{split}
\label{eq:compact}	
\end{equation}

It is clear that neither the initial state $\ket{\phi_i}= \hat{\cal O}\ket{0}$ nor the final state $\bra{\phi_f}$ in the matrix element 
\eqref{eq:mPPSO} are states of definite energy. The projection operator $\hat{P}_E$ resolves this problem by projecting onto the fixed energy 
states. This applies to both, the initial and the final states, since the energy $E$ is conserved in the transition amplitude and hence is the same
in the initial and the final states. This implies that $\hat{P}_E \hat{\cal O}\ket{0}$  selects the initial state with the energy equal to $E$ which
is injected into the vacuum state by the operator $\hat{P}_E \hat{\cal O}$ at the point $x=0$ -- in agreement with what we have already stated above.

The particle number, on the other hand, is not a conserved quantity, it is computed only for asymptotic free states and is equal to $n$ in the 
final state $\bra{\phi_f}\hat{P}_n$. In the initial state we want to have the particle number $n_i$ to be small, 1 or 2, to correspond a scattering process
`few~$\to$~many'. The selection of $n_i$ is achieved by a judicious choice of the operator ${\cal O}$ in the definition of the initial state.
We will see below that the requirement that the semiclassical approximation is applicable to the functional integral representation of the transition
rate in \eqref{eq:compact} would allow for the operators of the form,
\[ 
\hat{\cal O} \,=\, j^{-1} \, e^{j \phi(0)}\,,
\label{eq:opdef}
\]
where $j$ is a constant. 
To select the single-particle initial state $\langle 0| \phi(0)$, the limit $j\to 0$ will ultimately be taken in the computation of the
 probability rate \eqref{eq:compact} along with the semiclassical limit $\lambda \to 0$.\footnote{We will explain in 
 section~\ref{sec:validity} in the discussion below Eq.~\eqref{eq:resc} that the $j\to 0$ limit should be taken such that 
 $j/\lambda \sim 1$ to to guarantee that the number of initial particles is $\sim 1$ while the number of final state particles is 
 $n \sim 1/\lambda$. }
 Equation \eqref{eq:opdef} defines the local operator used by Son in \cite{Son:1995wz},
which we too will use (we will have more to say about this prescription in sections~\ref{sec:validity} and~\ref{sec:5.2.3}).

\medskip

To proceed with the determination of the multiparticle rate in \eqref{eq:compact}, we need expressions for the projection operators
$\hat{P}_E$ and $\hat{P}_n$. This is where the coherent states formalism is useful. The kernel of $\hat{P}_E$ is given by,
\begin{equation}
	P_E(b^*,a):= \braket{\{b\}|\hat{P}_E|\{a\}} \,=\,
	\int \frac{d\xi}{2\pi}\, \exp\left[- iE\xi+\int \dk\, b^*_\bec{k}a_\bec{k}e^{i\omk\xi} \right].
\label{eq:PEdef}	
\end{equation}
To derive this expression, consider applying the delta function,
\[
\delta(\hat{H}_0-E)\,=\,\int \frac{d\xi}{2\pi}\, e^{i(\hat{H}_0-E)\xi}\,,
\] 
to the coherent state $\ket{\{a\}}$,
\[
\delta(\hat{H}_0-E)\,\ket{\{a\}}\,=\, \int \frac{d\xi}{2\pi}\, e^{-iE\xi}\,\ket{\{a e^{i\omega t}\}}\,
\]
and then convoluting this with the state $\bra{\{b\}}$. 
Using \eqref{eq:abprod} we find,
\begin{equation}
\braket{\{b\}|\delta(\hat{H}_0-E)|\{a\}} \,=\,
\int \frac{d\xi}{2\pi}\, 	
	 \exp\left[- iE\xi+\int \dk\,b^*_\bec{k}a_\bec{k}e^{i\omk\xi} \right],
\label{eq:PEderiv}	
\end{equation}
which is equivalent to \eqref{eq:PEdef}.

Using the same line of reasoning we also get the kernel of the projection operator $\hat{P}_n$,
\begin{equation}
	P_n(b^*,a):= \braket{\{b\}|\hat{P}_n|\{a\}} \,=\,
	\int \frac{d\eta}{2\pi}\, \exp\left[- i n \eta+\int \dk\, b^*_\bec{k}a_\bec{k}e^{i  \eta} \right].
\label{eq:Pndef}	
\end{equation}
As seen in \eqref{eq:cscombo}, the kernel of a product of two operators is the convolution of their individual kernels, such that the combined energy and multiplicity projector is given by,
\begin{equation}
\label{eq:prodproj}
\begin{split}
	P_EP_n(b^*,a) =& \int d(\{c^*\},\{c\})\, e^{-\int \dk\,c^*_\bec{k}c_\bec{k}}P_E(b^*,c)P_n(c^*,a) \\
	=& \int d(\{c^*\},\{c\}) \,\frac{d\xi}{2\pi} \frac{d\eta}{2\pi}\,e^{-iE\xi-in\eta}
	\exp\left[\int \dk\left(-c^*_\bec{k}(c_\bec{k}-a_\bec{k}\,e^{i\eta})+b^*_\bec{k}c_\bec{k}e^{i\omega_\bec{k}\xi} \right) \right]
\\
	=& \int d(\{c\})\,\frac{d\xi}{2\pi} \frac{d\eta}{2\pi}\, e^{-iE\xi-in\eta}\delta(\{c\}-\{ae^{i\eta}\})
	\exp\left[\int \dk\,b^*_\bec{k}c_\bec{k}e^{i\omega_\bec{k}\xi}  \right]
\\
         =&\int \frac{d\xi}{2\pi} \frac{d\eta}{2\pi}\, \exp\left[ -iE\xi-in\eta +\int \dk\,b^*_\bec{k}a_\bec{k}e^{i\omega_\bec{k}\xi+i\eta} \right],         
	\end{split}
	\end{equation}
	where the delta function $\delta(\{c\}-\{ae^{i\eta}\})$ is shorthand for an infinite product of delta functions for the infinite set $\{c\}$ such that, after integration, $c_\bec{k}\to a_\bec{k} e^{i\eta}$ for all $\bec{k}$.
The expression on the last line of \eqref{eq:prodproj} can also be derived instantly without considering the convolution of two individual kernels,
by inserting the product of the two delta functions into the overlap $\braket{\{b\}|\{a\}}$.

After inserting the coherent state (over-)completeness relation \eqref{eq:csid}, the last line of our expression for the rate \eqref{eq:compact}
gives,
	\begin{equation}
	\label{eq:sig}
	\begin{split}
{\cal R}_(E)  
=& \int d(\{b^*\},\{b\})e^{-\int \dk\,b^*_\bec{k}b_\bec{k}}\braket{0|\hat{\cal O}^\dagger\hat{S}^\dagger\csbr\csbl\hat{P}_E\hat{P}_n\hat{S}\hat{\cal O}|0}\\
	=& \int d(\{b^*\},\{b\})e^{-\int \dk\,b^*_\bec{k}b_\bec{k}} \\
	\times& [SA(b^*,0)]^*\\
	\times& P_EP_nS {\cal O}(b^*,0),
	\end{split}
	\end{equation}
	where we have identified the two matrix elements as kernels of product operators in the coherent state formalism.
	
	Given that $\hat{\cal O}=\hat{\cal O}[\hat{\phi}(0)]$, we can simply absorb it into the path integral during the derivation of $S(b^*,a)$ seen in \refsec\ref{sec:skern}, 
	\begin{equation}
	\label{eq:SA}
	S{\cal O}(b^*,a) \,=\, \lim\limits_{t_f,t_i\to\pm\infty}\int d\phi_fd\phi_ie^{B_i(\phi_i;a)+B_f(\phi_f;b^*)}\int\mathcal{D}\phi \,
	{\cal O}[\phi]\,e^{iS[\phi]_{t_i}^{t_f}}.
	\end{equation}
	As in \refsec\ref{sec:skern}, the functional integral satisfies the boundary conditions in \eqref{eq:funcbcs}. The definitions of $B_i(\phi_i;a)$ and $B_f(\phi_f;b^*)$ are given in \eqref{eq:BTs}.
	
	We now turn to the incorporation of the projection operators. Following the logic used in deriving the product kernel $P_EP_n(b^*,a)$ in \eqref{eq:prodproj}, we deduce,
	\begin{equation}
	\label{eq:PPSA}
	\begin{split}
	P_EP_nS {\cal O}(b^*,a) =& \int d(\{c^*\},\{c\})e^{-\int \dk\,c^*_\bec{k}c_\bec{k}}P_EP_n(b^*,c)S{\cal O}(c^*,a) \\
	=& \int d(\{c^*\})\,\frac{d\xi}{2\pi} \frac{d\eta}{2\pi}\, e^{-iE\xi-in\eta}\delta(\{c^*\}-\{b^*e^{i\omk\xi+i\eta}\})S{\cal O}(c^*,a)\\
	=& \int \frac{d\xi}{2\pi} \frac{d\eta}{2\pi}\, e^{-iE\xi-in\eta}S{\cal O}(b^*e^{i\omk\xi+i\eta},a).
	\end{split}
	\end{equation}
We now have all the ingredients needed to write the master equation for $\mathcal{R}(n,E)$. Combining Eqs. \eqref{eq:sig} and \eqref{eq:PPSA}, we find,
	\begin{equation}
	\begin{split}
	\mathcal{R}(n,E)=& \int d(\{b^*\},\{b\})\,\frac{d\xi}{2\pi} \frac{d\eta}{2\pi}\,   \\
	\times& \exp\left[  -iE\xi -in\eta -\int \dk\, b^*_\bec{k}b_\bec{k}  \right]\\
	\times& [S{\cal O}]^*(b,0)\\
	\times& S{\cal O}(b^*e^{i\omk\xi+i\eta},0),
	\end{split}
	\end{equation}
where we have set $a=0$ to reduce the coherent state $\ket{\{a\}}$ to the vacuum as required by \eqref{eq:sig}. Making changes of variable,
	\begin{equation}
	b^*\to b^*e^{-i\omk\xi-i\eta},\qquad 
	\xi\to -\xi, \qquad
	\eta \to -\eta,
	\end{equation}
	gives,
	\begin{equation}
	\begin{split}
	\mathcal{R}(n,E) =& \int d(\{b^*\},\{b\})\,\frac{d\xi}{2\pi} \frac{d\eta}{2\pi}\,   \\
	\times& \exp\left[  iE\xi +in\eta -\int \dk\, b^*_\bec{k}b_\bec{k}e^{i\omega_\bec{k}\xi+i\eta}  \right]\\
	\times& [S{\cal O}]^*(b,0)\\
	\times& S{\cal O}(b^*,0).
	\end{split}
	\end{equation}

\medskip

\noindent Inserting the definition of $S{\cal O}(b^*,0)$ and the choice of operator $\hat{\cal O}$ in \eqref{eq:SA} finally yields the master equation for 
        $\mathcal{R}(n,E)$ in the form given in \cite{Son:1995wz}, which we write below specifiyng all integration variables in the functional integrals
        (and dropping factors of $1/(2\pi)$ and $1/j$):
\begin{equation}
\label{eq:master}
	\begin{split}
\mathcal{R}(n,E) =& \lim\limits_{t_f,t_i\to\pm\infty}\int d\xi d\eta\,db^*_{\bec{k}}\,d b_{\bec{k}}\,
 d\phi_i(\bec{x})\,d\phi_f(\bec{x})\,\mathcal{D}\phi(\bec{x},t)\, d\varphi_i(\bec{x}) \, d\varphi_f(\bec{x})\, \mathcal{D}\varphi(\bec{x},t) \\
	&\qquad \times \exp\left[  iE\xi +in\eta -\int \dk \,b^*_\bec{k}b_\bec{k}e^{i\omk\xi+i\eta} + \Xi \, \right],
	\end{split}
	\end{equation}
	with the functional $\Xi \,=\,  \Xi (\phi_i,\phi_f, \phi; \varphi_i,\varphi_f,\varphi; b^*_\bec{k},b_\bec{k};t_i,t_f)$ defined by,
\begin{equation}
\label{eq:master2}
\begin{split}
	\Xi \,=\,&
	 B_i(\phi_i;0)+B_f(\phi_f;b^*) +[B_i(\varphi_i;0)]^*+[B_f(\varphi_f;b^*)]^* \\
	 & \qquad \qquad \qquad + iS[\phi]_{t_i}^{t_f}-iS[\varphi]_{t_i}^{t_f}+j\phi(0)+j\varphi(0).
\end{split}	 
\end{equation}

\medskip

\noindent Equations \eqref{eq:master}-\eqref{eq:master2} specify the multi-dimensional (functional and ordinary) integral we need to compute or estimate 
in order to determine the rate for multiparticle production processes. We will do so by method of steepest descent, i.e. the semiclassical approximation,
and its validity will be justified in the following section by bringing the large parameter (the equivalent of $1/\hbar$ in the simple WKB method) 
out in front of all terms appearing in the exponent in  \eqref{eq:master}-\eqref{eq:master2}.

It will be useful to keep in mind the explicit forms of the four boundary terms. These follow from \eqref{eq:BTs} and are given below,
	\begin{equation}
	\label{eq:masterbt}
	\begin{split}
	B_i(\phi_i;0)&=-\dfrac{1}{2}\int \dk\, \omk \tilde{\phi}_i(\bec{k})\tilde{\phi}_i(-\bec{k}),\\
	B_f(\phi_f;b^*)&=-\dfrac{1}{2}\int \dk\, b^*_\bec{k} b^*_{-\bec{k}}e^{2i\omk t_f}-\dfrac{1}{2}\int \dk\, \omk \tilde{\phi}_f(\bec{k})\tilde{\phi}_f(-\bec{k})+\int \dk\sqrt{2\omk}b^*_\bec{k}\tilde{\phi}_f(\bec{k})e^{i\omk t_f},\\
	[B_i(\varphi_i;0)]^*&=-\dfrac{1}{2}\int \dk\, \omk \tilde{\varphi}_i(\bec{k})\tilde{\varphi}_i(-\bec{k}),\\
	[B_f(\varphi_f;b^*)]^*&=-\dfrac{1}{2}\int \dk\, b_\bec{k} b_{-\bec{k}}e^{-2i\omk t_f}-\dfrac{1}{2}\int \dk\, \omk \tilde{\varphi}_f(\bec{k})\tilde{\varphi}_f(-\bec{k})+\int \dk\sqrt{2\omk}b_\bec{k}\tilde{\varphi}_f(-\bec{k})e^{-i\omk t_f}
	\end{split}
	\end{equation}
Recall that tildas denote the spatial Fourier transformations of the fields defined in \eqref{eq:FTdef2}.

	\subsection{Application of steepest-descent method}
	
	\subsubsection{Discussion of the validity of steepest descent/semiclassical approach}
\label{sec:validity}

	In quantum mechanics, steepest descent methods are very useful, as one often obtains integrals of exponentials with a $1/\hbar$ prefactor in the exponent. The key to the validity of the method is that one can consider the $\hbar\to0$ limit. Of course $\hbar$ is a dimensionful parameter and one needs to identify the appropriate large dimensionless factor in front of the functions in the exponent that goes as $1/\hbar$.
	
In quantum field theory, the semiclassical approximation in the simplest scenarios is achieved by rescaling all fields in the action $S$ such that 
$S \propto 1/\lambda$ where $\lambda$ is the coupling constant.
The relevant limit is the weak-coupling limit $\lambda\to 0$. This reasoning holds for instanton calculations of Green functions and amplitudes in gauge theories~\cite{Rajaraman,Dorey:2002ik}. In this case one rescales the gauge fields $A_\mu \to gA_\mu$, where $g$ is the gauge coupling and, as a result, the microscopic action of the theory $S= \frac{1}{g^2} \int d^4x\, {\rm tr} F_{\mu\nu}F^{\mu\nu}\, \propto \frac{1}{g^2}$, which is the equivalent of $1/\lambda$. If scalar fields are 
also present in the theory, then one rescales them with $\sqrt{\lambda}$ and the relevant terms in the action scale as $1/\lambda$ which is taken to be
$\propto 1/g^2$ in the common weak-coupling limit $\lambda \to 0$ , $g^2 \to 0$. These semiclassical weak-coupling limits can be further combined 
with the large number of colours limit ($N_c \to \infty$) in certain scenarios, allowing one to compute all multi-instanton contributions to the correlators 
relevant in the context of the AdS/CFT correspondence as reviewed in \cite{Dorey:2002ik}.

\medskip

The main lesson concerning the applicability of the steepest descent approximation to the multiple integrals we want to evaluate, is that one needs to
arrange for all relevant terms appearing in the exponent of the integrand to contain the same large multiplicative factor. By \emph{relevant} terms we mean the terms that have a potential to influence the saddle-point solution, which will provide the dominant contribution to the integral. To be on the safe side, we can demand that \emph{all} terms in the exponent contain this large factor. Once this is achieved, we search for an extremum of the
function in the exponent -- called the stationary solution or the the saddle-point -- and expand all the integration variables in the integrand around this
extremum. Following such an expansion, one would usually compute the integral by integrating over the fluctuations around this extremum. This is equivalent to using a 
background perturbation theory in the background of the saddle-point solution. In reality, to obtain the leading-order result, it is sufficient to just compute
the exponent of the integrand on the saddle-point configuration. The leading-order corrections come from integrating over quadratic fluctuations 
around the saddle-point. These are Gaussian integrals and determine the prefactor in front of the exponent. Each subsequent order in fluctuations is suppressed by an extra power of $({\rm large\,parameter})^{-1/2} \ll 1$ on general dimensional grounds.

\medskip

In our case we have a priori three large dimensionless parameters, $1/\lambda$, $n$ and $E/m$. 
The first one is an internal parameter of the theory, while the second and the third are
process-dependent -- they arise from specifying the final state to contain $n\gg1$ particles at high energies $E\gg m$. In a sense, the entire rationale for developing the
coherent state approach that led to the expression for the rate in the form 
was to pull the dependence on $n$ and $E$ from the final state into the exponent of the rate. Essentially the quantity in the exponent on the 
second line of \eqref{eq:master}
can be thought of as an effective action which depends on three large parameters, $1/\lambda$, $n$ and $E$. Most important for the
validity of the steepest descent approach, is that no $n$- and $E$-dependence appears elsewhere, in particular not in the integration variables: the number of integrations (functional and ordinary ones) is fixed and independent of $n$, $E$ or $\lambda$. 

Now, for the application of the steepest descent method we need to have just one large parameter. For that reason the appropriate semiclassical limit 
is defined where $n$ and $1/\lambda$ are of the same order, such that their ratio is held fixed in the limit $n\to \infty$.
Indeed, it is easy to see that 
$n=\lambda n /\lambda$ is $ \sim 1/\lambda$ for $\lambda n={\rm fixed}.$ 
Similarly we have to hold $n \propto E/m$.
Thus the steepest descent approximation to the integral \eqref{eq:master} is justified in the
weak-coupling -- large-$n$ -- high-$E$ semiclassical limit:
\[ \lambda \to 0\,, \quad n\to \infty\,, \quad {\rm with}\quad
\lambda n = {\rm fixed}\,, \quad \varepsilon ={\rm fixed} \,.
\label{eq:limit}
\]
Here $\varepsilon$ denotes the average kinetic energy per particle per mass in the final state,
\[ \varepsilon \,=\, (E-nm)/(nm)\,.
\label{eq:varepsdef}
\]
Holding $\varepsilon$ fixed implies that in the large-$n$ limit we are raising the total energy linearly with $n$.
Note that there is no $E\to \infty$ appearing in the limit \eqref{eq:limit}. The variable $E/(nm)$ is traded for $\varepsilon$ using
\eqref{eq:varepsdef} and held fixed.
 
We further note that the perturbation theory in the background of the saddle-point solution has conceptually different conclusions from the usual perturbation theory in a trivial background. Even though the perturbative 
corrections in both cases are suppressed by powers of $\lambda$, in the case of the steepest descent method, these corrections cannot be enhanced 
by powers of $n$. As we mentioned already, in our approach $n$ and $E/m$ are large parameters of the same order as $1/\lambda$, and the hypothetical contribution $\sim \lambda n$ cannot appear as a perturbative order-$\lambda$ correction -- it should instead be a part of the leading-order
result. This is different from the usual perturbation theory in which $n$ can arise as a combinatorial enhancement of the order-$\lambda$ perturbative corrections. So it should not come as a surprise that the steepest descent, or equivalently the semiclassical method, is a non-perturbative 
computation, with controlled corrections in the semiclassical limit that are suppressed by powers of $1/n$, $\lambda$ and $m/E$.

\medskip

We now finally discuss the scaling of the exponent in \eqref{eq:master} with the large parameter.
For the semiclassical method to be applicable, all terms in the exponent must be the same order in $1/\lambda$
in the limit \eqref{eq:limit}. 
To achieve this we rescale the fields and coherent state variables, as well as the source $j$ coming from the operator ${\cal O}$ insertion
by $1/\sqrt{\lambda}$,
\[
\{\phi,\varphi,b^*_\bec{k},b_\bec{k}, j \}
\,\to \, \frac{1}{\sqrt{\lambda}}
 \{\phi,\varphi, b^*_\bec{k},b_\bec{k}, j \} .
 \label{eq:resc}
\]
Taking into account that $n\sim E \sim1/\lambda$, we see that the entire exponent in \eqref{eq:master} now 
scales as  $1/\lambda$ in the limit \eqref{eq:limit} as required for the validity of the steepest descent approach.
However, this scaling implies that the source term in the operator \eqref{eq:opdef} used to produce the initial state is $j/\sqrt{\lambda}$ and,
\[
\ket{\phi_i}\,=\,  e^{\frac{j}{\sqrt{\lambda}}\, \frac{\phi(0)}{\sqrt{\lambda}}}\, \ket{0}   \,=\, e^{\frac{j}{\sqrt{\lambda}}\, \phi(0)_{\rm rescaled}}\, \ket{0}   \,.
\label{eq:opdef2}
\]
This is somewhat problematic as the operator in terms of the rescaled $\phi$ now explicitly depends on $\lambda$.
The initial state is some semiclassical state with the mean particle number $\langle n_i \rangle \sim j/\lambda$ rather than 
being a single-particle state. As noted in the original papers \cite{Rubakov:1991fb,Son:1995wz}
that were developing this approach, this is the consequence of the non-semiclassical nature of the initial state with a single particle
or with few highly energetic particles rather than a large number of soft ones.
The resolution of the problem proposed in \cite{Rubakov:1991fb,Son:1995wz,Libanov:1997nt} is to continue applying the semiclassical 
i.e. the steepest descent approach to the integral in \eqref{eq:master} with the source $j/\sqrt{\lambda}$ where $j$ is a constant, and only after
establishing the saddle-point equations take the limit $j \to 0$.
In this case we effectively return to the single-particle initial state with $j/\lambda \sim 1$, but at the same time, the semiclassical method 
continues to be justified. Of course, this line of reasoning is not a proof, but at least it provides an unambiguous procedure for computation.
Furthermore, in this limit one ends up with an operator ${\cal O}$ that does not depend on $\lambda$ (or on $\hbar$ in quantum mechanics).
In the quantum mechanical case, it is known that the analogous semiclassical computation -- using the Landau WKB formulation -- gives the semiclassical exponent of the rate, $W=\log {\cal R}$, which does not
depend on the form of the operator used, in so far as the operator did not depend on $\hbar$ explicitly.

Perhaps the most important existing verification of this procedure is that following it Son has successfully reproduced in \cite{Son:1995wz}
the known results for the multiparticle rate at tree-level \cite{Libanov:1994ug} and in the resummed one-loop approximation 
\cite{Voloshin:1992nu,Smith:1992rq,Libanov:1994ug} without recourse to perturbation theory.
It was also demonstrated in \cite{Libanov:1995gh} based on a few calculable examples for $1\to n$ and $2\to n$ processes, that the semiclassical exponent $W=\log {\cal R}$
does not depend on the construction of the initial state and that the multiparticle amplitudes should be the same -- at the level of the exponent -- 
for all few-particle initial states. 

These computations 
were carried out in the regime of relatively low multiplicities where the fixed 
value of $\lambda n$ in \eqref{eq:limit}
is taken to be small. This is the regime where the comparison of the semiclassical method results~\cite{Son:1995wz,Libanov:1995gh}
with the tree-level and leading-order loop corrections in ordinary perturbation theory~\cite{Voloshin:1992nu,Smith:1992rq,Libanov:1994ug}
is meaningful. Of course the real usefulness of the semiclassical approach lies in applying it to the opposite regime of high multiplicities, where 
the rescaled multiplicity $\lambda n$ is taken to be large. This is the non-perturbative regime where currently no other predictions for the multiparticle
rates are known in QFT in 4 dimensions. 
Nevertheless, the semiclassical approach in the large $\lambda n$ limit can still be successfully tested in $(2+1)$ dimensions
against the known RG-resummed perturbative results \cite{Rubakov:1994cz}
in a regime where both approaches are valid. This was shown in \cite{Khoze:2018kkz}
and will be reviewed in section~\ref{sec:lowd}.

\medskip

From now on, we will take the $j\to 0$ prescription
as a constructive approach for applying the semiclassical method 
to the calculation of the $1 \to n$ processes following \cite{Son:1995wz}.
In summary: the semiclassical formalism is fully self-consistent for computing the multi-particle rate \eqref{eq:master} with the initial state
defined by \eqref{eq:opdef2}. To obtain the result for the probability rate of the $1 \to n$ processes we will take the limit $j\to 0$ 
after writing down the saddle-point equations that will follow from extremising the exponent in \eqref{eq:master}
in the next section.

\subsubsection{Finding the saddle-point}
\label{sec:findsp}
	
With all terms in the exponent in \eqref{eq:master} being of the same order with respect to the large semiclassical parameter
$1/\lambda$, we are ready to proceed with deriving the equations for its extremum. 
It is no longer necessary to use the rescaled fields \eqref{eq:resc}, as we are primarily interested in the leading-order semiclassical expression for the rate. Hence we will use the integral representation of the rate in the original form \eqref{eq:master}.
We also note that the saddle-point trajectory in the steepest descent method allows $\phi(x)$ to be complex,
 so from this point on we will have to take a little more care with the relationships between $\phi$ and $\phi^*$ in position and momentum space. 

Applying the steepest descent approach to the integral \eqref{eq:resc} we search for an extremum of,
\[
W\,=\, iE\xi +in\eta -\int \dk\, b^*_\bec{k}b_\bec{k}e^{i\omk\xi+i\eta} \,+\, \Xi(\phi_i,\phi_f, \phi; \varphi_i,\varphi_f,\varphi; b^*_\bec{k},b_\bec{k}).
\]
In principle, we should look for all extrema of this expression and then select the one which gives the dominant (i.e largest) contribution
to ${\cal R}_n(E)$ -- normally, this would be the one with the maximal value of $W$. More generally, one would sum over the contributions
to ${\cal R}_n(E) \propto e^{\,W}$ from all extrema. In what follows we will end up selecting a particular stationary point solution: the one with the highest symmetry between $\phi$ and $\varphi$ components, whose contribution gives the lower bound to the total
rate ${\cal R}_n(E)$.

The extrema or saddle-points are solutions of the equations $\delta_{\chi} W\,=0$, where the set
$\chi\,=\, \{\xi, \eta, \phi(\bec{x},t),\phi_i(\bec{x}),\phi_f(\bec{x}),\varphi(\bec{x},t),
\varphi_i(\bec{x}),\varphi_f(\bec{x}),b^*_{\bec{k}}, b_{\bec{k}}\}$ 
denotes all  integration variables.

Following \cite{Son:1995wz} we will look for a saddle-point solution for which $\xi$ and $\eta$ are purely imaginary (this corresponds to 
deforming the integration contours in $\xi$ and $\eta$ to pass through this complex saddle-point configuration -- the standard practice 
required in steepest descent). Keeping with Son's notation we change variables,
\begin{equation}
	\label{eq:ttheta}
	\xi =-iT,\qquad
	\eta = i\theta,
\end{equation}
and treat $T$ and $\theta$ as real variables.
We now vary $W$,
\[
W\,=\, ET - n\theta -\int \dk\, b^*_\bec{k}b_\bec{k}e^{\omk T-\theta} \,+\, \Xi(\phi_i,\phi_f, \phi; \varphi_i,\varphi_f,\varphi; b^*_\bec{k},b_\bec{k}),
\label{eq:Wdefn}
\]
with respect to,
\[
\chi\,=\, \{T, \theta, \phi(\bec{x},t),\phi_i(\bec{x}),\phi_f(\bec{x}),\varphi(\bec{x},t),
\varphi_i(\bec{x}),\varphi_f(\bec{x}),b^*_{\bec{k}}, b_{\bec{k}}
\} \,, \quad \frac{\delta W}{\delta \chi}\,=\, 0.
\]

Variations with respect to $T$ and $\theta$ give the equations for the $T$ and $\theta$ variables,
\eqref{eq:ttheta},
	\begin{align}
	\label{eq:sp5}
\partial_T W\,:&\qquad 
	E= \int \dk \,\omk b^*_\bec{k} b_\bec{k} e^{\omk T-\theta} \\
	\label{eq:sp6}
\partial_\theta W\,:&\qquad
	n= \int \dk \, b^*_\bec{k} b_\bec{k} e^{\omk -\theta}.
	\end{align}
Next we obtain the saddle-point equations for $\phi$, $\tilde{\phi}_i$, $\tilde{\phi}_f$ and $b^*_\bec{k}$,
\begin{align}
	\label{eq:sp1}
\frac{\delta W}{\delta \phi(x)}\,:&\qquad 
	\dfrac{\delta S}{\delta \phi(x)}\,=\, ij\delta^{d+1}(x)\\
	\label{eq:sp2}
\frac{\delta W}{\delta \tilde{\phi}_i(-\bec{k})}:&\qquad
	i\partial_{\,t_i} \tilde{\phi}_i(\bec{k})+\omk\tilde{\phi}_i(\bec{k})=0  \\
	\label{eq:sp3}
\frac{\delta W}{\delta \tilde{\phi}_f(-\bec{k})}:&\qquad
	i\partial_{\,t_f}\tilde{\phi}_f(\bec{k})-\omk\tilde{\phi}_f(\bec{k})+\sqrt{2\omk}\,b^*_{-\bec{k}}e^{i\omk t_f} =0 \\
	\label{eq:sp4}
\frac{\delta W}{\delta b^*_\bec{k}}:&\qquad
	-b_\bec{k}e^{\omk T- \theta}-b^*_{-\bec{k}}e^{2i\omk t_f}+\sqrt{2\omk}\, \tilde{\phi}_f(\bec{k})e^{i\omk t_f}=0 .
\end{align}
The first terms in \eqref{eq:sp2} and \eqref{eq:sp3} come from the boundary contributions to the action $S$ from total derivatives,
\begin{equation}
S_{\rm Boundary}[\phi_i,\phi_f]=\dfrac{1}{2}\int^{t_f}_{t_i}dt\int d^dx\, \partial_t(\phi\,\partial_t\phi)=\dfrac{1}{2}\int d^dx \,(\phi_f\partial_t\phi_f-\phi_i\partial_t\phi_i),
\label{eq:Bounddef}
\end{equation}
as explained in  Appendix~\ref{app:btsp} in more detail. The other terms arise rather straightforwardly from the rest of the expression in \eqref{eq:Wdefn}
	
	Unsuprisingly, equations analagous to \eqref{eq:sp1}-\eqref{eq:sp4} exist for $\varphi$, $\tilde{\varphi}_i$, $\tilde{\varphi}_f$ and $(b^*_\bec{k})^*$. Note that, \textit{a priori}, there is no need for $b_\bec{k}$ and $b_\bec{k}^*$ to be complex conjugates, nor is there any constraint on the complex phases of $\xi$ and $\eta$. Nevertheless, there exists a saddle-point for which $(b_\bec{k})^*=b^*_\bec{k}$, and $\xi$ and $\eta$ are purely imaginary (and thus $T$ and $\theta$ are purely real). We focus on this scenario, as Son does \cite{Son:1995wz}. 
With these assignments in mind, the final group of saddle-point equations give the equations for
the remaining field variables,
$\varphi$, $\tilde{\varphi}_i$, $\tilde{\varphi}_f$ and $b_\bec{k}$,
\begin{align}
\label{eq:sp1var}
\frac{\delta W}{\delta \varphi(x)}\,:&\qquad 
	\dfrac{\delta S}{\delta \varphi(x)}=-ij\delta^{d+1}(x)\\
	\label{eq:sp2var}
\frac{\delta}{\delta \tilde{\varphi}_i(-\bec{k})}:&\qquad
	-i\partial_{\,t_i} \tilde{\varphi}_i(\bec{k})+\omk\tilde{\varphi}_i(\bec{k})=0  \\
	\label{eq:sp3var}
\frac{\delta W}{\delta \tilde{\varphi}_f(-\bec{k})}:&\qquad
	-i\partial_{\,t_f}\tilde{\varphi}_f(\bec{k})-\omk\tilde{\varphi}_f(\bec{k})+\sqrt{2\omk}\, b_{\bec{k}}e^{-i\omk t_f} =0 \\
	\label{eq:sp4var}
\frac{\delta W}{\delta b_\bec{k}}:&\qquad
	-b^*_\bec{k}e^{\omk T-\theta}-b_{-\bec{k}}e^{-2i\omk t_f}+\sqrt{2\omk}\, \tilde{\varphi}_f(-\bec{k})e^{-i\omk t_f}=0 .
	\end{align}
It is not difficult to see that these equations \eqref{eq:sp1var}-\eqref{eq:sp4var} are satisfied by,
	\begin{equation}
	\label{eq:phivarphi}
	\varphi(t,\bec{x})=[\phi(t,\bec{x})]^*\quad \longrightarrow \quad
	\tilde{\varphi}(\bec{k})=[\tilde{\phi}(-\bec{k})]^*,
	\end{equation}
if $\phi$ satisfies its saddle-point equations \eqref{eq:sp1}-\eqref{eq:sp4}. 
We will focus on solutions for which \eqref{eq:phivarphi} holds from here on, which implies that we only need to solve the 
field equations \eqref{eq:sp1}-\eqref{eq:sp4}, and then trade the Lagrange multiplied variables $T$ and $\theta$ for the final state
energy and multiplicity, $E$ and $n$, using \eqref{eq:sp5}-\eqref{eq:sp6}.

\medskip	
	
Let us consider what the saddle-point equations imply for our scalar field, $\phi(x)$. Equation~\eqref{eq:sp1} gives the classical field equations with a singular point-like source at the origin $x=0$. We are searching for classical solutions in a $(d+1)$-dimensional theory that become
free fields at $t\to\pm\infty$ and thus the classical field in these limits must be a superposition of plane waves.\footnote{Indeed, in a
$(d+1)$-dimensional theory with $d\ge 2$, all time dependent solutions of non-linear equations of motion that are localised in space 
at time $t\sim 0$, must disperse and linearise at early and late times $t\to\pm\infty$. This property is satisfied, for example,
by spherically symmetric $O(d,1)$ finite-energy solutions
studied in Ref.~\cite{Farhi:1992pc}.}
Solving \eqref{eq:sp2} gives $\tilde{\phi}_i(\bec{k}) \sim e^{i\omk t_i}$ with no $e^{-i\omk t_i}$ components allowed. Using \eqref{eq:freeFT} to recover
the coefficient in front of $e^{i\omk t_i}$ we find,
\begin{equation}
\label{eq:bc1}
	\tilde{\phi}_i(\bec{k})\,=\,\dfrac{1}{\sqrt{2\omk}} \,a^*_{-\bec{k}}\,e^{i\omk t_i},
	\qquad t_i\to -\infty.
\end{equation}
This is the behaviour of $\tilde{\phi}(t,\bec{k})$ in the infinite past. 
The coefficient $a^*_{-\bec{k}}$ is an arbitrary Fourier component. 
Rearranging \eqref{eq:sp3} gives the behaviour in the infinite future,
\begin{equation}
	\label{eq:bc2}
	\tilde{\phi}_f(\bec{k})=\dfrac{1}{\sqrt{2\omk}}(b_\bec{k}e^{\omk T-\theta-i\omk {t_f}}+b^*_{-\bec{k}}e^{i\omk {t_f}}), \qquad t_f\to+\infty,
\end{equation}
which, as one would expect, satisfies \eqref{eq:sp4}. Thus Eqs.~\eqref{eq:sp2}-\eqref{eq:sp4} have simply provided boundary conditions 
at $t_i$ and $t_f$ for the solution ${\phi}(x)$ of the Euler-Lagrange equation \eqref{eq:sp1}.
Both boundary conditions correspond to a complex-valued saddle-point solution for ${\phi}(x)$, since the first
condition \eqref{eq:bc2} has $a^*_{\bec{k}}=0$, while the second boundary condition \eqref{eq:bc2} 
contains the factor $e^{\omk T-\theta}$ accompanying $b_\bec{k}$ that prevents the coefficients of $e^{\pm i\omk {t_f}}$ from being
complex conjugates of each other.
	
We can now compute the energy and the particle number on the saddle-point solution from its 
$t\to \pm \infty$ asymptotics \eqref{eq:bc1}-\eqref{eq:bc2}. At $t\to -\infty$ the energy and the particle number are vanishing since the 
corresponding solution contains only the $e^{i\omk t}$ harmonics. On the other hand at $t\to +\infty$, using the free-field solution \eqref{eq:bc2},
we find,
\[
E \,=\, \int \dk \,\, \omega_{\bf k}\, b_{\bf k}^* \, b^{}_{\bf k}\, e^{\omega_{\bf k}T-\theta}
\,, \qquad
n \,=\, \int \dk \,\, b_{\bf k}^* \, b^{}_{\bf k}\, e^{\omega_{\bf k}T-\theta}\,.
\label{eq:al4n}
\]
These are precisely the saddle-point equations \eqref{eq:sp5}-\eqref{eq:sp6}.
The energy of course is conserved by regular solutions at $t<0$ and at $t>0$ and changes discontinuously from $0$ to $E$ 
at the singularity at the origin $t=0={\bf x}$ induced by the $\delta$-function source in \eqref{eq:sp1}.

In other words, $E$ and $n$ are the energy and multiplicity of the solution $\phi$ for $t>0$. In the absence of the source, one expects the energy of the field to be conserved. Indeed, energy is conserved individually in the regions $t<0$ and $t>0$, where solutions contain no singularities and there is no source. However, at $t=0$, the point source will give a discontinuous jump in energy. This can be seen by looking at \eqref{eq:sp1}. The left-hand side reduces to an Euler-Lagrange term and so we have a second-order partial differential equation with a point source. We know from Green's function theory that we should expect the solution, $\phi$, to have a discontinuity in its first derivative in some direction at $x=0$. Suppose that this direction is the time direction such that by integrating \eqref{eq:sp1} over the region $-\epsilon\le t\le \epsilon$ for small $\epsilon$,
	\begin{equation}
	\int_{-\epsilon}^{+\epsilon}dt\dfrac{\delta S[\phi]}{\delta\phi(x)}=
	\left[ \dfrac{\partial\mathcal{L}}{\partial(\partial_t\phi)} \right]_{-\epsilon}^{+\epsilon}=(\partial_t\phi)_{+\epsilon}-(\partial_t\phi)_{-\epsilon}=\delta\dot{\phi}(0,\bec{x})=ij\delta^d(\bec{x}),
	\end{equation}
	with a dot indicating a time derivative, $\dot{\phi}=\partial_t\phi$. This gives an energy jump,
	\begin{equation}
	\begin{split}
\label{eq:Ejump}	
	\delta E =& \delta\left( \dfrac{1}{2}\int d^dx\dot{\phi}^2 \right)=\dfrac{1}{2}
	\int d^dx ([\dot{\phi}(+\epsilon,\bec{x})]^2-[\dot{\phi}(-\epsilon,\bec{x})]^2)
	\\
	=& \dfrac{1}{2}\int d^dx (\dot{\phi}_+-\dot{\phi}_-)(\dot{\phi}_++\dot{\phi}_-)
	=\int d^dx\dot{\phi}(0,\bec{x})\delta\dot{\phi}(0,\bec{x})=
	ij\dot{\phi}(0),
	\end{split}\end{equation}
	where $\dot{\phi}(0,\bec{x})$ is strictly the mean of the $t=\pm\epsilon$ values. Recall that the early-time assymptote \eqref{eq:bc1} has only positive frequency components and thus has zero energy. Therefore, the energy associated with the saddle-point field configuration undergoes a discontinous jump from $0$ to $E=\delta E=ij\dot{\phi}(0)$, when crossing $t=0$.

\subsubsection{The $j\to 0$ limit}
\label{sec:5.2.3}

After having found the defining equations for the saddle-point, we now want take the $j\to 0$ limit in order to obtain the rate 
for the $1^* \to n$ processes, as explained in section~\ref{sec:validity}.

Taking this limit amounts to more than just setting the source term to zero in the non-linear equations \eqref{eq:sp1} and \eqref{eq:sp1var}.
In fact, the solutions of the Euler-Lagrange equations without the source term must now become singular at the point $x=0$ 
in order to ensure ensure the jump in energy from $E=0$ at $t<0$ to $E\neq 0$ at $t>0$. This singular behaviour of the saddle-point solution 
is not an additional requirement, but a direct consequence of the saddle-point equations, which require the asymptotic behaviour 
\eqref{eq:bc1},\eqref{eq:bc2} with the jump in energy by $E$ in \eqref{eq:al4n}.
It follows from  \eqref{eq:Ejump} that the late-time energy is $E=ij\dot{\phi}(0)$. 
For $E$ to be fixed and non-vanishing, as is required for the scattering process of interest, we must require $\dot{\phi}(0)\to\infty$. In other words, the classical solution at the point $x=0$, as well as its derivative, 
are singular to ensure that,
\[
E\,=\, ij\dot{\phi}(x)|_{x=0}\,={\rm fixed}\,, \quad {\rm for} \quad
j\to 0 \quad {\rm with}\quad \dot{\phi}(x)|_{x=0}\to \infty\,.
\]
With these considerations in mind, we now take the limit $j\to 0$ in the saddle-point equations and in the exponent of the rate 
in \eqref{eq:Wdefn} and \eqref{eq:master2}.

	\subsubsection{Evaluation of integrand at saddle-point}
	\label{sec:5.2.4}
	
With the saddle-point equations found, we move onto imposing the saddle-point behaviour on the exponent  of the rate
in \eqref{eq:master}. The function in the exponent can be written as ({\it cf.} \eqref{eq:Wdefn} and \eqref{eq:master2}),
\begin{eqnarray}
W &=& ET - n\theta  \,+\,  iS[\phi] \,-\, iS[\varphi] 
\label{eq:Wdefn2}
 \\
&&+\, B_i(\phi_i;0)\,+\, B_f(\phi_f;b^*) \,+\, [B_i(\varphi_i;0)]^*\,+\, [B_f(\varphi_f;b^*)]^*
\, -\,\int \dk\, b^*_\bec{k}b_\bec{k}\,e^{\omk T-\theta}.
\nonumber
\end{eqnarray}
It is easy to see that the sum of the terms appearing on the second line in \eqref{eq:Wdefn2} is vanishing when evaluated on the 
saddle-point solution for ${\phi}_i$ and ${\phi}_f$ given in \eqref{eq:bc1}-\eqref{eq:bc2}, in the limit $t_i\to -\infty$ and 
$t_f\to +\infty$.
Indeed, 
\[ 
B_i(\phi_i;0) \,=\, 
-\dfrac{1}{2}\int \dk\, \omk \tilde{\phi}_i(\bec{k})\tilde{\phi}_i(-\bec{k})\,=\, \lim_{t_i\to-\infty}
\left(-\dfrac{1}{4} \int \dk\, a^*_{-\bec{k}}\, a^*_{\bec{k}}\, e^{2i\omk t_i} \right) \,=\, 0, 
\]
since only the negative frequency plane wave components are present in $\tilde{\phi}_i$.
We now evaluate the boundary term $B_f(\phi_f;b^*)$ at $t_f$ in the $t_f\to +\infty$ limit,
\begin{eqnarray}
	B_f(\phi_f;b^*)&=&
	-\dfrac{1}{2}\int \dk\, b^*_\bec{k} b^*_{-\bec{k}}e^{2i\omk t_f} 
	-\dfrac{1}{2}\int \dk\, \omk \tilde{\phi}_f(\bec{k})\tilde{\phi}_f(-\bec{k})
	+\int \dk\sqrt{2\omk}b^*_\bec{k}\tilde{\phi}_f(\bec{k})e^{i\omk t_f} \nonumber\\
 &\to&  0 
\,-\, \dfrac{1}{2} \int \dk\,\dfrac{\omk}{2\omk}\,2 b_\bec{k}e^{\omk T-\theta} b^*_{\bec{k}}
\,+\, \int \dk\sqrt{2\omk} \,  b^*_\bec{k} \dfrac{1}{\sqrt{2\omk}}b_\bec{k}e^{\omk T-\theta} \nonumber \\
&=& \dfrac{1}{2} \int \dk\, b_\bec{k}b^*_{\bec{k}}\,e^{\omk T-\theta},
\end{eqnarray}
and similarly, for $[B_i(\varphi_i;0)]^*$ we have the same result,
\[
[B_i(\varphi_i;0)]^*\,=\, \dfrac{1}{2} \int \dk\, b_\bec{k}b^*_{\bec{k}}\,e^{\omk T-\theta}.
\]
This implies that the sum of the boundary terms on the second line in \eqref{eq:Wdefn2} is vanishing, as 
already stated,
\[
B_i(\phi_i;0)\,+\, B_f(\phi_f;b^*) \,+\, [B_i(\varphi_i;0)]^*\,+\, [B_f(\varphi_f;b^*)]^*
\, -\,\int \dk\, b^*_\bec{k}b_\bec{k}\,e^{\omk T-\theta} \,=\, 0.
\]

Thus the expression in \eqref{eq:Wdefn2}, evaluated on the saddle-point solution, simplifies to,
\[
W \,=\, ET \,-\, n\theta \,+\,  iS[\phi] \,-\, iS[\phi]^* \,,
\label{eq:Wdefn3}
\]	
where we have identified $iS[\varphi] = iS[\phi]^*$ on our saddle-point solution.

Ultimately, as soon as the saddle-point solution $\phi(x)$ is found for all values of $t$, we obtain the saddle-point value of $\sigma(E,n)$ 
to exponential accuracy,
\begin{equation}
	{\cal R}_n(E)\,=\, e^{W(E,n)},
\label{eq:RnEWdef}	
\end{equation}
	with,
\begin{equation}
	\label{eq:W}
	W(E,n)=ET-n\theta - 2\mathrm{Im}S[\phi].
\end{equation}
	
Here the constant parameters $T$ and $\theta$ are the solutions of the corresponding saddle-point equations \eqref{eq:sp5}-\eqref{eq:sp6},
and $\phi(x)$ is the solution of the sourceless Euler-Lagrange equation, $\delta S/\delta \phi =0$, with the (initial and final) boundary conditions
\eqref{eq:bc1}-\eqref{eq:bc2}.

\medskip

It is also worth noting that 
the function $W(E,n)$ in \eqref{eq:W} is a function of $E$ and $n$ and does not depend explicitly on the 
$T$ and $\theta$ parameters.
$W(E,n)$ is in fact the Legendre transformation of $2\mathrm{Im}S(T,\theta)$,
where,
\begin{equation}
\label{eq:Legtrs}
	E=\dfrac{\partial\,  2\mathrm{Im}S}{\partial T},
	\qquad
	n=-\dfrac{\partial\, 2\mathrm{Im}S}{\partial \theta},
\end{equation}
and,
	\begin{equation}
	\dfrac{\partial W}{\partial E}=\,T, 
	\qquad 
	-\,\dfrac{\partial W}{\partial n}=\, \theta.
	\end{equation}
The equations \eqref{eq:Legtrs} defining $E$ and $n$ in terms of derivatives of the action of the classical field are in fact
equivalent to the already familiar equations for $E$ and $n$ in \eqref{eq:al4n}
computed on the asymptotics of $\phi$ at $t\to +\infty$.

\subsubsection{Summary of the approach in Minkowski spacetime}
\label{sec:sum1}

After the somewhat lengthy derivations in the previous sections it is worth summarising the 
resulting algorithm to compute the semiclassical rate~\cite{Son:1995wz}
in the context of the model \eqref{eq:L} with spontaneous symmetry breaking:
\begin{enumerate}
\item Solve the classical equation without the source-term,
\[
\frac{\delta S}{\delta h(x)}\,=\,0\,,
\label{eq:al1}
\]
by finding a complex-valued solution $h(x)$ with a point-like singularity at the origin
$x^\mu=0$ and regular everywhere else in Minkowski space. The singularity at the origin is selected by the location of the operator ${\cal O}(x=0)$.

\item Impose the initial and final-time boundary conditions,
\begin{eqnarray}
\lim_{t\to - \infty}\,h(x)  &=& v\,+\, 
\int \frac{d^d k}{(2\pi)^{d/2}} \frac{1}{\sqrt{2\omega_{\bf k}}}\,\, a^*_{\bf k}\, e^{ik_\mu x^\mu} \,,
\label{eq:al2}
\\
\lim_{t\to + \infty}\,h(x) &=& v\,+\, 
\int \frac{d^d k}{(2\pi)^{d/2}} \frac{1}{\sqrt{2\omega_{\bf k}}}\left( b_{\bf k}\,e^{\omega_{\bf k}T-\theta}
\, e^{-ik_\mu x^\mu}\,+\, b^*_{\bf k}\, e^{ik_\mu x^\mu}\right)\,.
\label{eq:al3}
\end{eqnarray}

\item Compute the energy and the particle number using the 
$t\to +\infty$ asymptotics of $h(x)$,
\[
E \,=\, \int d^d k \,\, \omega_{\bf k}\, b_{\bf k}^* \, b^{}_{\bf k}\, e^{\omega_{\bf k}T-\theta}
\,, \qquad
n \,=\, \int d^d k \,\, b_{\bf k}^* \, b^{}_{\bf k}\, e^{\omega_{\bf k}T-\theta}\,.
\label{eq:al4}
\]
At $t\to -\infty$ the energy and the particle number  are vanishing. 
The energy is conserved by regular solutions and changes discontinuously from $0$ to $E$ 
at the singularity at $t=0$.

 \item Eliminate the $T$ and $\theta$ parameters in favour of $E$ and $n$ using the expressions above.
 Finally, compute the function $W(E,n)$
  \[ 
W(E,n)  \,=\,
ET \,-\, n\theta \,-\, 2{\rm Im} S[h]
\label{eq:al5}
\]
on the set $\{h(x), T, \theta\}$ to obtain the semiclassical rate ${\cal R}_n(E) \,=\, \exp \left[ W(E,n)\right]$.
\end{enumerate}

           \subsubsection{Comment on more general saddle-points}
         
How can one be certain that only a single semi-classical solution dominates the 
multi-particle rate? To address this question let us recall the defining properties of the saddle-point solution we are after.

\begin{figure*}[t]
\begin{center}
\includegraphics[width=0.7\textwidth]{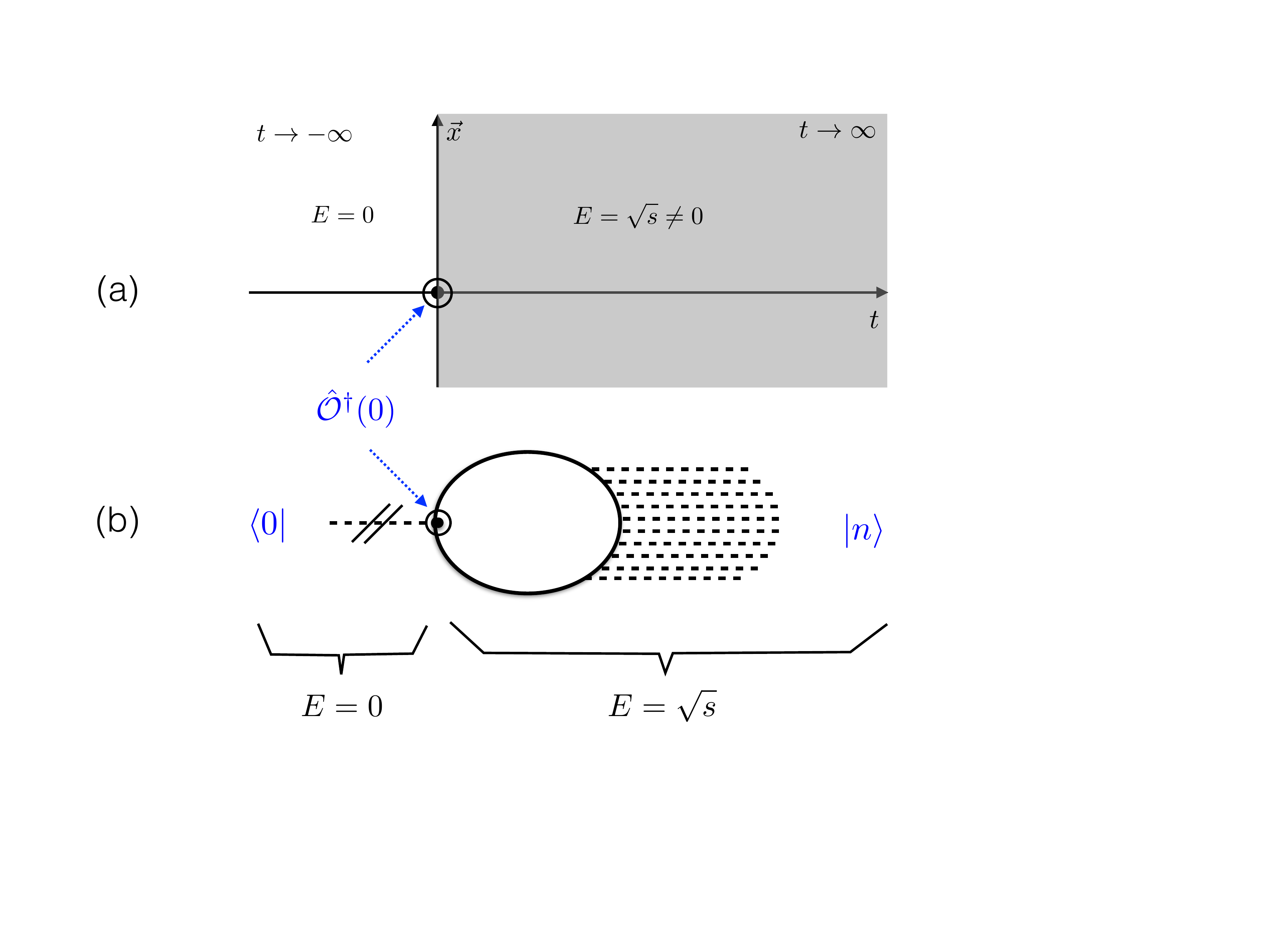}
\end{center}
\vskip-.5cm
\caption{{\bf Plot~(a)} shows a classical field configuration with a single jump in energy at the singular point at the origin
 $t=0=\vec{x}$ in
Minkowski space. {\bf Plot~(b)} depicts the contribution of such saddle-point configuration to the amplitude~\eqref{eq:ampl1PI} 
using a Feynman-diagram-type representation.
Saddle-point configurations with a single jump in energy contribute to the 1PI matrix elements, 
but not to the one-particle-reducible ones ({\it cf.}~Fig.~\ref{fig:two_sol}~(b)). }
\label{fig:one_sol}
\end{figure*}

Our specific solution to the boundary value problem in Minkowski space is characterised by a single
point-like singularity located at the origin, as shown in Fig.~\ref{fig:one_sol}~(a).
The energy $E$ of the solution is vanishing at all
 $t$ in the interval $-\infty <t<0$, and is non-vanishing and equal to $\sqrt{s}$ for $0<t<+\infty$. The solution 
 in Fig.~\ref{fig:one_sol}~(a)
 is singular at the origin, $x^\mu=0$. This is precisely the point where the operator ${\cal O}^\dagger$ is located 
 in the corresponding `Feynman diagram' contribution to the matrix element,
  \[ 
{\cal M}_{X\to n}^\dagger\,=\, {}^{\rm in}\langle X |n\rangle^{\rm out}_{\quad \rm 1PI}\,=\, 
 \langle 0|{\cal O}^\dagger(0)\,{\cal S}^\dagger|n\rangle_{\rm 1PI}\,,
 \label{eq:ampl1PI}
 \]
 as shown schametically in Fig.~\ref{fig:one_sol}~(b).
 The presence of this point-like singularity at the origin explains the jump 
 in the energy of the classical solution from $E=0$ to $E=\sqrt{s}$ when time passes from $t<0$ to $t>0$, and in 
Fig.~\ref{fig:one_sol}~(b) it corresponds to an injection of energy $E=\sqrt{s}$ by the local operator.

One can also consider multi-centred solutions, i.e. semi-classical saddle-points obtained by iterating 
 the solutions with a single singularity into more complicated saddle-points with multiple singularities. 
 These would result in multiple
 jumps in energy for each time the singularity is encountered. As such, these multi-centred saddle-points 
 would contribute to matrix elements with multiple insertions of local operators rather than the matrix element
 with a single ${\cal O}^\dagger$ in \eqref{eq:ampl1PI}. 
 
\begin{figure*}
\begin{center}
\includegraphics[width=0.8\textwidth]{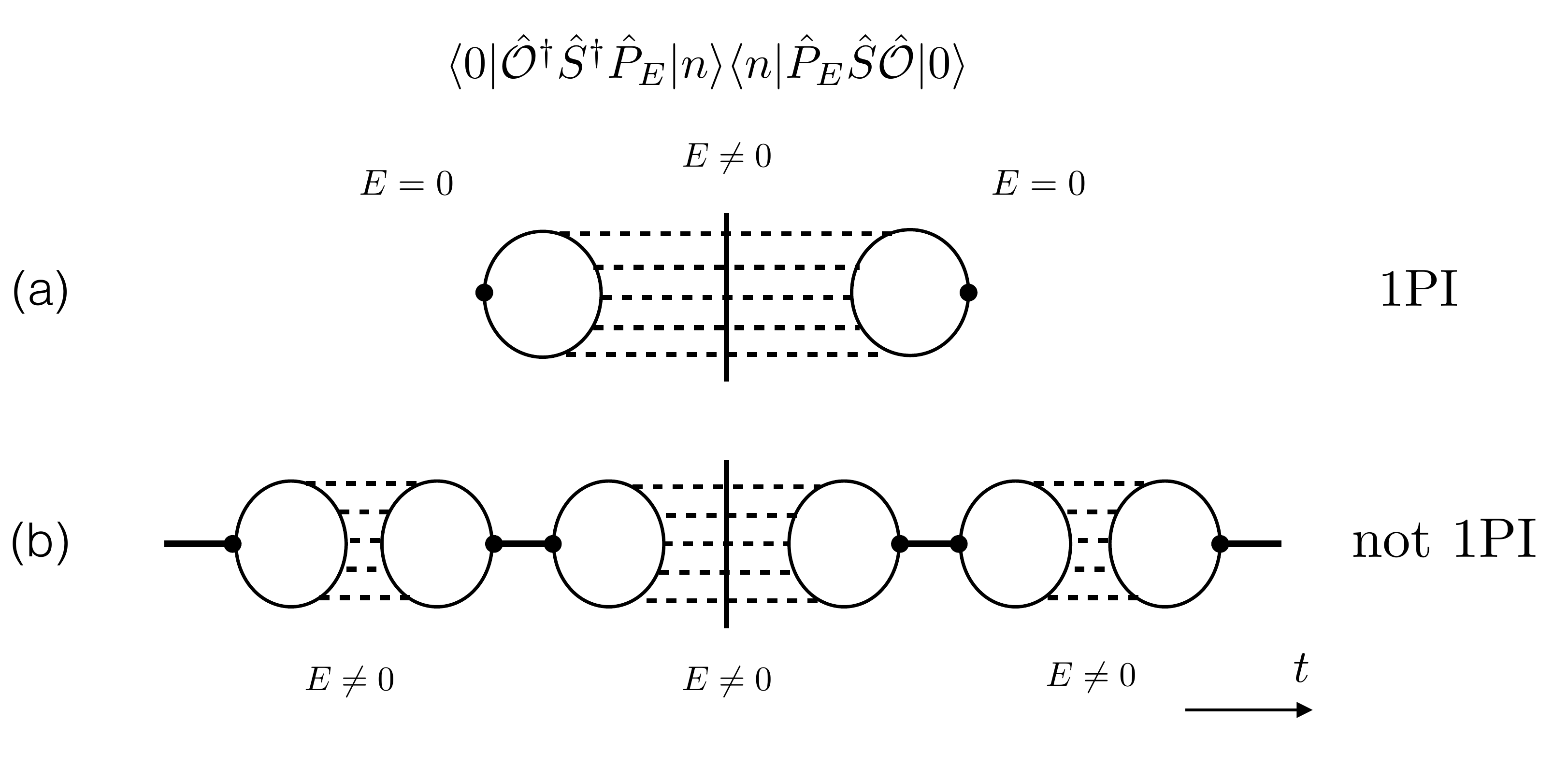}
\end{center}
\vskip-.5cm
\caption{Contributions to the $n$-particle rate~\eqref{eq:compact}. {\bf Plot~(a)} shows the one-particle-irreducible contribution
to $\braket{0|\hat{\cal O}^\dagger\hat{S}^\dagger\hat{P}_E\hat{P}_n\hat{S}\hat{\cal O}|0}_{ \rm 1PI}$ from a saddle-point 
configuration 
with a single energy jump.
{\bf Plot~(b)} shows one-particle-reducible contributions to the rate. They necessarily require multiple jumps from vanishing to 
non-vanishing energies and arise from saddle-point configurations with multiple 
singular points in Minkowski space.}
\label{fig:two_sol}
\end{figure*}

 Futhermore, by comparing contributions to the cross-section (i.e. to the matrix element squared) arising from
 the simple single-singularity solution in Fig.~\ref{fig:two_sol}~(a) to that of the multi-centred solution in Fig.~\ref{fig:two_sol}~(b), 
 one can see that the latter contribute to one-particle reducible, rather than 1PI matrix elements. 

 In this work we will concentrate on the contributions to \eqref{eq:ampl1PI} and will assume that the saddle-point solutions
 we will construct are the only
 saddle-points with a single point-like singularity in Minkowski space that contribute to these matrix elements. 
 If additional saddles of this type do exist, their contributions would have to be added to the ones we will be computing here.

\medskip
	\subsection{Reformulation of the boundary value problem}
\label{sec:5.3}
\medskip
To keep our discussion general, in subsections~\ref{sec:5.3.1} and~\ref{sec:5.3.2} we do not necessarily assume the existence of a 
spontaneously broken symmetry and return to a generic QFT case with the
scalar field denoted by $\phi(x)$.  Then in the mini-summary subsection~\ref{sec:5.3.3} we summarise the findings of this section
in the context of the theory of the scalar field $h(x)$ with the VEV. This follows the same presentational pattern as 
in the preceding section, where subsections~\ref{sec:validity}-\ref{sec:5.2.4} used a generic scalar $\phi(x)$ before presenting 
a summary in~\ref{sec:sum1} in terms of $h(x)$.

	\subsubsection{Extension to complex time}
	\label{sec:5.3.1}
	
\begin{figure*}[t]
\begin{center}
\includegraphics[width=0.85\textwidth]{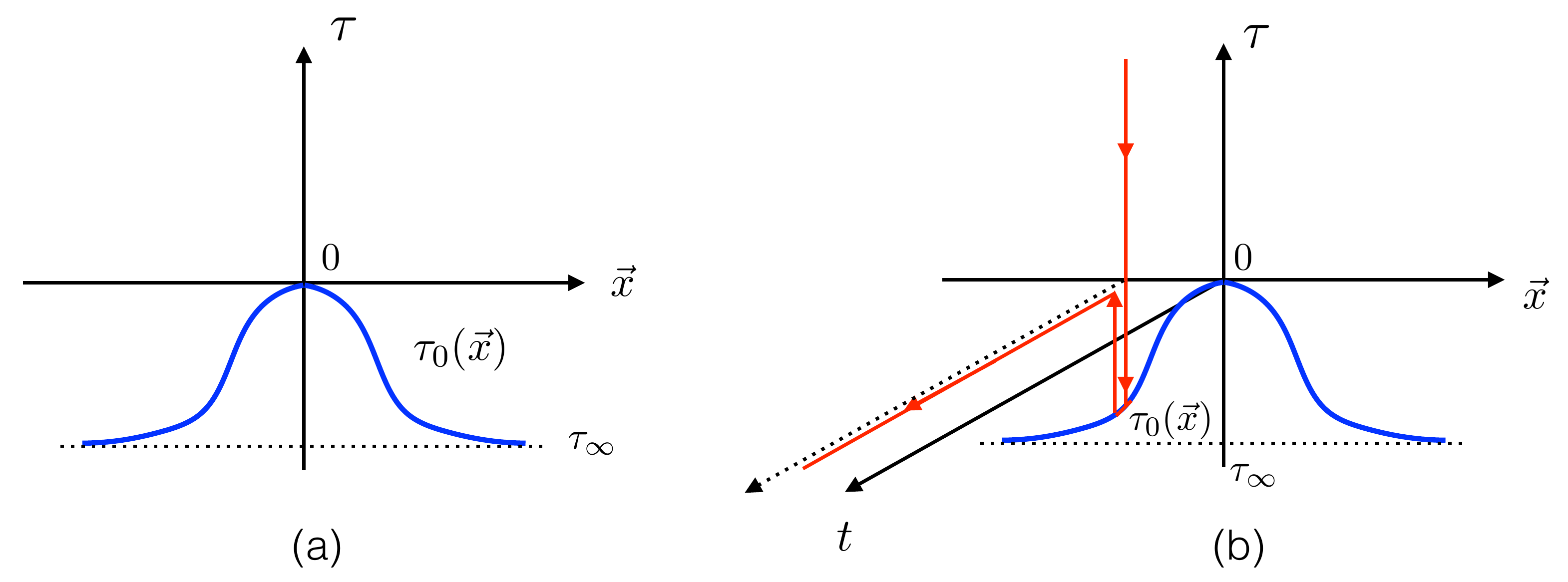}
\end{center}
\vskip-.5cm
\caption{{\bf Plot (a)}: The shape of the singularity surface $\tau=\tau_0(\bec{x})$ of the field configuration $h(x)$ is shown in blue.
{\bf Plot (b)} shows the the time evolution contour of Fig.~\ref{fig:contour}~(a), depicted in red, in the coordinate system $(t,\tau;\bec{x})$.}
\label{fig:tau}
\end{figure*}

In Minkowski space, we require that $\phi$ is regular everywhere except for the singularity at $x=0$. 
Ref.~\cite{Son:1995wz} complexifies the time coordinate, allowing for imaginary times, $\tau$, so that a general complex time, $\tc$, can be written as $t+i\tau$. Now $t=0$ is a $(d+1)$-plane in the $(d+2)$-dimensional $(\tc,\bec{x})$ space. As such, the point singularity at $(0,\bec{0})$ is in general extended to a $d$-dimensional singularity surface, $A$, parametised as 
$(i\tau_0(\bec{x}),\bec{x})$, with the constraint that $\tau_0(\bec{0})=0$. This constraint ensures that the correct Minkowski singularity structure is maintained. The time-evolution contour on the complex time plane is shown in red in  Figs.~\ref{fig:contour} and~\ref{fig:tau}~(b). 
The $d$-dimensional singularity surface is shown in blue in Fig.~\ref{fig:tau}~(a) in the $(d+1)$-dimensional Euclidean spacetime,
and in the $(t,\tau,{\bf x})$ $(d+2)$-coordinates  in Fig.~\ref{fig:tau}~(b).

We now look for the field configuration that satisfies the field equation and is singular on $A$. Following~\cite{Son:1995wz} 
we will search for the solution by breaking it into two parts: $\phi_1$ and $\phi_2$. Each of these is a classical solution that satisfies one of the boundary conditions in 
\eqref{eq:bc1} and \eqref{eq:bc2}.
The first part satisfies the Euclidean asymptotics,
\begin{equation}
\label{eq:early}
	\tilde{\phi}_1(\bec{k})\,=\,\dfrac{1}{\sqrt{2\omk}} \,a^*_{-\bec{k}}\,e^{-\omk \tau}\,\to\,0\,,
	\qquad \tau\to +\infty,
\end{equation}
whereas the second part satisfies the original Minkowski late-time limit,
\begin{equation}
\label{eq:late}
	\tilde{\phi_2}(\bec{k})=\dfrac{1}{\sqrt{2\omk}}(b_\bec{k}e^{\omk T-\theta-i\omk t}+b^*_{-\bec{k}}e^{i\omk t})\,, \qquad t\to+\infty.
\end{equation}
	For a given $\bec{x}$, we consider the time evolution of the solution 
	along the contour $C$ in complex time, which has three distinct parts in red in  Figs.~\ref{fig:contour} and~\ref{fig:tau}~(b):
\begin{enumerate}
\item $(i\infty,i\tau_0(\bec{x}))$: contour begins at infinite Euclidean time and comes down to meet the singularity surface, $A$. 
\item $(i\tau_0(\bec{x}),0)$: after point contact with $A$, return back to Minkowski time axis. Note that for $\bec{x}=\bec{0}$, this step vanishes as $\tau_0(\bec{0})=0$.
\item $(0, \infty)$: travel along Minkowski-time axis to late times.
\end{enumerate}
The first component, $\phi_1$, is defined on part (1) of the contour. It is a classical solution, satisfying the initial-time boundary 
condition \eqref{eq:early} at $\tau=+\infty$ and is singular at $\tau=\tau_0(\bec{x})$. The solution $\phi_1$ and the Euclidean action
evaluated on it at this segment of the contour are real-valued. Indeed, as we already noted in section~\ref{sec:class2}, 
classical evolution of the real-valued initial condition in \eqref{eq:early} along the $\tau$ axis results in a manifestly real field 
configuration along the first segment of the contour.
	
The second component, $\phi_2$, is a classical solution defined on the parts (2) and (3) of the contour 
in Fig.~\ref{fig:contour}. It is singular at $\tau=\tau_0(\bec{x})$,
where it is equal to $\phi_1$, and satisfies the final-time boundary condition \eqref{eq:late}. 
As explained in section~\ref{sec:class2}, the boundary condition~\eqref{eq:late}
requires that we keep the final segment (3) of the contour along the Minkowski time axis $t$, and the solution is necessarily 
complex-valued on this segment.

Both $\phi_1$ and $\phi_2$ can be obtained by starting from the boundary conditions \eqref{eq:early} and \eqref{eq:late} respectively; evolving them forward and backward in time, by solving the sourceless 
classical equations $\delta S/\delta \phi_{1,2}=0$; and formally matching $\phi_1(\bec{x})$ to $\phi_2(\bec{x})$ at some
\textit{a priori} arbitrary surface $A$ (defined as $\tau=\tau_0(\bec{x})$) where both $\phi_1(\bec{x})$ and $\phi_2(\bec{x})$ become singular. 
However, the combined field configuration
$\phi(x)=(\phi_1(x), \phi_2(x))$ on the contour $C$ is not yet the solution to the saddle-point equations.

 Note that there is a non-vanishing overlap in the range $(0\le \tau\le \tau_0(\bec{x}))$ at $t=0$, where both $\phi_1$ and $\phi_2$ 
 are defined. For a general surface $A$, the field configuration $\phi$ can still be discontinuous for all $\bec{x}$ at $t=0$. 
 However, we are interested specifically in the case where $\phi$ is only discontinuous at $\bec{x}=\bec{0}$, $t=0$, 
 which is the location of the source term and is the only source of the singularity/discontinuity of the field. That is, we require that,
	\begin{equation}
	\label{eq:cond}
	\phi_1(0,\bec{x})=\phi_2(0,\bec{x}),\qquad \forall\bec{x}\ne \bec{0}. 
	\end{equation}
If we can choose the surface $A$ such that this condition is satisfied, the combined field $\phi(x)$ will be the solution to the saddle-point equations.
Our next task is to explain how this can be achieved by extremising the action $S[\phi]$ over the singular surfaces.  
We will show that on the extremal surface, the requirement in \eqref{eq:cond} will be automatically satisfied, see 
Eq.~\eqref{eq:jdef2} below.
	
	The total action $iS[\phi]$, which (by standard convention) we write as the Euclidean action with a minus sign, 
	$iS[\phi]=-S_E[\phi]$, is the sum of the contributions from the three parts of the contour defined above,
	\begin{equation}
	\begin{split}
	\label{eq:ssum}
	iS[\phi]\,=&-S_E[\phi]\,=\,-S_E^{(1)}[\phi_1]-S_E^{(2)}[\phi_2]+iS^{(3)}[\phi_2]\\
	=&
	\int d^dx\int_{+\infty}^{\tau_0(\bec{x})} d \tau \, \mathcal{L}_E(\phi_1)
	+ \int d^dx \int^{0}_{\tau_0(\bec{x})} d\tau  \, \mathcal{L}_E(\phi_2)
	+i \int d^dx \int^{+\infty}_0 dt \, \mathcal{L}(\phi_2),
	\end{split}
	\end{equation}
	where $\mathcal{L}(\phi)$ is the usual Lagrangian as defined in \eqref{eq:lag} and,
	\begin{equation}
	\mathcal{L}_E(\phi)=\frac{1}{2}{(\partial_\tau\phi)^2}+\frac{1}{2}{|\nabla\phi|^2}+
	\frac{1}{2} m^2 \phi^2+\mathcal{L}_{\rm int}
	\,=\, \frac{1}{2}{(\partial_\tau\phi)^2}+\frac{1}{2}{|\nabla\phi|^2}+V(\phi),
	\end{equation}
is its Euclidean counterpart. 
Note that though $S^{(1)}_E$ and $S^{(2)}_E$ are infinite on the singularity surface
	i.e. at $\tau=\tau_0(\bec{x})$, their sum can be finite (at least on some surfaces) due to the differing integration directions for
	$S^{(1)}_E$ and $S^{(2)}_E$ in the vicinity of the singularity.
	
The imaginary part of the Minkowski action appearing in the expression for the rate in \eqref{eq:al5}, becomes the 
real part of the Euclidean action,
\[
{\rm Im}\, S \,=\, \frac{1}{2i}\left(S-S^*\right)\,=\, \frac{1}{2}\left(-iS+iS^*\right)\,=\, \frac{1}{2}\left(S_E-S_E^*\right)\,=\, 
{\rm Re}\, S_E\,.
\]

\medskip

	\subsubsection{Extremisation over singularity surfaces}
	\label{sec:5.3.2}

\medskip	

Here we will show that extremising the real part of the Euclidean action over all appropriate singularity surfaces will single out the desired singularity surface (i.e. that which satisfies the condition in \eqref{eq:cond}) and consequently yield the solution to the original boundary-value problem. By ``appropriate" we simply mean that $A$ must include the point $\tau=|\bec{x}|=0$ as previously stated. 
This reduction of the problem of finding the solution to the saddle-point equations to the extremisation over singular surfaces is a key element 
in the approach of  Ref.~\cite{Son:1995wz}.
	
To set up the problem we take the following steps:
	\begin{itemize}
		\item Since $\phi$ is infinite on the singularity surface, we regularise it by setting $\phi_1=\phi_2=\phi_0$ everywhere on $A$, with $\phi_0$ large but for now kept finite.
		\item Note that since $\phi_1$ and $\phi_2$ are different solutions (carrying different energies), their form in the vicinity of
		$A$ will be different. As such, we can denote the difference
\begin{equation}
		\label{eq:jdef}
		\partial_n(\phi_1-\phi_2) =  J(s_i) ,
\end{equation}
in terms of a function $J(s_i)$ defined on $A$.		
Here $\partial_n$ is the derivative in the direction normal to the singularity surface $A$, and $s_i$ (with $i=1,\ldots, d$) are coordinates on $A$, described by $x^\mu = x^\mu(s_i)$.
		\item Since the source-term $J(s_i)$ in \eqref{eq:jdef} is distributed over the surface, it follows that $\phi$ is actually the solution to,
		\begin{equation}
		\dfrac{\delta S_E[\phi]}{\delta \phi(x)}=J(x)= \int_A ds_i J(s_i)\delta(x^\mu-x^\mu(s_i)).
		\end{equation}
In other words, 
the source term is zero except for points $x^\mu(s_i)$ on the singularity surface, where it is given by the distribution $J(s_i)$. 	
\end{itemize}

With this framework in mind, we consider deforming the surface $A\to A'$ such that $x^\mu(s_i)\to x^\mu(s_i)+\delta x^\mu(s_i)$, with $\delta x^\mu(s_i)=n^\mu\delta x(s_i)$ and $n^\mu$ a unit normal vector. To match our boundary conditions we require that $x=0$ is included in both $A$ and $A'$, i.e. $\delta x^\mu|_{x=0}=0$.

We can now compute the variation of the real part of the action \eqref{eq:ssum} arising from varying the surface $A\to A'$.
It is given by the following simple formula~\cite{Son:1995wz}, which we will derive at the end of this section,
\begin{equation}
\label{eq:varSE}
	\delta\,\mathrm{Re}S_E[\phi]
	\,= \,\dfrac{1}{2}\int_A ds\left(
	(\partial_n\phi_1)^2-(\partial_n\phi_2)^2
	\right)
	\delta x(s).
\end{equation}
We now rearrange the expression in \eqref{eq:varSE} in the form,
\begin{equation}	
	\delta\,\mathrm{Re}S_E[\phi]
	\,= \, \dfrac{1}{2}\int_A ds[\partial_n(\phi_1+\phi_2)][\partial_n(\phi_1-\phi_2)]\delta x(s),
\end{equation}
and recognising $\partial_n(\phi_1-\phi_2)$ as $J(s)$ in \eqref{eq:jdef} and labelling $\phi=(\phi_1+\phi_2)/2$, we arrive at,
	\begin{equation}
	\delta\,\mathrm{Re}S_E[\phi]\,=\, \int_A ds (\partial_n\phi)J(s)\delta x(s).
	\end{equation}
On the extremised surface we have $\delta\,\mathrm{Re}S_E[\phi]=0$. Given that $\delta x|_{x^\mu=0}=0$, this is achieved by $J(x^\mu)=j_0\delta(x^\mu)$. In other words, the source becomes infinitely localised at $x=0$ and it follows from \eqref{eq:jdef} that,
\begin{equation}
		\label{eq:jdef2}
		\partial_n(\phi_1-\phi_2) \propto \delta^{(d+1)}(x) ,
\end{equation}
which satisfies the requirement \eqref{eq:cond}. In fact, as noted in~\cite{Son:1995wz} 
it follows from \eqref{eq:jdef2} and the fact that $\phi_1=\phi_2$ on $A$,
that when $A$ is the extremal surface, the 
two parts of the solution coincide on the entire range of the common domain, $\tau_0(\bec{x})<\tau<0$, with the exception of the point
at the origin $\tau=0=\bec{x}$.

This concludes the proof that the boundary-value problem is solved by extremising the real part of the Euclidean action over all singularity surfaces that include the point $\tau=|\bec{x}|=0.$ The solution $\phi(x)$ is obtained from the two branches, $\phi_1$ and $\phi_2$ that are matched on the 
extremal singular surface.
We also note that the matching condition  \eqref{eq:cond} implies that the Euclidean action integrals are real-valued (and positive)
on both Euclidean segments (1) and (2) of the contour.

\bigskip

In the remaining part of this section we derive the formula \eqref{eq:varSE}.
In varying the surface $A$, we expect the solutions to change, as well as the position of the surface,
\[
\begin{split}
\delta S \,=&\,\, S[\phi'_1(x) \,{\rm on}\, A']\,-\, S[\phi_1(x) \,{\rm on}\, A]\\
=& \left(S[\phi_1'(x) \,{\rm on}\, A']\,-\, S[\phi_1(x) \,{\rm on}\, A']\right) \,+\,
\left(S[\phi_1(x) \,{\rm on}\, A']\,-\, S[\phi_1(x) \,{\rm on}\, A]\right).
\end{split}
\]

Hence, there will be two contributions to $\delta S_E^{(1)}[\phi_1]$:
\begin{enumerate}
\item First contribution to $\delta S_E^{(1)}[\phi_1]$ comes from the change  $\phi_1\to \phi_1'= \phi_1+\delta\phi_1$
in field solution,
	\begin{equation}
	\label{eq:chf}
		-\int d^dx \int_{\infty}^{\tau_0(\bec{x})}d\tau \,\delta \left( \dfrac{1}{2}(\partial_\mu\phi_1)^2+V(\phi_1)  \right) \,=\,
		-\int _Ads(n^\mu\partial_\mu\phi_1)\delta\phi_1,		
       \end{equation}
which is obtained by using Gauss' theorem and the fact that $\phi_1(x)$ satisfies the Euler-Lagrange equation. We are left in \eqref{eq:chf}
with the boundary term, on the surface of $A$. Note that $ds$ is shorthand for the appropriate $d$-dimensional integration measure on the $d$-dimensional singularity surface, for surface coordinates $s_i$.
		
\item Second contribution to $\delta S_E^{(1)}[\phi_1]$ comes from the change in the position of the surface,
$x^\mu(s_i)\to x^\mu(s_i)+n^\mu\delta x(s_i)$.
It is given by,
	\begin{equation}
		-\int_A ds  \left[\left( \dfrac{1}{2}(\partial_\mu\phi_1)^2+V(\phi_1)  \right)\delta x(s)\right],
	 \label{eq:chf2}	
	\end{equation}
with details in Appendix~\ref{app:intrange}.	
	\end{enumerate}
Adding the contributions in \eqref{eq:chf} and  \eqref{eq:chf2}, we have a total variation of $S_E^{(1)}$ of,
	\begin{equation}
	\label{eq:change}
	\delta S_E^{(1)}[\phi_1]=-\int_A ds  \left[
	(n^\mu\partial_\mu\phi_1)\delta\phi_1 +
	\left( \dfrac{1}{2}(\partial_\mu\phi_1)^2+V(\phi_1)  \right)\delta x(s)\right].
	\end{equation}
	We now use the regularisation imposed above, noting that $\phi_1|_A=\phi_1'|_{A'}=\phi_0$,
	\begin{equation}
	\label{eq:trick}
	\delta\phi_1(x^\mu)=\phi_1'(x^\mu)-\phi_1(x^\mu)=\phi_1'(x^\mu)-\phi_1'(x^\mu+n^\mu\delta x)=-(n^\mu\partial_\mu\phi_1)\delta x(s),
	\end{equation}
	so that \eqref{eq:change} can be rewritten as,
	\begin{equation}
	\delta S_E^{(1)}[\phi_1]= \int_A ds  \left[
	\left((n^\mu\partial_\mu\phi_1)^2- \dfrac{1}{2}(\partial_\mu\phi_1)^2-V(\phi_1)  \right)\delta x(s)\right].
	\end{equation}
	Finally, recall that $\phi_1$ is constant on $A$ and thus the tangential derivative vanishes: we need only consider normal derivatives, $n^\mu\partial_\mu \phi_1= \partial_n\phi_1$. Therefore, we find,
	\begin{equation}
	\delta S_E^{(1)}[\phi_1]\,=\,  \int_A ds  \left[
	\left(\dfrac{1}{2}(\partial_n\phi_1)^2-V(\phi_1)  \right)\delta x(s)\right].
	\end{equation}
	Similarly, for $S_E^{(2)}[\phi_2]$ in \eqref{eq:ssum}, we find,
	\begin{equation}
	\delta S_E^{(2)}[\phi_2]\, =\, -\int_A ds  \left[
	\left(\dfrac{1}{2}(\partial_n\phi_2)^2-V(\phi_2  )\right)\delta x(s)\right].
	\end{equation}
	
Adding these two contributions together and noting that $\phi_1=\phi_2$ on $A$ (but not their normal derivatives), we find,
\begin{equation}
\label{eq:S12re}
	\delta S_E^{(1)}[\phi_1] + \delta S_E^{(2)}[\phi_2]\, =\,
\dfrac{1}{2}\int_A ds\left(
	(\partial_n\phi_1)^2-(\partial_n\phi_2)^2
	\right)
	\delta x(s).
\end{equation}	

	The case for $S^{(3)}[\phi_2]$ is a little different. This third part of the integration contour doesn't encounter the singularity surface and thus we only expect a contribution from the change in field solution, $\phi_2\to\phi_2+\delta\phi_2$,
	\begin{equation}
	-i\delta S^{(3)}[\phi_2]
	= -i\left[\int d^dx(\hat{t}^\mu\partial_\mu\phi_2)\delta\phi_2\right]^{t\to\infty}_{t=0}
	= -i\int d^dx\,\partial_t\phi_2\delta\phi_2\Bigl{|}_{t\to\infty},
	\end{equation}
	in a similar vein to \eqref{eq:chf}. The contribution from $t\to 0$ vanishes due to the equivalnet relation for $\delta\phi_2$ as for $\delta\phi_1$ in \eqref{eq:trick}, recalling that $\delta x|_{t=0}=0$. It will turn out that we do not need to consider this Minkowski boundary term as it is purely imaginary and thus does not appear in the $\mathrm{Im}S[\phi]\sim\mathrm{Re}S_E[\phi]$ term in \eqref{eq:W}. 
	
Consider the above term rewritten in terms of momentum-space fields,
\begin{equation}
\label{eq;delS3}
	-i \int \dx\,\partial_t\phi_2(x)\delta\phi_2(x)\Bigl{|}_{t\to\infty} \,=\,
		-i\int \dk\,\partial_t\tilde{\phi}_2(\bec{k})\delta\tilde{\phi}_2(-\bec{k})\Bigl{|}_{t\to\infty}.
\end{equation}
Inserting the late-time assymptotics \eqref{eq:late} of $\tilde{\phi}_2(\bec{k})$
and further noticing that, since $\tilde{\phi}_2(\bec{k})$ and $\tilde{\phi}'_2(\bec{k})$ obey the same asymptotics, so does 
$\delta\tilde{\phi}_2(\bec{k})$,
\begin{equation}
\label{eq:del-late}
	\delta\tilde{\phi_2}(\bec{k})=\dfrac{1}{\sqrt{2\omk}}(\delta b_\bec{k}e^{\omk T-\theta-i\omk t}+ \delta b^*_{-\bec{k}}e^{i\omk t})\,, \qquad t\to+\infty,
\end{equation}
we rewrite the expression in \eqref{eq;delS3} as follows,
\begin{equation}
\label{eq:S3again}
-i\delta S^{(3)}[\phi_2]\,=\, -i\int \dk\,\partial_t\tilde{\phi}_2(\bec{k})\delta\tilde{\phi}_2(-\bec{k})\Bigl{|}_{t\to\infty}
\,=\, 
-\int \dk\, (b_\bec{k} \delta b^*_\bec{k}- b^*_\bec{k} \delta b_\bec{k})e^{\omk T-\theta},
\end{equation}
which is purely imaginary.
	
Hence, \eqref{eq:S3again} does not contribute to the variation of the real part of the Euclidean action, which thus is given by the expression 
on the right-hand side of \eqref{eq:S12re}, 
confirming the formula for $\delta\,\mathrm{Re}S_E[\phi]$ in \eqref{eq:varSE}.

\subsubsection{Summary of the surface extremisation approach in complex time}
\label{sec:5.3.3}

Here we summarise the steps involved in solving the boundary value problem approach via
extremisation over singular surfaces~\cite{Son:1995wz}
in the context of the $(d+1)$-dimensional model \eqref{eq:L2} with SSB.
Note that all our considerations are general and that the expressions in the summary below can also be written in terms of the manifestly VEV-less field $\phi$ with the Lagrangian \eqref{eq:lag} using $\phi(x)=h(x)-v$.

\begin{enumerate}
\item Select a trial singularity surface located at $\tau=\tau_0(\bec{x})$.
The surface profile $\tau_0(\bec{x})$ is an ${\cal O}(d)$ symmetric function of $\bec{x}$ and is given by a local 
deformation of the flat singularity domain wall at $\tau_\infty$ with the single maximum touching the origin $(\tau, \bec{x})=0$
as shown in blue in Fig.~\ref{fig:tau}~(a).
Minkowski space is the $\tau=0$ slice of the $(t,\tau;\bec{x})$ space;
it intersects the singularity surface at a point located at the origin. 
Hence in Minkowski space the singularity of the field configuration $h$ is point-like and located 
at $t=0=\tau$ and $\bec{x}=0$ as required.

\item Deform the time evolution contour such that the paths in the Feynman path integral follow the contour 
on the complex plane $(t,\tau)$,
\[
\, \left[(0, \infty) \to (0,\tau_0(\bec{x}))\right] \, \oplus\,  \left[(0,\tau_0(\bec{x}))\to (0,0)\right] \, \oplus\, \left[(0,0) \to (\infty,0)\right] 
\,,
\label{eq:contour}
\]
as shown in Figs.~\ref{fig:contour}~(a) and~\ref{fig:tau}~(b).

\item Find a classical trajectory $h_1(\tau,\bec{x})$ on the first segment, $ +\infty > \tau > \tau_0(\bec{x})$, of the contour
\eqref{eq:contour} that satisfies the initial time (vanishing) boundary condition \eqref{eq:al2},
\[
\lim_{\tau\to + \infty}\, h_1(\tau,\bec{x})  \,-\, v 
\,\,\to \,\, 0\,,
\label{eq.E1}
\]
and becomes singular as $\tau \to \tau_0(\bec{x})$ so that
 $h_1(\tau,\bec{x})|_{\tau\to \tau_0(\bec{x})} \,\equiv\, \phi_0 \,\to\, \infty$.
 
\item Find another classical solution $h_2 (\tau,\bec{x})$ on the remaining part of the contour \eqref{eq:al2},
that satisfies the final time 
boundary condition \eqref{eq:al3},  
\[
\lim_{t\to + \infty}\,h_2(t,\bec{x}) \,-\, v  \,=\, 
\int \frac{d^d k}{(2\pi)^{d/2}} \frac{1}{\sqrt{2\omega_{\bf k}}}\left( b_{\bf k}\,e^{\omega_{\bf k}T-\theta}
\, e^{-ik_\mu x^\mu}\,+\, b^*_{\bf k}\, e^{ik_\mu x^\mu}\right)\,,
\label{eq.E3}
\]
and require that at $\tau \to \tau_0(\bec{x})$ the solution $h_2 (\tau,\bec{x})$ is singular and matches with $h_1$,
\[
h_2 (\tau_0,\bec{x}) \,= \, h_1(\tau_0,\bec{x}) \,=\, \phi_0\,\to\, \infty\,.
\label{eq.E2}
\]

\item For the combined configuration $h(x)$ to solve the classical equation~\eqref{eq:al1} on the entire contour~\eqref{eq:contour}
including at  $\tau = \tau_0(\bec{x})$, we need 
to extremise the action, 
\[
iS[h]\,=\,\int  d^d x \left( \int_{+\infty}^{\tau_0(\bec{x})} d\tau \, {\cal L}_{E}(h_1) \,+\,
\int_{\tau_0(\bec{x})}^0d\tau \, {\cal L}_{E}(h_2) \,+\,i 
 \int_{0}^\infty dt \, {\cal L}(h_2) \right)
 \label{eq:Se}
\]
over all singularity surfaces $\tau=\tau_0(\bec{x})$
containing the point $t=0=\bec{x}$. This determines the extremal surface $\tau=\tau_0(\bec{x})$.  

\item Finally, determine the semiclassical rate by evaluating 
\[ 
W(E,n)  \,=\,
ET \,-\, n\theta \,-\,  2{\rm Re} \,S_E [h]
\label{eq:al52}
\]
on the solution, using \eqref{eq:Se} for the action,  and expressions for $T$ and $\theta$ in terms of of $E$ and $n$ found from \eqref{eq:al4} as before.

\end{enumerate}

\noindent This is the general outcome of the semiclassical construction  of  Ref.~\cite{Son:1995wz}. 
One starts with the two individual solutions satisfying the boundary conditions \eqref{eq.E1}-\eqref{eq.E3} and then varies over the
profiles of the singular matching surface $\tau_0(\bec{x})$ to find an extremum 
of the imaginary part of the action  \eqref{eq:Se}. On the extremal surface not only the field configurations, but also their 
normal derivatives match $\partial_n (h_1\,-\,h_2)=0$ at all $\bec{x}$ except $\bec{x}=0.$ This implies that $h_1=h_2$ on the entire slice of the 
spacetime where they are both defined, i.e. for $\tau$ in the interval $[0,\tau_0]$, except at the point at the origin. Restricting to the
Minkowski space slice, i.e. at $\tau=0$, this implies $h_1(0,\bec{x})=h_2(0,\bec{x})$, as it should be. It does not mean however that the real part of the action in \eqref{eq:Se} vanishes, as the sum of the first two integrals can be viewed as encircling the singularity of the solution at $\tau_0$.

In summary, the highly non-trivial problem of searching for the appropriate singular field solutions $h(x)$ 
is reduced to a geometrical problem --
extremisation over the surface shapes $\tau_0(\bec{x})$ and accounting for the appropriate boundary conditions \eqref{eq.E1}-\eqref{eq.E3}.
This formulation of the problem is now well-suited for using the thin-wall 
approximation that will be described in section~\ref{sec:thin_w} and will allow us to address the  large $\lambda n$ 
regime, following~\cite{Khoze:2017ifq,Khoze:2018kkz}, where quantum non-perturbative effects become important. 

We proceed with the practical implementation of the steps 1-6 for the model \eqref{eq:L} in the following sections.

\bigskip
\section{Computation of the semiclassical rate}
\label{sec:main}
\medskip

We will now concentrate on the scalar field theory model with a non-vanishing vacuum expectation value \eqref{eq:L} in 3+1 dimensions.
Some of the material presented below, such as the general aspects of the approach, the computation of tree-level contributions, and 
the computation of quantum effects at $\lambda n \ll 1$, can be carried out in any scalar QFT with only minor modifications.
The case of the unbroken $\phi^4$ theory has already been
addressed in Ref.~\cite{Son:1995wz}, while we are predominantly interested in the broken theory \eqref{eq:L},
where the applications of the semiclassical method at small $\lambda n$ are new, though the resulting expressions are closely related to those
derived by Son in the unbroken theory.

In sections~\ref{sec:thin_w} and \ref{sec:beyond_tw} we focus on the $\lambda n \gg 1$ regime (previously
	unexplored in~\cite{Son:1995wz}), where quantum corrections provide the \emph{dominant} contribution to the 
	multiparticle rate. The computation we will present follows~\cite{Khoze:2018kkz} and is specific to the model of the type  
	\eqref{eq:L} with spontaneous symmetry breaking~\cite{Gorsky:1993ix,Khoze:2017ifq}.
	In this case the singular domain wall semiclassical configuration corresponds to a local minimum of the action (rather than a local 
	maximum or a saddle-point) and this will play a role in our construction.

\medskip

\subsection{Setting up the computation}
\label{sec:son_loops}

In this section we will specify and solve the boundary conditions in \eqref{eq.E1}, \eqref{eq.E3} at the initial and final times, 
deriving the coefficient functions $b^*_{\bf k}$  and $b_{\bf k}\,e^{\omega_{\bf k}T-\theta}$ 
in \eqref{eq.E3}. We will then determine the $T$ and $\theta$ parameters 
and compute the general expression for the exponent of the rate $W(E,n)$ in \eqref{eq:al5}.

In the limit $\varepsilon= 0$, the scattering amplitude is on the multiparticle threshold, the final state
momenta are vanishing and one would naively assume that the classical solution describing this limit is uniform in space. 
This is correct for the tree-level solution but not for the solution incorporating quantum effects. In the latter case, the correct
and less restrictive assumption is that the presence of the singularity at $x=0$ deforms the flat surface of singularities near its location,
as shown in Fig.~\ref{fig:tau}.
From now on we will concentrate on the physical case where $\varepsilon$ is non-vanishing and
non-relativistic, $0< \varepsilon \ll 1$. At the same time, the parameter $\lambda n$ is held fixed and arbitrary. It will ultimately be taken to be large. 

The initial-time boundary condition \eqref{eq.E1} dictates that the solution $h_1(t_{\mathbb{C}}=i\tau,\bec{x})-v $ must vanish with exponential accuracy 
as $e^{-m\tau}$ in the limit $\tau \to \infty$. The final-time boundary condition \eqref{eq.E3} of the finite-energy solution $h_2 (x)$ requires the solution to be singular on the
singularity surface $\tau_0(\bec{x})$.
Following Son, without loss of generality, we can search for $h_2$ in the form,
\begin{eqnarray}
h_2(t_{\mathbb{C}},\bec{x}) 
&=& v\, \left(\frac{ 1\,+\,e^{im (t_{\mathbb{C}}-i\tau_\infty)}}
{1\,-\, e^{im (t_{\mathbb{C}}-i\tau_\infty)}}\right)\,+\, \tilde{\phi}(t_{\mathbb{C}},\bec{x})
\,.
\label{clas_sol2A}
\end{eqnarray}
The first term on the right-hand side is the $\bec{x}$-independent field configuration  $h_0(t_{\mathbb{C}})$
that we discussed in section~\ref{sec:class1}.
It is an exact classical solution 
\eqref{clas_sol2} with the surface of singularities at $ t_{\mathbb{C}} = i \tau_{\infty}$, which is a 3-dimensional plane 
spanned by ${\bf x}$,
as shown in Fig.~\ref{fig:kink}.
The second term, $\tilde{\phi}(t_{\mathbb{C}},\bec{x})$,
describes the deviation of the singular surface from the $\tau_{\infty}$-plane. 
This deviation, $\tau_0({\bf x}) -\tau_\infty$, is locally non-trivial around $\bec{x}=0$ and vanishes at 
$\bec{x}\to \infty.$ There is no loss of generality in \eqref{clas_sol2A} because the configuration $\tilde{\phi}(t_{\mathbb{C}},\bec{x})$
is so far completely unconstrained. 

Now we can start imposing the boundary conditions  \eqref{eq.E3} at $t\to +\infty$ on the expression \eqref{clas_sol2A}.
On the final segment of the time evolution contour, $t (1+i\delta_+)$ as $t\to +\infty$, the first term in \eqref{clas_sol2A} can be Taylor-expanded 
in powers of $e^{im t(1+i\delta_+)}$
and linearised (since $\delta_+$ is positive) giving,
\[
\lim_{t\to + \infty} h_0(x) \,-\, v \, =\, 
2v \, e^{m\tau_\infty}\, e^{imt} \,.
\label{eq.Eh0}
\]
For the second term in \eqref{clas_sol2A} we write the general expression involving the positive-frequency and the negative frequency 
components in the Fourier transform,
\[
\lim_{t\to + \infty}\, \tilde{\phi}( t, {\bf k})
  \,=\, 
\frac{1}{\sqrt{2\omega_{\bf k}}}\left(f_{\bf k}\, e^{-i\omega_{\bf k} t}\,+\, g_{-\bf k}\, e^{i\omega_{\bf k}t}\right)\,.
\label{eq.Etilphi}
\]

We will now show that for the solution in the non-relativistic limit, $\epsilon \ll 1$, the boundary conditions \eqref{eq.E3}
will require that $g_{-\bf k} = 0$ and impose a constraint on the coefficient function $f_{\bf k}$, so that,
\begin{eqnarray}
g_{-\bf k} &=& 0\,, \label{eq:gk0} \\
f_{\bf k=0} &=& \frac{n \sqrt{\lambda}}{(2 \pi m)^{3/2}} \, e^{-m\tau_\infty}\,. \label{eq:fk0}
\end{eqnarray}

To derive \eqref{eq:gk0}-\eqref{eq:fk0} we proceed by combining the asymptotics \eqref{eq.Etilphi} 
with the Fourier transform of \eqref{eq.Eh0} 
and write down the full solution in \eqref{clas_sol2A} in the form,
\[
\lim_{t\to + \infty}\, h_2( t, {\bf k}) \,-\, v \, =\, 
\frac{1}{\sqrt{2\omega_{\bf k}}}\left(f_{\bf k}\, e^{-i\omega_{\bf k} t}\,+\, 
\left\{ g_{-\bf k} \,+\,
2v \sqrt{2\omega_{\bf k}}\, e^{m\tau_\infty}\,(2\pi)^{3/2}\, \delta^{(3)}({\bf k})
\right\}e^{i\omega_{\bf k}t}\right).
\label{eq.Efull}
\]
Comparing with the the final-time boundary condition \eqref{eq.E3} we read off the expressions for the coefficient functions,
\begin{eqnarray}
 b_{\bf k}\,e^{\omega_{\bf k}T-\theta} &=& f_{\bf k} 
 \label{eq:Eb}
 \\
  b^*_{\bf k}&=& g_{-\bf k} \,+\, 2v \sqrt{2m}\, e^{m\tau_\infty}\,(2\pi)^{3/2}\, \delta^{(3)}({\bf k}) \,.
\label{eq:Ebd}
\end{eqnarray}
We will now make an educated guess that the parameter $T$ will be infinite in the limit $\varepsilon \to 0$.
In fact we will soon derive that $T= 3/(2 m\varepsilon)$, so this assumption will be justified \textit{a posteriori}. 
We can then re-write \eqref{eq:Eb} as,
\[
 b_{\bf k}\,=\,  f_{\bf 0}\, e^{-\omega_{\bf k}T}\, e^{\theta}\,.
\]
In the limit where $\varepsilon \to 0$ (and thus $T\to\infty$) the factor $e^{-\omega_{\bf k}T}$ can be thought of as the 
regularisation of a momentum-space delta-function:
it cuts-off all non-vanishing values of ${\bf k}$ by minimising $\omega_{\bf k}$, thus reducing ${\bf k}$ to zero. 
Therefore, we set $f_{\bf k}$ to $ f_{\bf 0}$
in the equation above. 

Furthermore, since the function $b_{\bf k}$ is proportional to the (regularised) delta-function, its complex conjugate $ b^*_{\bf k}$
 must be too. This implies that the coefficient function $g_{-\bf k}$ in \eqref{eq:Ebd} 
must be zero \cite{Son:1995wz}, which verifies \eqref{eq:gk0}, so that \eqref{eq.Etilphi} becomes,
\[
\lim_{t\to + \infty}\, \tilde{\phi}( t, {\bf k})
  \,=\, 
\frac{1}{\sqrt{2\omega_{\bf k}}}  \,\, f_{\bf k}\, e^{-i\omega_{\bf k}t}\,.
\label{eq.Etilphi2}
\]

We have obtained the expression for the coefficient function $b_{\bf k}$ (and its complex conjugate) and 
also obtained a symbolic identity involving the parameters $T$, $\theta$  and the delta-function,
\[
b_{\bf k}\,=\, f_{\bf 0}\, e^{-\omega_{\bf k}T}\, e^{\theta} \,=\, 2v \sqrt{2m}\, e^{m\tau_\infty}\,(2\pi)^{3/2}\, \delta^{(3)}({\bf k}) \,=\, b^*_{\bf k}\,.
\label{eq:Esymb}
\]
This symbolic identity should be interpreted as follows. In the limit of strictly vanishing $\varepsilon$,
all these terms are proportional to the delta-function. Away from this limit, i.e. in the case of processes near
the multiparticle threshold where $0<\varepsilon\ll 1$, the function $\delta^{(3)}({\bf k})$ appearing in the third term above is not the strict delta-function, but a narrow peak with the singularity regulated by $\varepsilon$. This can be derived by
allowing the surface 
$\tau_\infty$ in the first term in \eqref{clas_sol2A} to not be completely flat at small non-vanishing $\varepsilon$, but to have a
tiny curvature $2\varepsilon/3\ll 1$ \cite{Son:1995wz}, thus leading to a regularised expression for $\delta^{(3)}({\bf k})$ in the
final term in \eqref{eq.Efull}.

To proceed, we integrate the two middle terms in \eqref{eq:Esymb} over $d^3k$,
\[
 f_{\bf 0}\, e^{\theta}\, \int d^3 k \, e^{-\omega_{\bf k}T} \,=\, 2v \, e^{m\tau_\infty}\,(2\pi)^{3/2}\,.
\label{eq:inteq}
\]
The integral on the left hand side of \eqref{eq:inteq},
\[
\int d^3 k \, e^{-\omega_{\bf k}T} \,=\, 4\pi\, m^3\, e^{-mT} \, \int_0^\infty dx\, x^2 \, e^{-mT(\sqrt{1+x^2}-1)}\,,
\]
where $x= k/m$. Note that this integral is dominated by $x\sim mT$, which at large $T$ allows us to 
simplify this as,
\[
4\pi\, m^3\, e^{-mT} \, \int_0^\infty dx\, x^2 \, e^{-mTx^2/2} \,=\, 4\pi\, m^3\, e^{-mT} \, \frac{\sqrt{\pi/2}}{(mT)^{3/2}}\,.
\]
We can now solve the equation~\eqref{eq:inteq} for $ f_{\bf 0}$ and 
 find that at large $T$,
\[
 f_{\bf 0}\,=\, \frac{4}{\sqrt{\lambda}}\, (T)^{3/2}\, e^{mT-\theta+m\tau_\infty}\,.
\label{eq:Esymb2}
\]

We can now compute the particle number $n$ and the energy $E$ in the final state using equations \eqref{eq:al4}
and the now known coefficient functions \eqref{eq:Esymb} along with \eqref{eq:Esymb2}. We find,
\begin{eqnarray}
n &=& 
\int d^3 k \,\, b_{\bf k}^* \, b_{\bf k}\, e^{\omega_{\bf k}T-\theta} 
\,=\,  \int d^3 k \,\, b_{\bf k}^* \, f_{\bf 0} 
\,=\, \frac{16}{\lambda} \, (2 \pi mT)^{3/2}\,e^{mT-\theta+2m\tau_\infty}
\label{eq:1stint}
\end{eqnarray}
and
\begin{eqnarray}
mn\varepsilon \,=\, E-mn &=&
 \int d^3 k \, \frac{{\bf k}^2}{2}\, b_{\bf k}^* \, b_{\bf k}\, e^{\omega_{\bf k}T-\theta} \nonumber\\ 
 &=&
 \int d^3 k \, \frac{{\bf k}^2}{2}\, b_{\bf k}^* \, f_{\bf 0}\,=\, 
\frac{16}{\lambda} \, (2 \pi mT)^{3/2}\,e^{mT-\theta+2m\tau_\infty}\, \frac{3}{2T}
\label{eq:2ndint}
\end{eqnarray}
It turns out that it was sufficient to know just the value of $f_{\bf k}$ at ${\bf k}=0$ to evaluate the integrals above,
due to the fact that $b_{\bf k}^*$ and $b_{\bf k}$ are sharply peaked at ${\bf k}=0$ as dictated by \eqref{eq:Esymb}.

Dividing the expression on the right hand side of \eqref{eq:2ndint} by the expression in \eqref{eq:1stint} we find,
\[
T\,=\, \frac{1}{m}\,\frac{3}{2}\, \frac{1}{\varepsilon}\,.
\label{eq:Tfound}
\]
The second parameter $\theta$ is found to be,
\[ 
\theta\,=\,
-\, \log\frac{\lambda n}{4}\,+\, \frac{3}{2}\log\frac{3\pi}{\varepsilon}\,+\, 2m\tau_\infty\,+\, \frac{3}{2}\,\frac{1}{\varepsilon}\,.
\label{eq:Thfound}
\]
We now finally substitute these parameters into the equation \eqref{eq:al52} for the `holy grail' function $W(E,n)$, and find, 
\begin{eqnarray}
W(E,n) &=&
ET \,-\, n\theta \,-\, 2{\rm Re} S_{E}[h]\,=\,mn(1+\varepsilon)T\,-\, n\theta \,-\, 2{\rm Re} S_{E}[h]
\nonumber\\\nonumber\\
&=& n\, \log\frac{\lambda n}{4}\,+\, n\left(\frac{3}{2} \log\frac{\varepsilon}{3\pi}+1\right)\,-\,  2nm\,\tau_\infty\,-\, 2{\rm Re} S_{E}[h]\,.
\label{eq:Eal5a}
\end{eqnarray}
We also note that the expression for $f_{\bf 0}$ found in \eqref{eq:Esymb2} evaluated with $T$ and $\theta$ given by 
\eqref{eq:Tfound}-\eqref{eq:Thfound}, reproduces the equation \eqref{eq:fk0}, which was our second constraint on the general
form of the solution 
$h_2(t_{\mathbb{C}},\bec{x})$ in \eqref{clas_sol2A}.

Before interpreting the expression \eqref{eq:Eal5a} for the `holy grail' function,
 we would like to separate the terms appearing on the right-hand side 
 into those that depend on the location and shape of the singularity surface $\tau_0(\bec{x})$, and those that do not.
The first two terms in \eqref{eq:Eal5a} have no dependence on the singularity surface; the third term, $2nm\,\tau_\infty$, depends
on its location at $\tau_\infty$.
The final term, $2{\rm Re} S_{E}$, is obtained by taking the real part of the three integrals appearing in \eqref{eq:Se}.
The first two integrals are along the Euclidean time $\tau$ segments of the contour and are real-valued,
\[
2{\rm Re} \, S^{(1,2)}_{E}\,=\,2\, \int  d^3 x \left[ -\, \int_{+\infty}^{\tau_0(\bec{x})} d\tau \, {\cal L}_{E}(h_1) \,-\,
\int_{\tau_0(\bec{x})}^0d\tau \, {\cal L}_{E}(h_2) \right]\,,
 \label{eq:Se12R}
\]
while the remaining integral along the third segment of the contour appears to be purely imaginary. This last statement is almost correct,
as it applies to the bulk contribution of the Minkowski-time integral $\int_{0}^\infty dt \, {\cal L}(h_2)$, but not to the boundary contribution 
at $t\to \infty$. The full contribution from the third segment of the contour is,\footnote{The expression \eqref{eq:Se3R} for the boundary contribution 
to the Minkowski action is also in agreement with the construction in \cite{Son:1995wz} and \cite{Libanov:1997nt}.}
\begin{eqnarray}
2{\rm Re} \, S^{(3)}_{E} &=& 2\, \int  d^3 x \left[ -\,i 
 \int_{0}^\infty dt \, \int d^3x\,  \partial_t\left(\tilde \phi \, \partial_t h_2\right) \right] \nonumber\\
 &=&
 -\,\int d^3 k \,\, b_{\bf k}^* \, b^{}_{\bf k}\, e^{\omega_{\bf k}T-\theta}\,\,=\, -\,n
 \,.
 \label{eq:Se3R}
\end{eqnarray}

Accounting for the effect of the boundary contribution \eqref{eq:Se3R}
we can write the expression for the rate \eqref{eq:Eal5a} in the form:
\[
W(E,n) \,=\, n \left(\log\frac{\lambda n}{4}\,+\, \frac{3}{2} \log\frac{\varepsilon}{3\pi}\,+\,\frac{1}{2}\right)
\,-\,  2nm\,\tau_\infty\,-\, 2{\rm Re} \,S^{(1,2)}_{E}(\tau_0)\,.
\label{eq:EalWfin1}
\]

This is a remarkable formula in the sense that the expression on the right-hand side of \eqref{eq:EalWfin1} cleanly separates into
two parts. The first part,
$n \left(\log\frac{\lambda n}{4}\,+\, \frac{3}{2} \log\frac{\varepsilon}{3\pi}\,+\,\frac{1}{2}\right)$, does not depend on the shape of the singularity surface $\tau_0(\bec{x})$ and coincides with the known
tree-level result for the scattering rate in the non-relativistic limit $0<\varepsilon\ll 1$, as we will demonstrate below. The 
entire dependence of $W(E,n) $ on $\tau_0(\bec{x})$ is contained in the last two terms in \eqref{eq:EalWfin1}, which correspond to the purely quantum
contribution in the $\varepsilon \to 0$ limit.

The tree-level contribution to $W$ is well-known; it was computed using the resummation of
Feynman diagrams by solving the tree-level recursion relations \cite{Libanov:1994ug} and integrating over the phase-space. In the model
\eqref{eq:L}, the tree-level result to the order $\varepsilon^1$ was derived in \cite{Khoze:2014kka} and reads,
\[
W(E,n; \lambda)^{\rm tree}  \,=\,n\,\left( f_1(\lambda n)\,+\, f_2(\varepsilon) \right)\,,
\label{eq:Wtree}
\]
where
\begin{eqnarray}
\label{f0SSB}
f_1(\lambda n)&=&  \log\left(\frac{\lambda n}{4}\right) -1\,, 
\\
\label{feSSB}
f_2(\varepsilon)|_{\varepsilon\to 0}&\to& f_2(\varepsilon)^{\rm asympt}\,=\, 
\frac{3}{2}\left(\log\left(\frac{\varepsilon}{3\pi}\right) +1\right) -\frac{25}{12}\,\varepsilon\,.
\end{eqnarray}
First ignoring the order-$\varepsilon^1$ terms in the tree-level contribution, we see that the perturbative result
is correctly reproduced by the 
first two terms in the semiclassical expression on the right-hand side of \eqref{eq:EalWfin1},
\[
W(E,n)^{\rm tree} \,=\, n \left(\log\frac{\lambda n}{4}-1\right)\,+\, \frac{3n}{2}\left( \log\frac{\varepsilon}{3\pi}+1\right)
\,.
\label{eq:EalWfinfin}
\]
Schematically, the contribution $n\log \lambda n \subset W^{\rm tree}$ comes from squaring the tree-level amplitude on threshold
and dividing by the Bose symmetry factor,
$\frac{1}{n!}\, (n! \lambda^{n/2})^2 \sim n!\lambda^n\sim e^{n\log \lambda n}$, while the contribution $\frac{3}{2}n\log \varepsilon$
comes from the non-relativistic $n$-particle phase space volume factor $\varepsilon^{\frac{3n}{2}} \sim e^{\frac{3}{2}n\log \varepsilon}$. 
[We refer the interested reader to Refs.~\cite{Libanov:1994ug,Khoze:2014kka} for more details on the derivation of $W(E,n)^{\rm tree}$
directly in perturbation theory.]

\medskip

The apparent agreement between the first term in the expression on the right-hand side of  \eqref{eq:EalWfin1}  and the result of an independent tree-level perturbative calculation \eqref{eq:EalWfinfin}, provides a non-trivial consistency check of the semiclassical formalism that led us to \eqref{eq:EalWfin1}.

Furthermore, it was shown in \cite{Son:1995wz} that the tree-level results are also correctly reproduced by the semiclassical result to order-$\varepsilon^1$.
It would also be interesting to pursue such 
terms at the quantum level, but this is beyond the scope of this paper.
We will neglect all ${\cal O}(\varepsilon)$ terms as they are vanishing in the $\varepsilon \to 0$ limit.

\medskip

We can finally re-write the expression \eqref{eq:EalWfin1} for the rate $W(E,n)$ in the form \cite{Son:1995wz},
\[
W(E,n) \,=\, W(E,n; \lambda)^{\rm tree}  \,+\, \Delta W(E,n; \lambda)^{\rm quant}\,,
\label{eq:EalWfin2}
\]
where the quantum contribution is given by
\begin{eqnarray}
\Delta W^{\rm quant} &=&
\,-\,  2nm\,\tau_\infty\,-\, 2{\rm Re} \,S^{(1,2)}_{E}
\nonumber\\ 
&=& 2nm\,|\tau_\infty|\,+\, 2 \int  d^3 x \bigg[ \int_{+\infty}^{\tau_0(\bec{x})} d\tau \, {\cal L}_{E}(h_1) \,+\,
\int_{\tau_0(\bec{x})}^0d\tau \, {\cal L}_{E}(h_2) \bigg]
\label{eq:EWq} \\
&=& 2nm\,|\tau_\infty|\,-\, 2 \int  d^3 x \bigg[ \int^{+\infty}_{\tau_0(\bec{x})} d\tau \, {\cal L}_{E}(h_1) \,-\,
\int_{\tau_0(\bec{x})}^0d\tau \, {\cal L}_{E}(h_2) \bigg] \nonumber.
\end{eqnarray}
Here we have used the fact that $\tau_{\infty}$ is manifestly negative (as the singularity surface away at $\bec{x} \neq 0$ is by construction assumed to be located at negative $\tau$) to indicate that $-2nm\,\tau_\infty$ is a positive-valued contribution $+2nm\,|\tau_\infty|$.

The problem of finding the singularity surface $\tau_0(\bec{x})$ that extremises the expression \eqref{eq:EWq} has a simple
physical interpretation \cite{Son:1995wz,Gorsky:1993ix,Khoze:2017ifq}: 
it is equivalent to finding the shape of the membrane $\tau_0(\bec{x})$
at equilibrium, which has the 
surface energy  ${\rm Re} \,S^{(1,2)}_{E}$ and is pulled at the point $\bec{x}=0$ by a constant force equal to $nm$.
Note that even before the extremisation of \eqref{eq:EWq}
with respect to $\tau_0(\bec{x})$, both configurations $h_1(x)$ and $h_2(x)$ are tightly constrained. They are required to be
solutions of the classical equations; they have to have satisfy the correct boundary conditions in time, and consequentially, their energy is fixed: $h_1$ has $E=0$ and $h_2$ has $E=nm$ (in the $\varepsilon \to 0$ limit). These conditions 
constrain the extremisation of \eqref{eq:EWq} with respect to $\tau_0(\bec{x})$.

\medskip
\subsection{Computation of quantum effects at large $\lambda n$}
\label{sec:thin_w}
\medskip

In this and the following sections we will closely follow the calculation in Ref.~\cite{Khoze:2018kkz}.
We view the saddle-point 
field configuration in the model \eqref{eq:L} as a domain wall solution separating vacua with  different VEVs ($h \to \pm v$) on different
sides of the wall. Our scalar theory with spontaneous symmetry breaking clearly supports
such field configurations.
The solution is singular on the surface of the wall, which has thickness $\sim 1/m$. The effect of the `force'
$nm$ applied to the domain wall is to pull the centre of the wall upwards and gives it a profile $\tau_0(\bec{x})$, 
as depicted in Fig.~\ref{fig:tau}. 
The Euclidean action on the solution characterised by the domain wall at $\tau_0(\bec{x})$, becomes a functional of the surface function,
$S_E[\tau_0(\bec{x})]$. The shape of the surface will
be straightforward to determine by extremising the action $S_E[\tau_0(\bec{x})]$, which we will compute in the thin-wall approximation, over
all surface profile functions $\tau_0(\bec{x})$. The validity of the thin wall approximation will be be justified in the limit $\lambda n \to \infty$.
The idea of using the thin-wall approximation in the large $\lambda n$ limit
was pursued earlier by Gorsky and Voloshin in Ref.~\cite{Gorsky:1993ix}, where it was applied to the standard regular bubbles of the
false vacuum. These were interpreted as intermediate physical bubble states in the process $1^* \to {\rm Bubble} \to n$. Conceptually, 
this is different from
our approach, where the thin-wall solutions are singular points on the deformed contours of the path integral in Euclidean time. In our setting they cannot be interpreted 
as physical  macroscopic states in real Minkowski time representing an intermediate state in the $1^* \to n$ process.

From now on we concentrate on the large $\lambda n$ regime, where the semiclassical rate is non-preturbative.

\medskip
\subsubsection{Classical fields and singularity surfaces}
\label{sec:cl-surf}
\medskip

Our first task is to implement the realisation of the singular field configuration $h(x)$ in terms of domain walls with thin-wall singular surfaces.
The $h_1$ branch of the solution is defined on the first part of the time-evolution contour, i.e. the imaginary time interval
$+\infty > \tau \ge \tau_0(\bec{x})$.
It is given by,
\[
h_1(\tau,\bec{x}) \,=\, h_{0E} (\tau-\tau_0(\bec{x})) \,+\, \delta h_1(\tau,\bec{x})\,.
\label{clas_sol111Aag}
\]
The first term on the right-hand side of \eqref{clas_sol111Aag} is the familiar singular domain wall,
\[
h_{0E} (\tau-\tau_0(\bec{x})) \,=\, v \left(\frac{ 1\,+\,e^{-m (\tau-\tau_0(\bec{x}))}}
{1\,-\, e^{-m (\tau-\tau_0(\bec{x}))}}\right)\,,
\label{clas_sol222inf}
\]
with its centre (or position) at $\tau = \tau_0(\bec{x})$.
This profile is similar to the one depicted in Fig.~\ref{fig:kink}. The field configuration interpolates between $h=+v$ at $\tau \gg  \tau_0(\bec{x}) $ and  $h=-v$ at $\tau \ll  \tau_0(\bec{x})$,
and is singular on the 3-dimensional surface $\tau=\tau_0(\bec{x})$. Since $\tau_0(\bec{x})$ depends on the spatial variable,
the correction $\delta h_1(\tau,\bec{x})$ is required in \eqref{clas_sol111Aag} to ensure that the entire field configuration $h_1(x)$ 
satisfies the classical equations. The $\delta h_1$ term vanishes on the singularity surface; in fact it is straightforward to show that 
$\delta h_1 \sim (\tau-\tau_0(\bec{x}))^3$ near the singularity surface by solving the linearised classical equations for $\delta h_1$ 
in the background of the singular $h_0$ \cite{Son:1995wz}.
The initial time condition on $h_1$ is
\[
\lim_{\tau\to \infty} h_1 (x)\,=\, v + {\cal O}(e^{-m\tau})\,,
\]
which also guarantees that $\delta h_1(x)\to 0$ exponentially fast at large $\tau$.
Hence, in computing the action integral of $h_1(x)$ in the thin-wall approximation, where the main contribution comes from $\tau$ in the vicinity
of $\tau_0(\bec{x})$, it will be a good approximation to neglect $\delta h_1(x)$ and 
use,
\[
{\rm thin\,  wall}\, : \quad h_1(\tau,\bec{x}) \,\approx\, h_{0E} (\tau-\tau_0(\bec{x})) \,.
\label{eq:thwapp1}
\]

Now consider the second branch of the solution, $h_2(x)$. We search for solutions of the form required by Eq.~\eqref{clas_sol2A},
\[
h_2(t_{\mathbb{C}},\bec{x}) \,=\, h_0 (t_{\mathbb{C}}) \,+\, \tilde{\phi}(t_{\mathbb{C}},\bec{x})
\,,
\label{clas_sol2Aag}
\]
The first term on the right-hand side of \eqref{clas_sol2Aag} is the uniform in space and singular on the plane $\tau=\tau_{\infty}$
classical configuration
\[
 h_0 (t_{\mathbb{C}}) \,=\, v\, \left(\frac{ 1\,+\,e^{im (t_{\mathbb{C}}-i\tau_\infty)}}
{1\,-\, e^{im (t_{\mathbb{C}}-i\tau_\infty)}}\right)
\,.
\label{clas_sol22inf}
\]
In the previous section we derived the asymptotic form for the second term, $\tilde{\phi}(t_{\mathbb{C}},\bec{x})$,  
appearing on the right-hand side of \eqref{clas_sol2Aag}: for the final part of the time-evolution contour, where 
$t_{\mathbb{C}} = t \to +\infty$ we have,
\[
\lim_{t\to + \infty}\, \tilde{\phi}( t, {\bf x})
  \,=\, \int\frac{ d^3 k}{(2\pi)^{3/2}}\frac{1}{\sqrt{2\omega_{\bf k}}}  \,\, f_{\bf k}\, e^{-i\omega_{\bf k}t} \,.
  \label{eq:tilphiinf}
\]
This is in agreement with
Eqs.~\eqref{eq.Eh0} and \eqref{eq.Etilphi2} and its characteristic feature is that it contains only the negative frequency components
(at large $t$).
The coefficients of positive frequency components that were present in $\tilde{\phi}( t, {\bf x})$ at earlier times, closer to the origin at 
$t\sim 0$ become 
suppressed as the real time variable $t$ grows and ultimately disappear for a sufficiently large positive $t$.
We are now going to assume that the asymptotic expression \eqref{eq:tilphiinf} which is valid in the $mt \gg 1$ regime on or near
the real time axis, in fact also continues to hold when $\tilde{\phi}( t+i\tau, {\bf x})$ moves in the $\tau$ direction, i.e. perpendicular
to the real time contour at large fixed value of $t$.
More precisely we expect that the equation \eqref{eq:tilphiinf} generalises to the complex time variable $t_{\mathbb{C}}$ 
and holds as long as the real time coordinate $t$ is large ($t\gg 1/m$),
\[
\lim_{ t \to + \infty}\, \tilde{\phi}( t_{\mathbb{C}}, {\bf k})\,=\,
 \frac{1}{\sqrt{2\omega_{\bf k}}}  \,\,  f_{\bf k}\, e^{-i\omega_{\bf k}t_{\mathbb{C}}} \,=\,
 \frac{1}{\sqrt{2\omega_{\bf k}}}  \,\,  f_{\bf k}\, e^{\omega_{\bf k}\tau} \, e^{-i\omega_{\bf k}t} 
   \,.
  \label{eq:tilphiinf2}
\]
As always,  $t_{\mathbb{C}}=t+i\tau$, and for concreteness we will take 
the $\tau$ component to be negative, i.e. we will only need this expression for shifting downwards from the real time 
contour at large $t$.

\medskip

 \begin{figure*}[t]
\begin{center}
\includegraphics[width=0.95\textwidth]{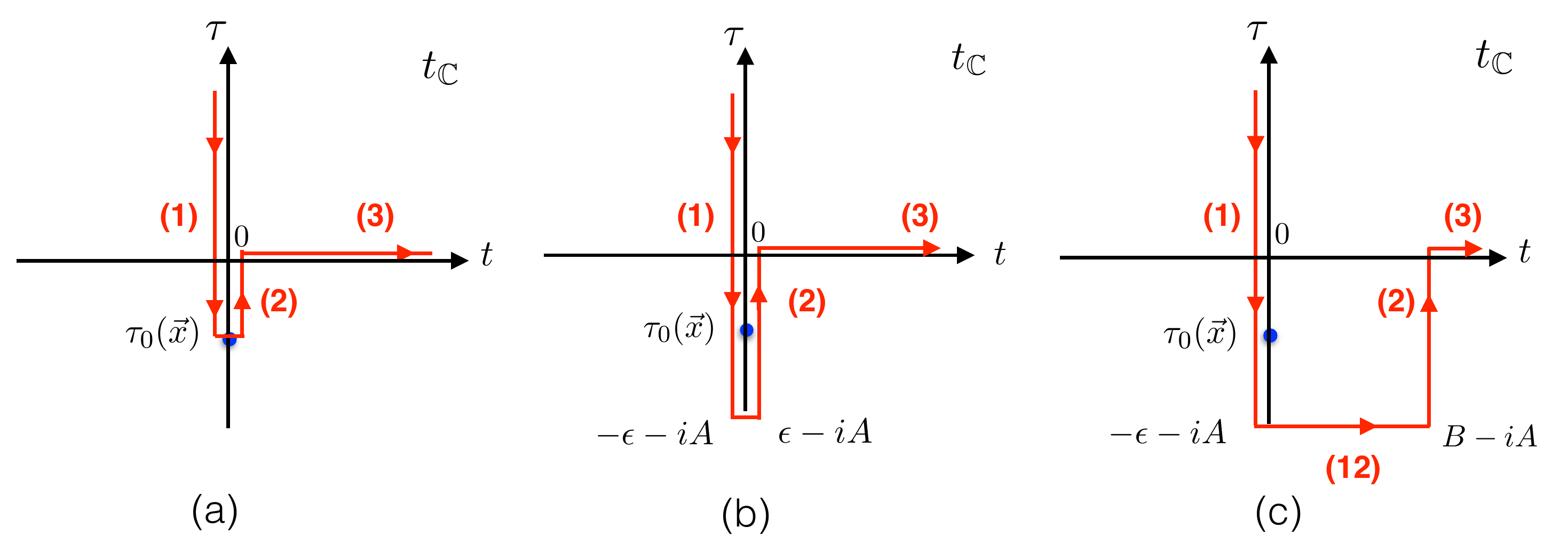}
\end{center}
\vskip-.5cm
\caption{Deformations of the time evolution contour in $t_{\mathbb{C}}$.  {\bf Plot (a)} shows the
original contour that touches the sigularity located at $t=0$, $\tau=\tau_0(\bec{x})$.
{\bf Plot (b)} gives the resolved contour, now surrounding the singularity with the vertical segments of the contour shifted infinitesimally by
 $\pm i \epsilon$ and descending to $\tau = -A$.
{\bf Plot (c)} shows now a finite deformation of the vertical part (2) of the contour the right. We use large shift values, $-A\gg 1/m$ and $B\gg 1/m$ to justify the thin wall approximation. Consequitive contour segments are denoted (1), (12), (2) and (3).}
\label{fig:contour3}
\end{figure*}

We now turn to the evaluation of the Euclidean action integrals appearing in \eqref{eq:Se12R} and \eqref{eq:EWq}. 
On the first segment
of the contour, indicated as (1)  in Fig.~\ref{fig:contour3}~(a), the classical field configuration is $h_1(x)$, while on the
segment (2) of the contour in Fig.~\ref{fig:contour3}~(a), the field is $h_2(x)$, hence,
\[
{\rm Fig.~\ref{fig:contour3} (a)}: \quad
-\,{\rm Re} \, S^{(1,2)}_{E}\,=\, \int  d^3 x \left[ \int_{+\infty}^{\tau_0(\bec{x})} d\tau \, {\cal L}_{E}(h_1) \,+\,
\int_{\tau_0(\bec{x})}^0d\tau \, {\cal L}_{E}(h_2) \right].
 \label{eq:Se12R2}
\]
The two individual integrals in \eqref{eq:Se12R2} are singular at the integration limit $\tau=\tau_0(\bec{x})$. 
However, their sum is expected to be finite, which is also known from the Landau-WKB approach in Quantum Mechanics \cite{landau1981quantum}. 

Instead of reaching the singularity and then cancelling 
the resulting infinite contributions 
at $\tau\to \tau_0(\bec{x})$, we advocate a more practical approach and deform the integration contour to encircle the singularity,
as shown in the contour deformation from Fig.~\ref{fig:contour3} (a) to  Fig.~\ref{fig:contour3} (b).
The contour is shifted infinitesimally by $t=-\epsilon$
in the first integral in \eqref{eq:Se12R2} and by $t=+\epsilon$ in the second. 
Since the integration contour in Fig.~\ref{fig:contour3} (b) passes on either side of the singularity at $\tau=\tau_0(\bec{x})$, the action
integrals and the solutions themselves are finite. One can extend the integration contours down to $\tau=-\infty$ or to 
any arbitrary large negative value $\tau=-A$.
At $\tau=-A$, where $\tau$ is well below the final singularity surface $\tau_\infty$, the two contours are joined. As a result,  
the action integrals now read:
\[
{\rm Fig.~\ref{fig:contour3} (b)}: \quad
-\,{\rm Re} \, S^{(1,2)}_{E}\,=\,
\int_{\,+\infty-i\epsilon}^{\,-A-i\epsilon} d\tau \, L_{E}[h_1] \,+\,
\int_{\,-A+i\epsilon}^{\,0+i\epsilon}d\tau \, L_{E}[h_2] \,,
 \label{eq:Se12R22}
\]
where $L_{E}=\int d^3x\, {\cal L}_{E}$, and each of the two integrals in \eqref{eq:Se12R22} is finite.
The first integral in \eqref{eq:Se12R22} depends on the classical branch $h_1(x)$, and in the thin wall
approximation \eqref{eq:thwapp1}
we will be able to evaluate it as the functional of the surface $\tau_0(\bec{x})$ using the $h_{0E}$ profile in \eqref{clas_sol222inf}.

The second integral in \eqref{eq:Se12R22} is evaluated on the classical configuration $h_2(x)$. It is given by 
\eqref{clas_sol2Aag}, where the correction $\tilde{\phi}( t_{\mathbb{C}}, {\bf x})$ to the classical profile 
$h_0 (t_{\mathbb{C}}) $ in \eqref{clas_sol22inf} is known at large values of
$t$, see Eq.~\eqref{eq:tilphiinf2}. To make use of these expressions for $h_2(x)$ we continue 
shifting the contour to the right by a 
constant value $B$ as shown in Fig.~\ref{fig:contour3} (c).
The resulting contributions to the Euclidean action from the integration contour in Fig.~\ref{fig:contour3} (c)
are given by the following integrals,
\[
{\rm Fig.~\ref{fig:contour3} (c)}: \quad
-\,{\rm Re} \, S^{(1,12,2)}_{E}\,=\,
\int_{\,+\infty-i\epsilon}^{\,-A-i\epsilon} d\tau \, L_{E}[h_1] \,+\,
i\int_{(12)}d t  \, L[h_2] \,+\,
\int_{\,-A+iB}^{\,0+iB}d\tau \, L_{E}[h_2] \,.
 \label{eq:Se12R22c}
\]
An obvious consequence of the thin wall approximation is that the middle integral on the right hand side
of \eqref{eq:Se12R22c} vanishes for $A$ sufficiently far below $\tau_\infty$ since in this case we are sufficiently deep
into the $h_2 = -v$ domain, the field configuration is constant there and the action on the (12) segment of the contour vanishes,
$\int_{(12)}d t  \, L[h_2]\,=\, 0$.

 \begin{figure*}[t]
\begin{center}
\includegraphics[width=0.65\textwidth]{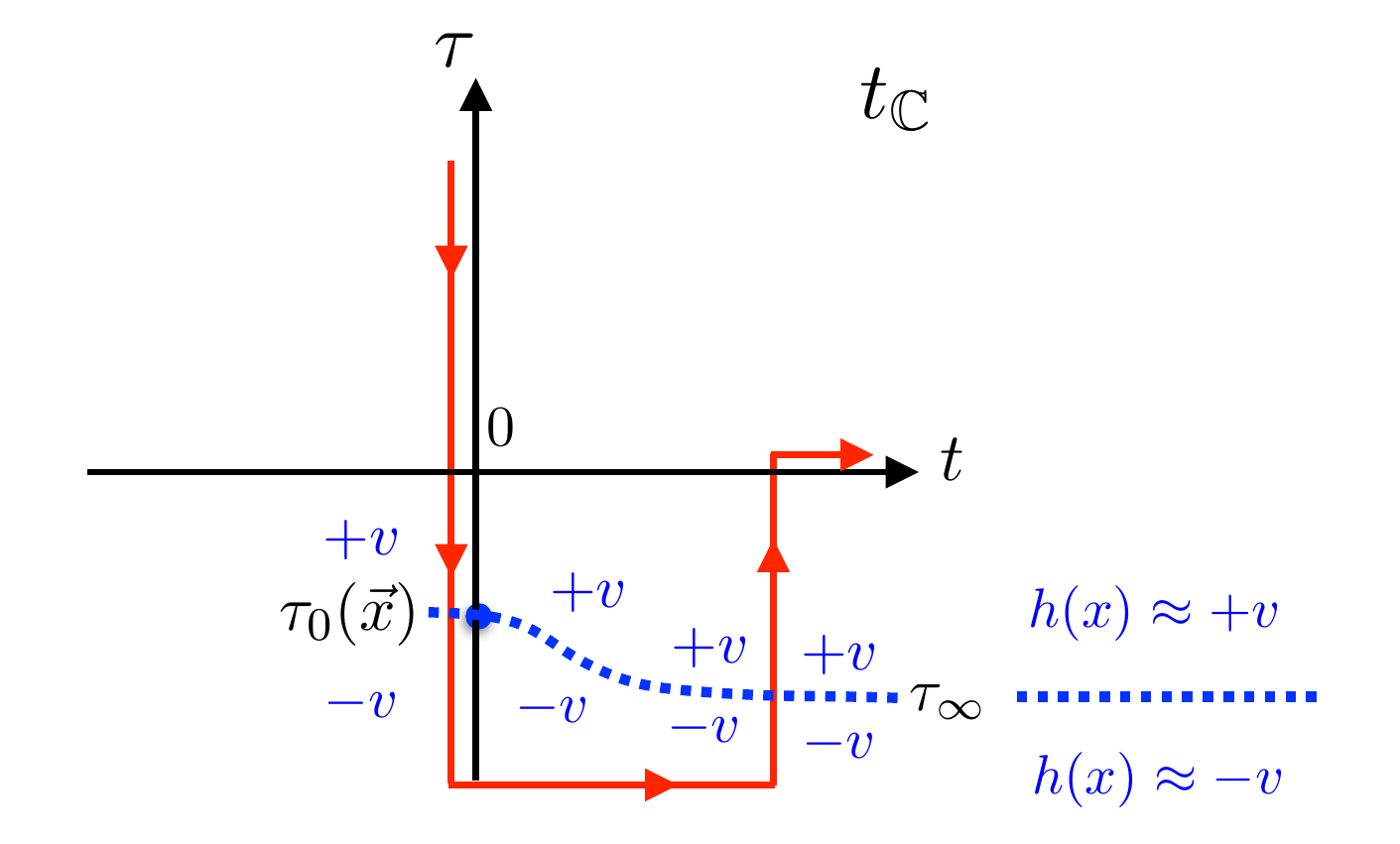}
\end{center}
\vskip-.5cm
\caption{The same complex-time evolution contour as in Fig.~\ref{fig:contour3} (c). The boundary separating 
the domains $h(x) \to +v$ and 
$h(x)\to -v$ for the classical solution in a thin wall approximation is shown as the dotted blue line.The singularity of the solution is 
at the point $t=0$, $\tau=\tau_0({\bf x})$, as depicted by a blue blob on the dotted line of the inter-domain boundary.}
\label{fig:contour_fin}
\end{figure*}

\medskip

Next, we can readily evaluate the last integral in \eqref{eq:Se12R22c}. It arises from segment (2) of the contour
in Fig.~\ref{fig:contour3} (c), which is the integral over the imaginary time component $d\tau$ and is
situated at a fixed value of real time at ${\rm Re} \, t_{\mathbb{C}} =B \gg 1/m$. 
Hence we can use the 
asymptotic expression~\eqref{eq:tilphiinf2} for  $\tilde{\phi}( t_{\mathbb{C}}, {\bf x})$ on this segment of the contour,
so that the entire solution $h_2(x)$ is given by,
\[
h_2^{\rm segment\, (2)}\,=\,   v\, \left(\frac{ e^{-im B -m(|\tau|-|\tau_\infty|)}\,+\,1}
{e^{-im B -m(|\tau|-|\tau_\infty|)}\,-\,1}\right)
\,+\,  \int\frac{ d^3 k}{(2\pi)^{3/2}}\,
 \frac{1}{\sqrt{2\omega_{\bf k}}}  \,\, f_{\bf k}\, e^{-\omega_{\bf k}|\tau|} \, e^{-i\omega_{\bf k}B} 
   \,.
  \label{eq:tilphiinf22}
\]
Note that on this segment of the contour $t=B$, $-A\le \tau \le 0$, hence 
$0\le |\tau|\le A$ and $0<|\tau_\infty|\ll A$.
In the large $\lambda n$ limit, we will find in the following section that in fact  $0\ll |\tau_\infty|$, 
and that the 
only non-trivial contribution in the thin wall limit on this segment of the contour will come from
the first term on the right hand side of \eqref{eq:tilphiinf22}. The location of the wall
 separating the two $\pm v$ domains of the field configuration is depicted in 
 Fig.~\ref{fig:contour_fin}.
The $\tilde{\phi}$ term on its own cannot
contribute to the action integral since it contains only the negative frequencies. Furthermore,
its overlap with the $h_0$ configuration 
 at $\tau \approx \tau_\infty$ is exponentially suppressed by $e^{-m|\tau|_\infty} \ll 1$.  Hence we are left with,
\[
{\rm thin\,  wall}\, : \quad h_2(\tau,\bec{x}) \,\approx\, h_{0E} (\tau-\tau_\infty) \,.
\label{eq:thwapp2}
\]
This equation is applicable on segment (2) of the contour in Fig.~\ref{fig:contour3}~(c), where 
 the argument $\tau$ of both functions in \eqref{eq:thwapp2} is understood as $\tau - iB$.
 
 Equations \eqref{eq:thwapp1} and \eqref{eq:thwapp2} give us the required precise implementation of the thin wall
 approximation that we will apply in what follows.
 In both cases, the field configurations ($h_1$ in \eqref{eq:thwapp1} and $h_2$ in \eqref{eq:thwapp1}) are approximated
 in the thin wall approach
 by the Brown's solution profile $h_{0E}$. The important difference between the two cases, however, is that the  
 domain wall in \eqref{eq:thwapp1} is the ${\bf x}$-dependent surface $\tau_0({\bf x})$, while in the case of the
 $h_2$ configuration in \eqref{eq:thwapp2}, the domain wall is at $\tau_\infty$ and is spatially-independent.
 As the result, the the first integral on the right hand side of our expression for the action in \eqref{eq:Se12R22c},
 is a functional of the domain-wall surface $\tau_0({\bf x})$,
 \[
 S^{(1)}_{E}\,=\,
\int^{\,+\infty-i\epsilon}_{\,-A-i\epsilon} d\tau \, L_{E}[h_1]\,=\, S_{E}[\tau_0(({\bf x})]\,,
 \label{eq:Se12R22c1}
\]
while the the third integral in \eqref{eq:Se12R22c} is evaluated on the uniform in space solution \eqref{eq:thwapp2}
and is a constant,
 \[
 S^{(2)}_{E}\,=\,
-\, \int_{\,-A+iB}^{\,0+iB}d\tau \, L_{E}[h_2] \,=\, -\,{\rm const}\,.
 \label{eq:Se12R22c2}
\]
In both cases, on segment (1) and segment (2) of the contour, the field configurations are regular,
since, by construction, the contour avoids the singularity by the $-i \epsilon$ shift in the first integral and by the
$+iB$ shift in the second.
 
We now proceed to compute the integral in \eqref{eq:Se12R22c2}.
This integral is evaluated on the field configuration,
\[
h_{2} (\tau+iB) \,=\, v \left(\frac{ 1\,+\,e^{-m (\tau-\tau_\infty+iB)}}
{1\,-\, e^{-m (\tau-\tau_\infty+iB)}}\right)\,,
\label{clas_sol22infh1}
\]
and can be calculated exactly\footnote{For simplicity we extend the integration limits 
along the vertical axis to $\pm \infty$. Given the narrow width of the wall, any changes due to this extension are negligible.}, 
giving,
\[
\int^{+\infty+iB}_{-\infty+iB} d\tau \int  d^3 x {\cal L}_{E}(h_2) \,=\,
 \mu \int_0^R 4\pi r^2 dr\,=\, 
\mu\,  \frac{4\pi}{3} \, R^3\,.
 \label{eq:Se1act}
\]
Since the field is uniform in space, to ensure that the $\int d^3x$ is finite,
 we used the finite volume regularisation with finite spatial radius $R$.     
 The infinite-volume limit, $R\to \infty$, will be taken at the end of the calculation, 
after combining the two action integrals in \eqref{eq:Se12R22c1} and \eqref{eq:Se1act}.
The parameter $\mu$ appearing on the right-hand side of \eqref{eq:Se1act}
is the surface tension on the bubble solution \eqref{clas_sol22infh1},
\[
\mu\,=\, \int_{-\infty+i\epsilon}^{+\infty+i\epsilon} d\tau \left( \frac{1}{2} \left(\frac{d h}{d\tau}\right)^2 +
 \frac{\lambda}{4} \left( h^2 - v^2\right)^2 \right) \,=\,  \frac{m^3}{3 \lambda}\,.
 \label{eq:mu}
\]
It can easily be checked (e.g. by use of the residue theorem)
that the value of $\mu$ does not depend on the numerical value of $iB$ in the shift of the integration contour:
any value of $iB\neq 0$ that shifts the contour such that it does not pass directly through the singularity at $\tau_\infty$
will suffice.
This shift-independence argument also applies to the integral on the fist segment of the contour where the shift is $-i\epsilon$.

\medskip

Let us summarise our construction up to this point. We have derived the expression for the contribution of quantum effects \eqref{eq:EWq}
to the semiclassical
rate $W$ \eqref{eq:EalWfin2} in the form,
\[
\frac{1}{2} \Delta W^{\rm quant}\,=\, 
nm\,|\tau_\infty|\,-\,  
\underbrace{\int^{\,+\infty+i\epsilon}_{\,-\infty-i\epsilon} d\tau \, L_{E}(h_1;\tau_0(\bec{x}))}_{\equiv \, S_{E}[\tau_0(\bec{x})]} 
\,+\,\, \frac{4\pi}{3} \,\mu R^3
\,.\label{eq:halfEWqnew}
\]
We note that no extremisation of the rate with respect to the surface $\tau=\tau_0(\bec{x})$ has been carried out so far.
The expression in \eqref{eq:halfEWqnew} is the general formula equivalent to the expression in \eqref{eq:EWq}.
It will be now extremised with respect to the domain wall surface $\tau_0(\bec{x})$. 
The constant term $\frac{4\pi}{3} \,\mu R^3$ will be cancelled with its counterpart arising from the action integral in \eqref{eq:halfEWqnew} 
before the infinite-volume limit is taken.

\medskip

Following from the discussion at the end of section~\ref{sec:son_loops}, the shape of the singular surface, $\tau_0(\bec{x})$, should be determined by
extremising the function $\Delta W^{\rm quant}$ in the exponent of the multiparticle probability rate.
This is equivalent to
searching for a stationary (i.e. equilibrium surface) configuration described by the `surface energy' functional, given by the right hand side of  \eqref{eq:halfEWqnew}.
Finding the stationary point corresponds to balancing the surface energy of the stretched surface, given by the integral 
$S_{E}[\tau_0(\bec{x})]$ in \eqref{eq:halfEWqnew},
against the force $nm$ that stretches the surface $\tau_0(\bec{x})$ by the
amount $|\tau_\infty|$. The third term on the right hand side of \eqref{eq:halfEWqnew} plays no role in the extremisation procedure over $\tau_0(\bec{x})$ and gives a positive-valued constant 
contribution to $\frac{1}{2} \Delta W^{\rm quant}$ that will be cancelled against its counterpart in $S_{E}[\tau_0(\bec{x})]$.
The overall result will be finite, as expected in the infinite volume limit. 

\medskip
\subsubsection{Extremal singular surface in the thin wall approximation}
\label{sec:thin_surf}
\medskip

The action $S_{E}[\tau_0(\bec{x})]$ can now be written as an integral over the domain wall surface $\tau_0(\bec{x})$
in the thin-wall approximation. This is equivalent to stating that
 the action is simply the surface tension of the domain wall $\mu$, as computed in \eqref{eq:mu},
 multiplied by the area. The 3-dimensional area of a curved surface in 3+1 dimensions 
 has infinitesimal element $4\pi \mu \, r^2 \sqrt{(d\tau)^2 +(dr)^2}$. Hence, the action reads,
 \[ 
 S_{E}[ \tau_0(r)] \,=\,
\int_{\tau_\infty}^0 d\tau \,4\pi \mu \, r^2 \sqrt{1+\dot r^2}\,\,\equiv\,  \int_{\tau_\infty}^0 d\tau \, L (r,\dot r)
\,,
\label{eq_thinw}
\]
where $r=|\bec{x}|$ and  $\dot r= dr/d \tau$. The integral depends on the choice of the domain wall surface 
$\tau_0(\bec{x})$
implicitly via the $\tau$-dependence of $r(\tau)$ and $\dot r(\tau)$, which are computed on the domain wall.
 
Since  $L (r,\dot r)$ can be interpreted as the Lagrangian, one can introduce the  Hamiltonian function 
defined in the standard way\footnote{In Euclidean space
$L=K-V$ and $H=-K+V$, where $K$ and $V$ are the kinetic and potential energies respectively.} as the Legendre transformation,
\[
H(p,r)\,=\, L (r,\dot r) \,-\, p\, \dot r\,,
\label{Hdef}
\]
where the momentum $p$, conjugate to the coordinate $r$, is 
\[
p\,=\, \frac{\partial L (r,\dot r) }{\partial \dot r}\,\,\,=\, 4\pi\, \mu \frac{r^2 \dot r}{\sqrt{1+\dot r^2}}
\label{eq:defp1}
\]
On a classical trajectory $r=r(\tau)$ that satisfies the Euler-Lagrange equations for $L (r,\dot r)$,
the Hamiltonian is time-independent ($dH/d \tau=0$) and is given by the energy $E$ of said classical trajectory.\footnote{One should not confuse the energy of the classical trajectory $r=r(\tau)$ -- which is essentially the Euclidean surface energy of the domain wall --
with the energy of the classical solutions $h_1$ and $h_2$. Both energy variables are denoted as $E$, but the energy of the domain wall
at the stationary point will turn out to be $E=mn$, while the energy of the corresponding field configuration $h_1$ was $E=0$.}
Hence, on a stationary point of  $S_{E}[ \tau_0(r)]$ with energy $E$, we can rewrite the action as,
\[ 
S_{E}[ \tau_0(r)]_{\rm stationary} \,=\, -\tau_{\infty}\,E
\,+\,
 \int_{\tau_\infty}^0 d\tau \, (L - H) \, 
\,=\,  -\,E\tau_{\infty} \,+\, \int_{R}^0 p(E) \,dr
\,.
\]
In the equation above
we have added and subtracted the constant energy of the solution ($E=H$) and used the fact that $L-H=p \dot{r}$. 
The lower and upper integration limits are consequently set to $r(\tau_\infty)=R$ and $r(0)=0$. 
The expression above gives us the action functional $S_{E}[ \tau_0(r)]$ on a trajectory $r(\tau)$, or equivalently $\tau=\tau_0(r)$, 
which is a classical trajectory (i.e. an extremum of the action for a fixed energy $E$).
Equivalently, for the stationary point of the expression in \eqref{eq:halfEWqnew} we have,
\[
\frac{1}{2} \Delta W^{\rm quant} \,=\, 
(E-nm)\tau_\infty\,-\,   \int_{R}^0 p(E) \,dr
\,+\, \frac{4\pi}{3} \,\mu R^3
\,.\label{eq:halfEWqst}
\]
Extremization of this expression with respect to the parameter $\tau_\infty$ gives $E=nm$, thus selecting this
energy for the classical trajectory as required,
\[
\frac{1}{2} \Delta W^{\rm quant}_{\quad\rm stationary} \,=\, 
  -\, \int_{R}^0 p(E) \,dr
\,+\, \frac{4\pi}{3} \,\mu R^3 \,, \qquad E=nm 
\,.\label{eq:halfEWqfi}
\]

To evaluate \eqref{eq:halfEWqfi} we need to determine the dependence of the momentum of the classical trajectory on its energy.
To find $p(E)$, we start by writing the expression for the energy, $E= L-p\dot{r}$, in the form,
\[
E\,=\, 4\pi \mu \, r^2 \sqrt{1+\dot r^2} \,-\, 4\pi\, \mu \frac{r^2 \dot r}{\sqrt{1+\dot r^2}}\,=\,
4\pi\, \mu \frac{r^2}{\sqrt{1+\dot r^2}}\,,
\label{eq:Econs}
\]
and then compute the combination $E^2 + p^2$ using the above expression and \eqref{eq:defp1},
\[
E^2\,+\, p^2\,=\, \left(4\pi \mu \,r^2 \right)^2 \left( \frac{1}{1+\dot r^2}\,+\, \frac{\dot r^2}{1+\dot r^2}\right)\,=\,
 \left(4\pi \mu \,r^2 \right)^2\,.
 \]
This gives the desired expression for the momentum $p=p(E)$,
\[
p(E,r) \,=\,-\,  4\pi \, \mu \, \sqrt{r^4- \left(\frac{E}{4\pi\mu}\right)^2}\,,
\label{eq:conjp}
\] 
where we have selected the negative root for the momentum
in accordance with the fact that $p(\tau) \propto \dot{r}$ (as follows from \eqref{eq:defp1})
and that $r(\tau)$ is a monotonically decreasing function.

Substituting this into expression \eqref{eq:halfEWqfi} we have,
\[
\frac{1}{2} \Delta W^{\rm quant} \,=\, 
-  \int_{R}^{r_0} p(E) \,dr
\,+\, \frac{4\pi}{3} \,\mu R^3\,=\, 
-\, \int^{R}_{r_0}  
4\pi \, \mu \, \sqrt{r^4-r_0^4} \,dr
\,+\, \frac{4\pi}{3} \,\mu R^3
\,.\label{eq:pdrN}
\]
The minimal value of the momentum (and the lower bound of the integral in \eqref{eq:pdrN}) is cut-off at the critical radius $r_0$,
\[
r_0^2 \,=\, \frac{E}{4\pi\mu} \,.
\label{r0def}
\]
Note that the contribution to the integral \eqref{eq:pdrN} on the interval $0\le r \le r_0$ will be considered in \refsec\ref{sec:beyond_tw}; let us temporarily ignore it.

The integral on the right-hand side of \eqref{eq:pdrN} is evaluated as follows,
\[
\int_1^{R/r_0}
 \sqrt{x^4-1} \,\,dx\,=\, \left[\frac{1}{3}\, x\, \sqrt{x^4-1}\,- \, \frac{2}{3} \, i\,  {\rm EllipticF}[{\rm ArcSin}(x), -1]\right]^{x=R/r_0}_{x=1}
 \nonumber
 \]
 where the {\it Mathematica} function ${\rm EllipticF}[z,m]$ is also known as the elliptic integral of the first kind, $F(z|m)$.
In the $R/r_0 \to \infty$ limit, the integral simplifies to,
\begin{eqnarray}
(-4\pi \mu r_0^3)\, \int_1^{R/r_0}  \sqrt{x^4-1} \,\,dx &\to& 
-\, \frac{4\pi}{3} \,\mu R^3
\,+\, 4\pi \mu r_0^3\, \sqrt{4\pi}\,
\frac{1}{3}\, \frac{\Gamma(5/4)}{\Gamma(3/4)}
\nonumber\\
&=& -\, \frac{4\pi}{3} \,\mu R^3
\,+\, \frac{E^{3/2}}{\sqrt{\mu}}\, 
\frac{1}{3}\, \frac{\Gamma(5/4)}{\Gamma(3/4)}\,.
\label{eq:pdrNev}
\end{eqnarray}
Note that, 
\[
4 \pi \mu r_0^3 \,=\, \frac{E^{3/2}}{\sqrt{\mu}}\frac{1}{\sqrt{4\pi}}\,=\,
n \sqrt{\lambda n} \,\frac{\sqrt{3}}{\sqrt{4\pi}}\,.
\label{eq:ident}
\]
We see that the large volume constant term $\frac{4\pi}{3} \,\mu R^3$ cancels between the expressions in \eqref{eq:pdrNev}
and \eqref{eq:pdrN}, as expected. The final result for the thin-wall trajectory contribution to the quantum rate is given by,
\[
 \Delta W^{\rm quant} \,=\, 
\frac{E^{3/2}}{\sqrt{\mu}}\, 
\frac{2}{3}\, \frac{\Gamma(5/4)}{\Gamma(3/4)} \,=\, 
\, \frac{1}{\lambda} \, (\lambda n)^{3/2}\, \frac{2}{\sqrt{3}}\,
\frac{\Gamma(5/4)}{\Gamma(3/4)}\,\simeq\, \, 0.854\,  n \sqrt{\lambda n}
\,.\label{eq:pdrNfinalr0}
\]
We note that this expression is positive-valued, that it grows in the limit of $\lambda n \to \infty$, and that it has
the correct scaling properties for the semiclassical result, i.e. it is of the form $1/\lambda$ times a function of $\lambda n$.

Our result \eqref{eq:pdrNfinalr0} and the derivation we presented, followed closely the construction in \cite{Khoze:2018kkz,Khoze:2017ifq}.
The expression \eqref{eq:pdrNfinalr0} 
is also in agreement with the formula derived much earlier in Ref.~\cite{Gorsky:1993ix}, based on a somewhat different semiclassical reasoning
involving regular thin-wall bubble configurations in Euclidean and Minkowski time.

Importantly, the thin-wall approximation is justified in the $\lambda n \gg 1$ limit on the extremal surface in the regime where $r(\tau)>r_0$,
as originally noted in \cite{Gorsky:1993ix}.
The thin-wall regime corresponds to the spatial radius of the bubble (i.e. the spatial extent of the $O(3)$ symmetric configuration at a fixed $\tau$)
being much greater than the thickness of the wall,
$r \gg 1/m$. For the classical configuration at hand, the radius is always greater than the critical radius,
\[
r m\, \ge\, r_0 m\,=\, m\left( \frac{E}{4\pi \mu}\right)^{1/2} \propto\, \left( \frac{\lambda\, E}{m}\right)^{1/2} =\,
 \sqrt{\lambda n}\,\,\gg\, 1\,,
\]
where we have used the value for the energy  $E=nm$ on our solution.

\medskip
\subsection{Singular surfaces at $r\le r_0$: beyond thin walls}
\label{sec:beyond_tw}
\medskip

\begin{figure*}[t]
\begin{center}
\includegraphics[width=0.6\textwidth]{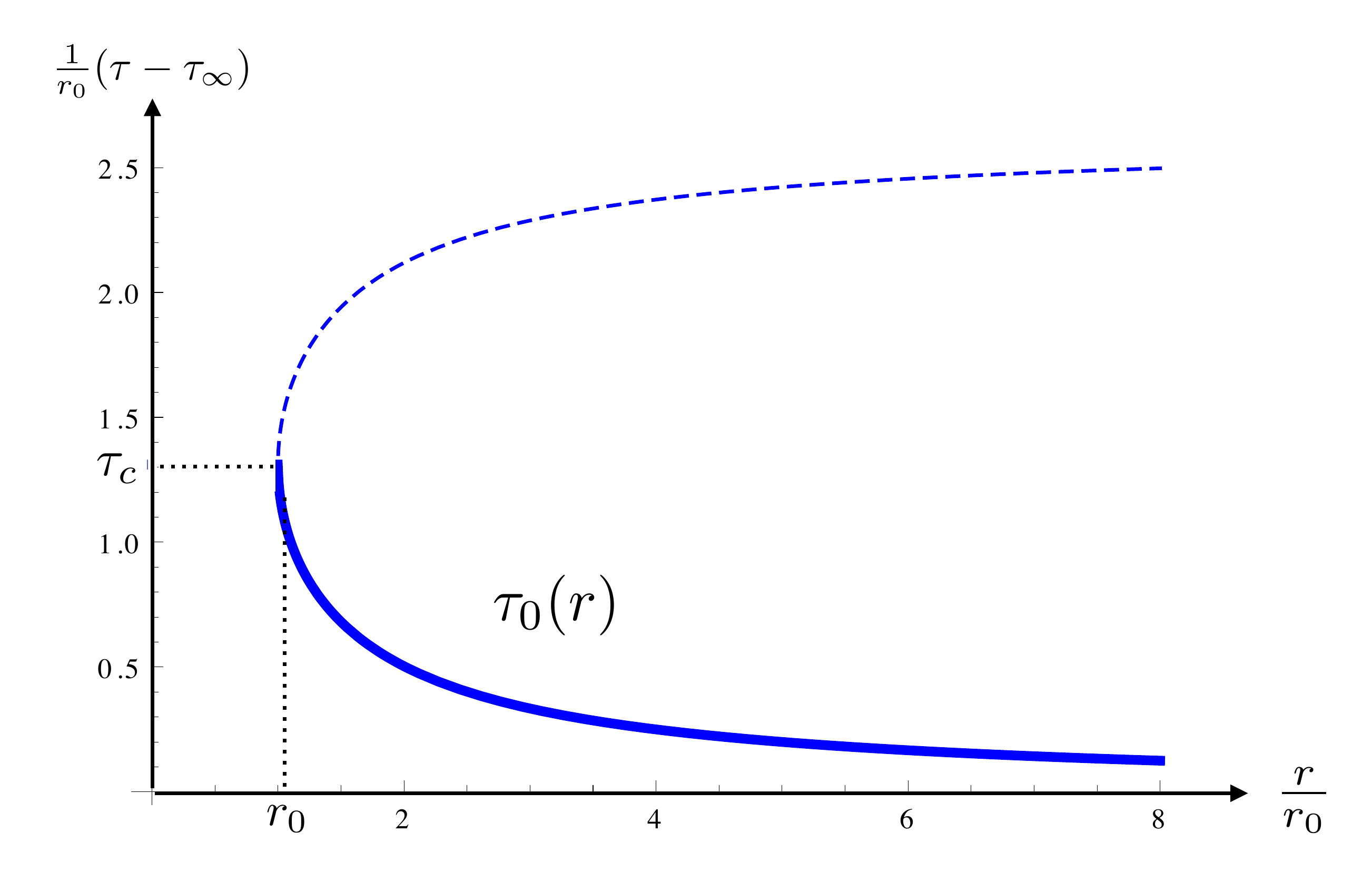}
\end{center}
\vskip-.5cm
\caption{Extremal surface $\tau=\tau_0(r)$ of the thin-wall bubble solution \eqref{eq:profile1}. Solid line denotes the 
bubble wall profile of the bubble radius $r$ above the critical radius $r_0$. The dashed line corresponds to the branch
of the classical trajectory beyond the turning point at $r_0$. }
\label{fig:profile}
\end{figure*}

What happens with the extremal surface in the regime $0\le r(\tau)\le r_0$? 
To address this question, first let us determine the classical trajectory $r(\tau)$ -- or equivalently the wall profile $\tau=\tau_0(r)$ of the classical solution -- on which the rate $W$ was computed in \eqref{eq:pdrNfinalr0}. To find it, we simply integrate the equation for the conserved energy
\eqref{eq:Econs}
on our classical solution,
\[
E\,=\, 4\pi\, \mu \frac{r^2}{\sqrt{1+\dot r^2}}\,,
\] 
or, equivalently, the expression $(r/r_0)^4 \,=\, 1+ \dot{r}^2$. One finds,
\[
\int_{\tau_{\infty}}^{\, \tau} d\tau\,=\,-\, \int_R^{\,r} \frac{dr}{\sqrt{\left(\frac{r}{r_0}\right)^4-1}}\,,
\]
which after integration can be expressed in the form,
\[
\tau(r) \,=\, \tau_\infty  \,+\, r_0 \left(\frac{\Gamma^2(1/4)}{4\sqrt{2\pi}}\,+\, 
{\rm Im} \left({\rm EllipticF}[{\rm ArcSin}(r/r_0), -1]\right)
\right)\,.
\label{eq:profile1}
\]
This classical trajectory gives the thin-wall bubble classical profile for $r_0 < r(\tau) <\infty$, which is the result 
\eqref{eq:pdrNfinalr0} for the quantum contribution to the 
rate $\Delta W^{\rm quant}$. This trajectory is plotted in Fig.~\ref{fig:profile}.

What happens when the radius of the bubble $r(\tau)$ approaches the critical radius $r_0$ 
\eqref{r0def} where the momentum \eqref{eq:conjp} vanishes?
Recall that in the language of a mechanical analogy we are searching for an equilibrium (i.e. the stationary point 
solution) where the surface $\tau_0(r)$ is pulled upwards (in the direction of $\tau$)
by a constant force $E=nm$ acting at the point $r=0$. This corresponds to finding an extremum -- in our case the true minimum -- 
of the expression in \eqref{eq:halfEWqnew}, which we rewrite now in the form,
\[
\frac{1}{2} \Delta W^{\rm quant}\,=\, 
\underbrace{E\,|\tau_\infty|}_{{\rm force}\times{\rm height}} -\,\,  \underbrace{\mu \int d^{2+1} {\rm Area}}_{\rm surface\, energy}
\,.\label{eq:halfEWqNNew}
\]
Sufficiently far away from the origin (where the force acts), the surface is nearly flat and does not extend in the $\tau$ direction.
As $r$ approaches the origin from larger values, the surface becomes increasingly stretched in the $\tau$
direction. At the critical radius $r_0$, the surface approaches the shape of a cylinder $R^1 \times S^2$, with $R^1$
along the $\tau$ direction. 

\begin{figure*}[t]
\begin{center}
\includegraphics[width=0.7\textwidth]{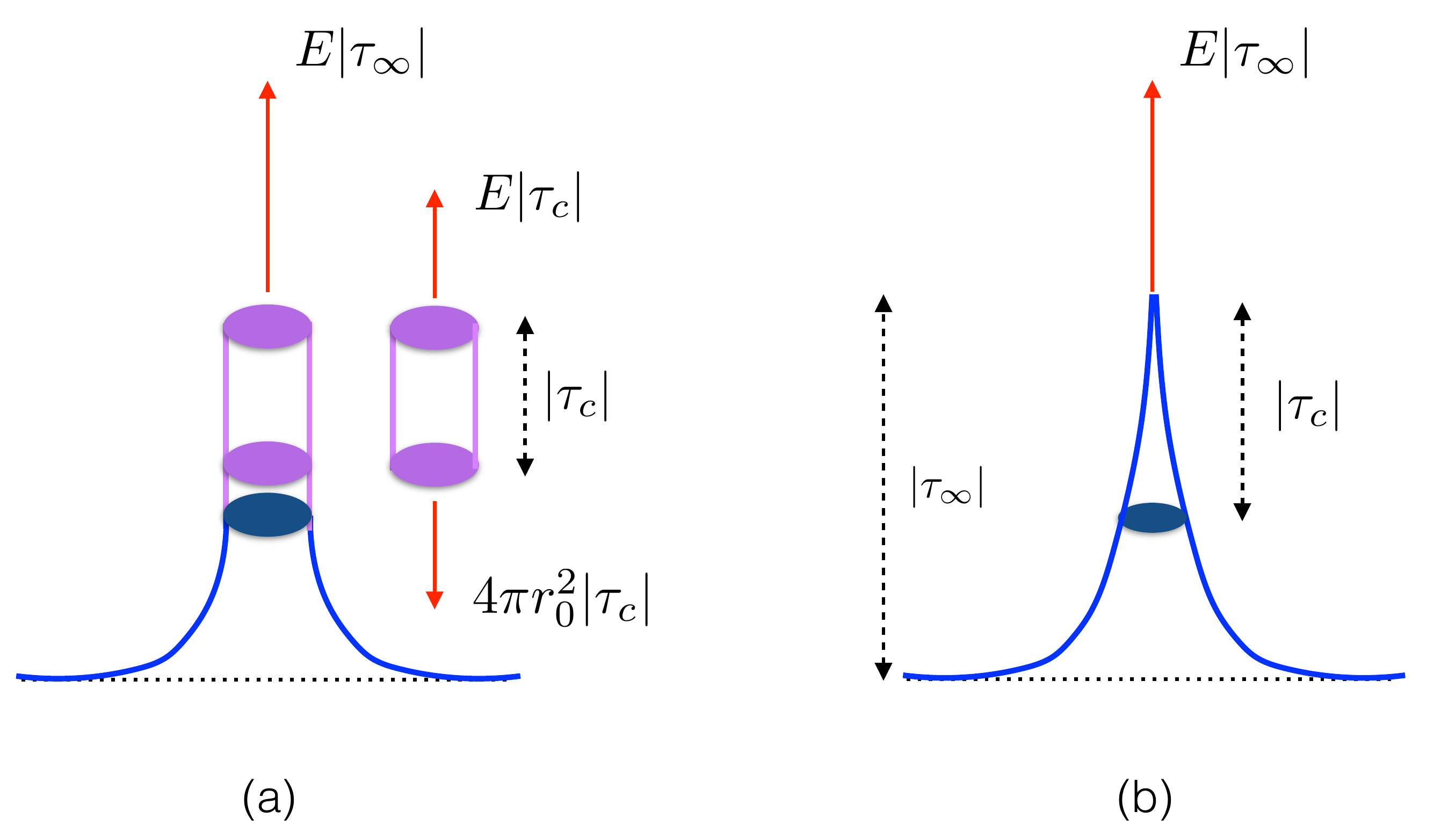}
\end{center}
\vskip-.5cm
\caption{Stationary surface configuration obtained by gluing two branches. {\bf Plot (a)}  shows the surface in the thin-wall
approximation, which glues the original solution \eqref{eq:profile1} to the infinitely stretchable cylinder solution of \eqref{eq:pdrNfinalr01}.
{\bf Plot (b)} depicts its more realistic implementation where the infinite cylinder is replaced by a cone as a consequence of allowing 
the surface tension $\mu$ to increase with $|\tau|$ in the regime where the highly stretched surface effectively becomes a 1-dimensional spring. }
\label{fig:prof2}
\end{figure*}

Up to the critical point $\tau_c$ where $r=r_0$, the force and the surface tension must balance each other,
\[
E\,|\tau_\infty-\tau_c| \,-\, \left(\int_{\tau_\infty}^{\tau_c} d\tau \,4\pi \mu \, r^2 \sqrt{1+\dot r^2}
\,-\, \frac{4\pi}{3} \,\mu R^3\right)=0\,.
\label{eq:pdrNfinalr01}
\]
The right-hand side was of course calculated in Eqs.~\eqref{eq:pdrN} and \eqref{eq:pdrNfinalr0}.
However, when the critical point $r_0$ is reached at $\tau_c$, the balance of forces becomes trivial,
\[
E\,|\tau_c| \,-\,  4\pi \,\mu \,r_0^2  \, |\tau_c| \,=\, 0\,.
\label{eq:vanishing}
\]
Clearly, the branch of the classical trajectory shown as the dashed line in Fig.~\ref{fig:profile} is unphysical in the sense that it does not 
describe the membrane pulled upwards with the force $E=mn$. Furthermore, the membrane surface does not satisfy the boundary condition 
that $\tau_0=0$
at $r=0$ since it currently does not even extend to $r< r_0$.
The vanishing of the expression \eqref{eq:vanishing} is the consequence of the definition of the critical radius in \eqref{r0def}. 
As soon as the radius $r(\tau)$ approaches the critical radius $r_0$,
the radius freezes (since $p\propto d_\tau r = 0$); 
the two terms in \eqref{eq:vanishing} become equal, $E=\,\mu \, 4\pi \,r_0^2$, and remain so at all times above the critical time $\tau_c$.
The thin-wall profile becomes an infinitely stretchable
cylinder,
as shown in Fig.~\ref{fig:prof2}~(a), giving no additional contribution to $ \Delta W^{\rm quant}$ on top of \eqref{eq:pdrNfinalr01}.

The freely-stretched cylinder in Fig.~\ref{fig:prof2}~(a) is an idealised approximation to the more realistic configuration that would be 
realised in practice in our mechanical analogy. One can consider what this realistic mechanical solution would look like. 
Let us define the quantity $d(r)$,
\[ 
d\,=\, \tau +\tau_\infty\,, \qquad 0\le d\le |\tau_{\infty}|\,.
\label{eq:ddef}
\]
as the coordinate along the vertical axis in Fig.~\ref{fig:prof2}
measuring the height of the surface stretched by the `force' $E|\tau_\infty|$ as a function of $r$. At the base
of the surface we have $d\simeq 0$.\footnote{Recall
that the tip of the surface is at $\tau=0$, where $d=|\tau_\infty|$, and that the surface's  base is at a negative 
$\tau=\tau_\infty=-|\tau_\infty|$, which corresponds to $d=0$.}
The surface wall profile is nearly flat in the $\tau$ direction. As $d$ increases from $0$, the radius $r(\tau)$ grows smaller, 
following the profile of the thin-wall
solution contour in the lower part of Fig.~\ref{fig:prof2}. As $r$ approaches the critical radius $r_0$, the surface becomes almost parallel to the $d$ (or $\tau$)
direction. Such a surface behaves more like a spring along the $\tau$ coordinate. For the strict thin-wall approximation, the surface tension $\mu$ is assumed to be
a constant. However, in the case of the spring, it should instead be the Young's elastic modulus, $k_{\rm \, Young}$ that takes a constant value. Hence, for a highly-stretched surface in the $\tau$ direction, we should introduce some dependence on 
$d$ \eqref{eq:ddef}  into the surface tension via,
\[
\mu(d) \,=\, \mu_0 \,(1\,+\, \hat{k}\, d) \,,
\label{eq:newmu}
\]
where $\hat{k}\ll 1$ is a dimensionless constant.
In the limit $d\to 0$, the surface tension  $\mu(d)  \to \mu_0$, where $\mu_0= \, \frac{m^3}{3\lambda}$
is the same constant contribution to the surface tension as we computed earlier in \eqref{eq:mu} in the strict thin-wall case. The corresponding Young's modulus of the stretched surface would be $k_{\rm \,Young}= \mu_0\,\hat{k}$.
Equation \eqref{eq:newmu} describes a small deviation from the standard thin wall approximation, where the surface tension is now dependent on the stretching of the surface. This expression can be thought of as the zeroth and first order terms in the Taylor expansion
of the function $\mu(\tau +\tau_\infty)$.

The result of this improvement of $\mu$ is that the balance between the two terms in \eqref{eq:vanishing} continues to hold. However, for an adiabatic approximation of nearly constant $\mu$, with $\hat{k} d \ll 1$, it is now in the form,
\[
\left(E\,-\, \mu(d) \cdot 4\pi \,r(d)^2\right) d\,=\, 0\,, \quad {\rm where} \,\, d\ge |\tau_c| \,.
\label{eq:vanishing2}
\]
For every infinitesimal increase in the vertical coordinate $d$ above $|\tau_c|$, the radius $r(d)$ becomes a little smaller than its value $r_0$ at the base of the cylinder
in Fig.~\ref{fig:prof2}~(a). As a result, the cylinder gets narrower as $d$ increases and turns into the cone-like shape shown in Fig.~\ref{fig:prof2}~(b). 
The actual choice of the modification of the surface tension expression, such as in \eqref{eq:newmu}, is of course determined by the field configurations
themselves; it can be seen as a part of the extremisation procedure. For an adiabatically-slowly varying $\mu$ (such that the contribution from the cone
to $W$ is negligible), the overall contribution $\Delta W^{\rm quant}$ is dominated by the surface at $r>r_0$ in the large $\lambda n$ limit.
In this case we conclude that,
\[
 \Delta W^{\rm quant} \,=\, 
\, \frac{1}{\lambda} \, (\lambda n)^{3/2}\, \frac{2}{\sqrt{3}}\,
\frac{\Gamma(5/4)}{\Gamma(3/4)}\,\simeq\, \, 0.854\,  n \sqrt{\lambda n}
\,.\label{eq:pdrfinal}
\]


 \begin{figure*}[t]
\begin{center}
\includegraphics[width=0.6\textwidth]{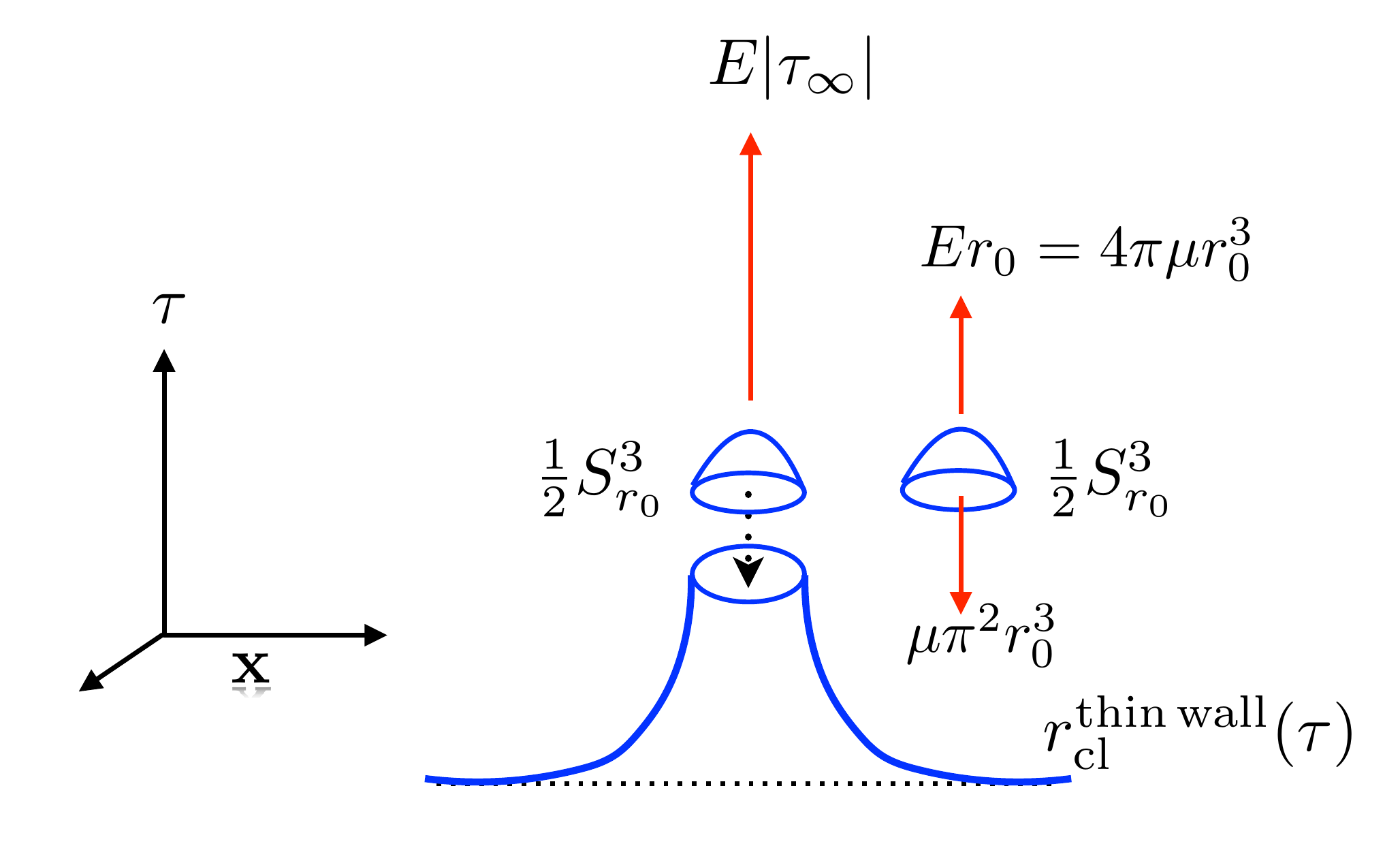}
\end{center}
\vskip-.5cm
\caption{Surface obtained by `completing' the thin wall classical solution $r_{\rm cl}^{\rm thin\,wall}$ with the upper half of a three-sphere $S^3_{r_0}$, of radius $r_0$ 
in $(\bec{x},\tau)$, at $0\le |\bec{x}|\le r_0$. The force stretching the surface is $E=4 \pi \mu r_0^2$.  }
\label{fig:sphere}
\end{figure*}

\subsubsection{Further discussion of $r\le r_0$ region}

Let us discuss the apparent loss of the classical strictly-thin-wall solution at $r < r_0$ in a little more detail.
An important point we want to emphasise is that in our model with spontaneous symmetry breaking, the $d$-dimensional domain wall surface 
separating two distinct vacuum domains
at $\pm v$, is a local \emph{minimum} of the Euclidean action. The infinite extent of the surface, which reaches the boundary of space ($r\to R$ or $r \to \infty$ when the cut-off $R$ is removed), is important. It is different from the Coleman's bounce 
solution~\cite{Coleman}, which is known to be a
 local maximum, or 
more precisely the saddle-point of the action. The bounce has a membrane of finite extent;
in the $O(4)$ symmetric case, it is the surface of the $S^3$ spherical bubble separating $h \simeq \pm v$ on the inside/outside of the sphere, and the radius of the bubble is the negative mode for the bounce solution.
 However, in the case of the infinite domain wall surface, the domain wall is a topologically-stable configuration and hence a local minimum.\footnote{A 
 recent discussion of domain walls and their stability can be found in the textbook~\cite{Shifman}.}
While the action on the
wall with infinite extent in $d$ dimensions contains an infinite constant, $S=\frac{4\pi}{3} R^3 \to \infty$, this contribution is subtracted
in our construction, as dictated by Eq.~\eqref{eq:halfEWqnew}.
The force being applied to the domain wall
in the $\tau$-direction stretches and curves the surface of the wall to balance the action of the force. The resulting stationary solution for the 
stretched surface in this mechanical analogy is a stable solution when the required boundary conditions, $\tau_0(r=R)=\tau_\infty$ and
$\tau_0(r=0)=0$, are satisfied.

We conclude that there must exist a stable classical solution for the domain wall surface with the boundary conditions imposed 
at infinite and zero radii. But what we have learned from Figs.~\ref{fig:profile}~and~\ref{fig:prof2}~(a) is that this solution can be described 
by the strict thin-wall
configuration only for $r>r_0$. At the values of $r(\tau)$ below the critical radius $r_0$, the solution corresponding to the minimum of the action
requires a deviation from the strict thin wall limit. One approach to achieve this is to allow for the $\tau$-dependent surface tension, 
as we already explained. Alternatively, we can continue using the thin wall configurations with constant $\mu$ and attempt to complete our solution at
$0\le r(\tau)\le r_0$. We treat these completions as \emph{trial configurations} approximating the true minimal solution. The benefit of this approximation 
is that it does not require any precise knowledge of how the deviation from the thin wall regime is realised.

A simple completion of the solution is to add the surface of an upper half of the three-sphere $S^3_{r_0}$, of radius $r_0$, to our thin wall classical solution,
\[ 
r_{\rm trial}^{\rm thin\,wall}(\tau)\,=\, 
\begin{cases}
r_{\rm cl}^{\rm thin\,wall}(\tau)
 & :\,\,{\rm for}\,\,  r_0 \le r< \infty\\
\frac{1}{2} \,S^3_{r_0}(\bec{x},\tau) & :\,\,{\rm for}\,\,  0\le r\le r_0\, ,
\end{cases}
\label{eq:pol}
\]
as shown in Fig.~\ref{fig:sphere}.
It is easy to deduce the contribution to the semiclassical rate from the surface of $\frac{1}{2} \,S^3_{r_0}$; it is given by,
\[
(1/2)\, \delta W^{\rm quant}_{\frac{1}{2}S^3} \,=\, 
 E r_0 \,-\, 
\mu \pi^2 r_0^3\,=\, 
 4\pi \mu r_0^3\, (1-\pi/4)
\,\simeq\, \, 0.1\,  n \sqrt{\lambda n}
\,.
\label{eq:pdrS3}
\]
The force $E$ is expressed in terms of the critical radius, as before,
 $E=4\pi \mu r_0^2$. The extent of the $S^3_{r_0}$ half sphere in the $\tau$-direction is $r_0$, and the action of this surface is,
 \[
 \frac{1}{2}\, S[S^3_{r_0}]\,=\,  \frac{1}{2}\, 2\pi^2 r_0^3\,=\, \pi^2 r_0^3\,,
 \]
 which together with the identity \eqref{eq:ident} justifies the formula \eqref{eq:pdrS3}.
 
 The contribution in \eqref{eq:pdrS3} should be added to the expression for $\Delta W^{\rm quant}$
 in \eqref{eq:pdrfinal} that came from the thin wall classical solution at $r>r_0$. 
 We note that the expression in \eqref{eq:pdrS3}
 is positive, signalling that the force gives a slightly greater contribution than the surface energy of the half-sphere, increasing the overall rate in
 \eqref{eq:pdrS3}. 
 
 In general, the completion of the classical solution by the $O(4)$ symmetric thin wall surface in \eqref{eq:pdrS3}
 does not gives the minimum of the action but rather a trial configuration. For the true minimum, the rate $\Delta W^{\rm quant}$ would only be larger.
 For example a cone-like $O(3)$ symmetric completion, such as the one depicted in Fig.~\ref{fig:prof2}~(b) would
 give an even larger contribution. However, to evaluate it, one would be required to use a beyond-the-thin-wall description 
 of the surface. For our purposes, it is sufficient to approximate the true classical stationary point by the thin wall configuration 
in \eqref{eq:pol} with the rate given by  \eqref{eq:pdrfinal} with a small positive correction  \eqref{eq:pdrS3} arising from the $O(4)$ 
completion of the classical surface at $0\le r\le r_0$. In total we have,
\[
 \Delta W^{\rm quant} \,=\, 
 0.854\,  n \sqrt{\lambda n} \,+\, \delta W^{\rm quant}\, >\,  1\cdot n \sqrt{\lambda n}
\,.\label{eq:pdrfinal22}
\]
Equation \eqref{eq:pdrfinal22} is the main result of this section.
To be on the conservative side we can always ignore the positive $\delta W^{\rm quant}$ contribution and continue using the
expression in \eqref{eq:pdrfinal} for the contribution of quantum effects to the function $W$ in the exponent of the semiclassical rate.
In summary, our result for the semiclassical approximation to the rate reads,
\begin{equation}
	{\cal R}_n(E)\,=\, e^{W(E,n)}\,, \quad 
	W\,\simeq\, n \left(\log\frac{\lambda n}{4}-1\right)\,+\, \frac{3n}{2}\left( \log\frac{\varepsilon}{3\pi}+1\right)\,+\, 
	0.854\,  n \sqrt{\lambda n}\,,
\label{eq:RnEfin}	
\end{equation}
reproducing the result in~\cite{Khoze:2018kkz,Khoze:2017ifq}.
The expression \eqref{eq:RnEfin} was derived in the near-threshold high-energy high-multiplicity limit,
\[ \lambda \to 0\,, \quad n\to \infty\,, \quad {\rm with}\quad
\lambda n = {\rm fixed}\gg 1\,, \quad \varepsilon ={\rm fixed}\ll 1 \,,
\label{eq:limitL}
\]
where final state particles are non-relativistic so that $\varepsilon$ 
is treated as a fixed number much smaller than one. 
The overall energy  and the final state multiplicity are related linearly via
$E/m \,=\, (1 + \varepsilon) \, n \simeq n \gg 1 $.
Clearly, for any small fixed value of $\varepsilon$ one can choose a sufficiently large value of $\lambda n$, such that the function 
$W(\lambda n, \varepsilon)$ in  \eqref{eq:RnEfin} is positive. 
These semiclassical expressions imply that at sufficiently large particle multiplicities, the 
expression ${\cal R}_n(E)$ grows exponentially with $n$ and consequentially with the energy $E$.

\medskip
\section{Comparison with the Landau WKB method in QM}
\label{sec:Landau}
\medskip

The semiclassical approach in quantum mechanics (QM) is known as the WKB method. It is commonly used and provides a powerful non-perturbative formalism for solving quantum mechanics problems. In their classic volume \cite{landau1981quantum}, Landau and Lifshitz formulate the WKB approach 
using singular classical solutions analytically continued to complex space. In this sense the Landau WKB approach 
in QM~\cite{Landau2,landau1981quantum} has many similarities 
with the semiclassical approach in quantum field theory \cite{Son:1995wz} that we are using in this work. 
Considerations in early literature 
of various generalisations of the Landau-WKB singular configurations approach to multi-dimensional systems can be found in
Refs.~\cite{IP,Voloshin:1990mz,Khlebnikov:1992af,Diakonov:1993ha,Gorsky:1993ix,Libanov:1997nt}.

\medskip

The main purpose of this section is to compare the the semiclassical approach 
in quantum field theory, covered in sections \ref{sec:son} and \ref{sec:main},
to the Landau WKB in quantum mechanics, and to discuss the differences between these two semiclassical realisations.

\medskip
\subsection{Matrix elements in the Landau WKB formulation}
\label{sec:LWKB}
\medskip

Landau and Lifshitz consider a matrix element of some physical operator, $\hat{\cal O} (\hat q)$, where $\hat{q}$ is the position operator for a 
1-dimensional quantum system with potential $U(q)$. Such a system is of course governed by the
time-independent Schr$\ddot{\mathrm{o}}$dinger equation,
\begin{equation}
-\dfrac{\hbar^2}{2m}\dfrac{d^2}{dq^2}\,\psi(q)\,+\,U(q)\psi(q)\,=\, E\psi(q),
\end{equation}
where $m$ is the mass and $E$ is the energy associated with positional wavefunction $\psi(q)$. The matrix element of $\hat{\cal O}$ between two states of energy $E_1$ and $E_2$ can be written as,
\begin{equation}
\label{eq:A12}
O_{12} = \braket{E_1|\hat{\cal O}|E_2} \, = \int_{-\infty}^{\infty} \psi_1^*(q)\, \hat{\cal O} (\hat q)\,\psi_2(q) dq.
\end{equation}

\begin{figure*}[t]
	\begin{center}
		\includegraphics[width=0.5\textwidth]{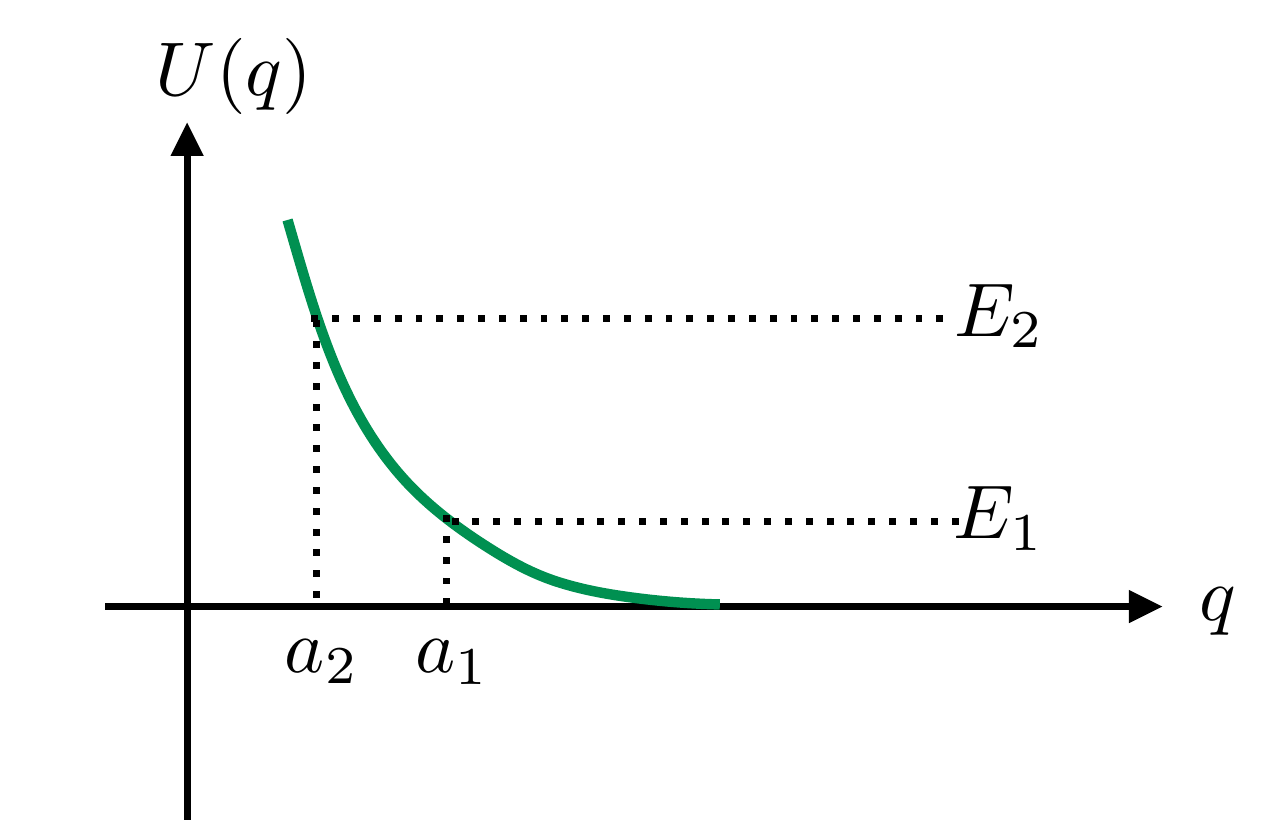}
	\end{center}
	\vskip-.5cm
	\caption{Quantum mechanical potential $U(q)$ with two energy eigenstates $E_2>E_1$.
		The turning points are $q=a_1$ and $q=a_2$ where
		$U(q_i)=E_i$.  The wavefunctions $\psi_1(q)$ and $\psi_2(q)$ are given in \eqref{eq:psi1}
		and \eqref{eq:psi2} in the classically forbidden regions $q<a_1$ and $q<a_2$ to the left of the potential barrier.}
	\label{fig:QMpot}
\end{figure*}

When the potential varies on length scales much larger than a wavelength, WKB methods can be used to approximate the wavefunctions, $\psi_i$, for the states of energies $E_i$ by solving the  
Schr$\ddot{\mathrm{o}}$dinger equation by iterations in the small parameter $\hbar$. At the leading WKB order
the wavefuctions  are given by \cite{landau1981quantum}, 
\begin{eqnarray}
\label{eq:psi1}
\psi_1(q) &\simeq& \frac{C_1}{2\sqrt{|p_1|}}\, e^{-\frac{1}{\hbar}\left| \int_{a_1}^{q} p_1 dq\right|}\,,
\qquad {\rm for}\,\, q<a_1,\\
\label{eq:psi2}
\psi_2(q) &\simeq&  \frac{iC_2}{2\sqrt{|p_2|}}\, e^{+\frac{1}{\hbar}\left| \int_{a_2}^{q} p_2 dq\right|}\,,
\qquad {\rm for}\,\, q<a_2,
\end{eqnarray}
where we have assumed the ordering of the energies, $E_1<E_2$ to select the 
signs of the roots in the exponents.
In the above expressions $p_i$ denote the classical momenta,
\[
p_i \,=\, \sqrt{(2m)(E_i-U(q))},
\]
and $C_{1,2}$ are constants.
The expressions \eqref{eq:psi1}-\eqref{eq:psi2} are written in the classically forbidden regions, $q< a_1$ and $q<a_2$ for the wavefunctions 
$\psi_1(q)$ and $\psi_2(q)$ respectively, where $a_1$ and $a_2$ are the turning points, $U(a_i)=E_i$ as shown in 
Fig.~\ref{fig:QMpot}.
Importantly, the roots of the wavefunctions are selected in such a way that the wavefunction $\psi_1$ with the smaller energy, $E_1<E_2$,
has the negative-valued exponent, while the wavefunction $\psi_2$ with the higher energy has the positive exponent.

These approximations \eqref{eq:psi1}-\eqref{eq:psi2} break down in the vicinity of the classical turning points $q\simeq a_1$. 
Landau and Lifshitz explain how to avoid these regions by deforming the integration contour into the upper half plane of complex $q$, 
away from the classical turning points.
Then it follows from \eqref{eq:A12}, \eqref{eq:psi1}-\eqref{eq:psi2} 
that the matrix element is dominated by the contribution from the singularity of the potential: i.e. the point, $q_0$, where $U(q_0)\to\infty$,
\[
\label{eq:expbeh2}
O_{12} \,\simeq\,
\exp\left\{  
-\dfrac{1}{\hbar} \left| \int^{q_0} p_1 dq\right| \,+\, \dfrac{1}{\hbar} \left| \int^{q_0} p_2 dq\right|
\right\}.
\]
This WKB formula is often presented in the form \cite{landau1981quantum}, 
\[
\label{eq:expbeh}	
O_{12} \,\simeq\,
\exp\left\{  
-\dfrac{1}{\hbar}\mathrm{Im}\left[
\int^{q_0}\sqrt{2m(E_2-U(q))}dq
\,-\, \int^{q_0}\sqrt{2m(E_1-U(q))}dq
\dfrac{}{}\right]
\right\},
\]
which requires an additional clarification for selecting the sign of the imaginary part, or equivalently of the square roots in \eqref{eq:expbeh}.

The lower limits of the $q$ integrations in the expressions \eqref{eq:expbeh2}-\eqref{eq:expbeh} 
are unimportant to the leading WKB accuracy as the dominant
contributions come from the vicinity of the singularity at $q\sim q_0$ near the upper limit of the integrals.

Notice that the WKB exponent has no dependence on the operator $\hat{\cal O}$. 
Indeed the precise form of $\hat{\cal O}$ may change the overall prefactor, but it does not affect the bulk behaviour. The above estimate 
is derived for \emph{generic} operators that contain no explicit dependence on $\hbar$.
Furthermore, the generic operators above should not include ``bad" choices of $\hat{\cal O}$  that would give an unusually small coefficient 
in front of the exponent in \eqref{eq:expbeh}. A simple example is $\hat{\cal O}=$ constant, for which $O_{12}$ vanishes due to the orthogonality of $\psi_1$ and $\psi_2$.

\medskip

It will be useful to summarise the characteristic features of the WKB result \eqref{eq:expbeh2}:
\begin{enumerate}
	\item The WKB matrix element is dominated by the singularity of the potential, $U(q)\to \infty$ as $q\to q_0$. 
	Since the energies $E_{1,2}$ are finite, the singularity is located in the classically forbidden region $U(q)>E_{1,2}$,
	where the coordinates $q$ are analytically continued on deformed contours and momenta $p$ are complex-valued.
	\item For generic operators with no exponential dependence on $\hbar$, the leading order WKB expression for their matrix elements 
	\eqref{eq:expbeh2} does not depend on the specific choice of the operator. 
	
	However, this is where the similarities between the QM WKB and the QFT semiclassical methods end.
	
	\item The sign prescription in the definition of the imaginary part in the WKB exponent in \eqref{eq:expbeh} is fixed by the expression in
	\eqref{eq:expbeh2}. It is easy to verify that for every $q$ in the classically forbidden region, 
	the absolute value of the complex momentum $|p_1|$ is always greater 
	than $|p_2|$ since $U(q)-E_1 > U(q)-E_2$ for the energy ordering $E_1<E_2$.
	Hence the 
	resulting matrix elements computed using the WKB approximation in quantum mechanics \eqref{eq:expbeh2}-\eqref{eq:expbeh}
	are always exponentially suppressed, unlike what we saw in the QFT calculation in the previous section.
\end{enumerate}

\medskip

\noindent{\tt Example}:  Landau-Lifshitz section 51 Problem 1~\cite{landau1981quantum}.

\noindent {Calculate the exponential factor of the matrix elements in the potential $U(q)=U_0 e^{-\alpha q}$.}

\medskip

\noindent Solution: $U(q)$ becomes infinite only for $q\to -\infty$, hence the singularity is the point $q=q_0=-\infty$.
Using \eqref{eq:expbeh2} we write
\begin{eqnarray}
&O_{12} \simeq
\exp\left\{  
-\frac{1}{\hbar} \left| \int_{a_1}^{-\infty} \sqrt{2m(U-E_1)} \,dq\right| \,+\, 
\frac{1}{\hbar} \left| \int_{a_2}^{-\infty} \sqrt{2m(U-E_2)} \,dq\right|
\right\} 
\nonumber \\
&\qquad =
\exp\left\{  
+\frac{1}{\hbar}  \int_{a_1}^{-\infty} \sqrt{2m(U_0 e^{-\alpha q}-E_1)} \,dq \,-\, 
\frac{1}{\hbar}  \int_{a_2}^{-\infty} \sqrt{2m(U_0 e^{-\alpha q}-E_2)} \,dq
\right\}
\label{eq:expbeh2E1}	
\end{eqnarray}	
where $a_1$ and $a_2$ are the turning points, $a_1>a_2>q_0=-\infty$, similar to what is shown in Fig.~\ref{fig:QMpot}.
It is convenient to introduce the velocity variables $v_1$ and $v_2$ and use them instead of the energies $E_1$ and $E_2$,
\[
v_1^2\,=\, \frac{2E_1}{m}\,, \qquad v_2^2\,=\, \frac{2E_2}{m}\,,
\]
and rewrite the integrals on the right-hand side of \eqref{eq:expbeh2E1} ($i=1,2$) as
\[
\label{eq:inti}
\frac{1}{\hbar}\,\int_{a_i}^{-\infty}dq\, \sqrt{2m(U_0 e^{-\alpha q}-E_i)}  \,=\, 
\frac{mv_i}{\hbar}\,\int_{a_i}^{-\infty}dq\, \sqrt{\frac{2U_0}{m^2v_i^2}\,e^{-\alpha q}-1} \,.
\]
Making the change of integration variables from $q$ to $y$ via,
\[
y\,=\, -\alpha q \,+\, \log\frac{2U_0}{m^2v_i^2}
\,,
\]
the integral in \eqref{eq:inti} becomes
\begin{eqnarray}
-\, 
\frac{m}{\hbar}\frac{v_i}{\alpha}\,\int_{0}^{+\infty} dy\, \sqrt{e^y-1}  &=& -\,\frac{2m}{\hbar}\frac{v_i}{\alpha}\,\left(\sqrt{e^{y_0}-1}
- {\rm Arctan}\left(\sqrt{e^{y_0}-1}\right)
\right)|_{y_0\to \infty} \nonumber\\
&=& -\,\frac{2\sqrt{2U_0}}{\hbar \alpha}\,\exp \left[{\frac{\alpha}{2}|q_0|}\right]|_{|q_0|\to \infty} \,+\, \frac{\pi m}{\hbar\alpha} v_i
\,. \label{eq:inti2}
\end{eqnarray}
Finally, we note that in the difference between the two integrals in \eqref{eq:expbeh2E1} the infinite terms cancel 
and we are left with the finite contribution coming from the second term in \eqref{eq:inti2} for $i=1,2$. The WKB result for the matrix element
to exponential accuracy
reads,
\[
O_{12}\, \simeq\, e^{-\, \frac{\pi m}{\hbar \alpha} (v_2-v_1)}\,.
\label{eq:exlast}
\]
Since the velocities are ordered in the same way as the energies, $v_2>v_1$, we conclude that the matrix element is exponentially suppressed.
We also note that the exponent in \eqref{eq:exlast} is  independent of the choice of the operator $\hat{\cal O}(q)$.

\medskip

This example confirms the general conclusion we already reached in item (3.) above,  that in quantum mechanics the WKB approximation by construction can only result in exponentially-suppressed 
matrix elements and can never give exponentially growing probability rates. This is of course fully expected: given that ordinary quantum mechanics with
a Hermitian Hamiltonian is a unitary theory, probabilities must be conserved and cannot exceed unity.

What is then the technical difference between the WKB formulation in quantum mechanics and the semiclassical method 
in quantum field theory? Both formalisms compute matrix elements of certain operators and both use singular complex-valued configurations 
to find the dominant contributions. Nevertheless the field theoretical formulation does not have a built-in restriction to disallow positive-valued 
functions $W$ in the exponent of the semiclassical rate. This is evidenced by the function $W$ in our result \eqref{eq:RnEfin},
becoming positive at sufficiently large values of $\lambda n$. 
The critical difference between the QM and QFT formulations is that in QFT we have an entire \emph{surface} of singularities 
rather than isolated 
singular point(s) in QM.

\medskip
\subsection{The role of the singular surface in QFT}
\label{sec:surf}
\medskip

For concreteness when discussing the semiclassical method in quantum field theory we will continue using the model \eqref{eq:L} with SSB,
which in terms of the shifted field  $\phi(x)=h(x)-v$ in $(d+1)$~dimensions has the Lagrangian description,
\[
{\cal L} (\phi) \,=\, \frac{1}{2}\, \partial^\mu \phi \,\partial_\mu \phi\,-\, \frac{1}{2} m^2 \phi^2 
\,-\, \lambda v \phi^3 \,-\, \frac{\lambda}{4}\, \phi^4
\label{eq:L2phi}
\]
with the mass $m=\sqrt{2 \lambda}v$. It is also useful to denote the potential energy density as ${\cal V}(\phi)$, and its interacting part as
${\cal V}_{\rm int}(\phi)$,
\[
\begin{split}
\label{eq:Vphical}
&{\cal V}(\phi) \,=\,\, \frac{1}{2} m^2 \phi^2 
\,+\, \lambda v \phi^3 \,+\, \frac{\lambda}{4}\, \phi^4\,, \\
&{\cal V}_{\rm int}(\phi) \,= \lambda v \phi^3 \,+\, \frac{\lambda}{4}\, \phi^4\,.
\end{split}
\]
We recall that the classical trajectory $\phi(x)=h(x)-v$ used in the semiclassical method in sections~\ref{sec:son}~and~\ref{sec:main}
is infinite on a $d$-dimensional singularity
surface in $(\tau,\bec{x})$ that touches
the origin $x=0$ in Minkowski space. Correspondingly, the value of the potential ${\cal V}(\phi)$ also becomes infinite 
at the singularity, which is analogous to the statement about the singularity of the potential $U(q)$ in quantum mechanics,
\[
\begin{split}
&{\rm QFT}: \qquad \lim_{\tau\to \tau_0(\bec{x})} {\cal V}( \phi(x)) \,=\, \infty\,,
\\
&{\rm QM}: \qquad \,\,\,\,\, \lim_{q\to q_0} U( q) \,=\, \infty\,.
\end{split}
\]

However, the potential in QFT is singular on the $d$-dimensional surface,  $\tau=\tau_0(\bec{x})$ found by extremising the Euclidean action over all 
appropriate shapes of trial singular surfaces; while in QM there are no spatial dimensions, $q$ is a function of time only, $q=q(\tau)$ 
and the potential $U(q)$ is singular at 
a point\footnote{Without loss of generality we can use translational invariance to set it at the time $\tau=0$.} $q(0)=q_0$.
Hence, there are no surfaces to extremise over in QM and consequently no
dependence on the surface shape $\tau_0(\bec{x})$ or the value of $|\tau_{\infty}|$, which were of critical importance in QFT,
as manifested by Eqs.~\eqref{eq:EWq}, \eqref{eq:halfEWqnew},
\[
\frac{1}{2} \Delta W^{\rm quant}\,=\, 
nm\,|\tau_\infty|\,-\, \, 
S_{E}[\tau_0(\bec{x})]
\,+\, 
S_{E}[{\rm flat}]
\,.\label{eq:halfEWqnew2}
\]
In fact, the very reason why the contributions of quantum corrections to the rate in the field theory case 
are not forced to be exponentially suppressed is due to the positive-valued contribution of the term $nm\,|\tau_\infty|$
on the right-hand side of \eqref{eq:halfEWqnew2}.
It is easy to verify that the action integral of the surface stretched in the $\tau$-direction ({\it cf.}~\eqref{eq_thinw}),
\[ 
S_{E}[ \tau_0(r)] \,=\,
\int_{\tau_\infty}^0 d\tau \,4\pi \mu \, r^2 \sqrt{1+\dot r^2}\,=\, \int dr\,4\pi \mu \, r^2 
\sqrt{(d\tau/dr)^2 +1}
\,,
\label{eq_thinw22}
\]
is always greater in absolute value than the action of the flat surface ({\it cf.}~\eqref{eq:Se1act}),
\[
S_{E}[{\rm flat}]  \,=\,
\mu \int_0^R 4\pi r^2 dr\,=\, 
\mu\,  \frac{4\pi}{3} \, R^3\,,
\label{eq:Se1act22}
\]
and the semiclassical exponent of the rate $\frac{1}{2} \Delta W^{\rm quant}$ in \eqref{eq:halfEWqnew2} can be positive only because of the
presence of the contribution $nm\,|\tau_\infty|$, which is large and positive in the high multiplicities limit $\lambda n \gg 1$ thanks to the non-trivial stretching of the surface $|\tau_\infty|\neq 0$. As we already explained, this effect is impossible in ordinary quantum mechanics.

\medskip

Before concluding this section we would like to list the differences between the semiclassical method we are using in 
the $(d+1)$-dimensional QFT model \eqref{eq:L2phi} and a naive attempt to apply the same method to a quantum mechanical model 
with the same Lagrangian in $(0+1)$ dimensions.

\begin{enumerate}
	\item As there are no spatial degrees of freedom in $(0+1)$ dimensions, there is no phase space to integrate over. Hence, one would need to compute just the square of the matrix element. One can continue using the coherent space representation for the final states, as in \eqref{eq:sig},
	but the integration over the final states $1\,=\,\int d(\{b^*\},\{b\})e^{-b^*b}$ involves the ordinary rather than functional integrals over $b$ and $b^*$.
	
	\item With no phase space for final states in QM, it is impossible to project simultaneously on states of fixed energy $E$ and fixed occupation number $n$.
	In QM unlike QFT, $E$ and $n$ are related,
	\[
	E_n \,=\, n m (1+\varepsilon_n)\,,
	\]
	where the quantity $\varepsilon_n$ is fixed in a given QM model and is not a free parameter. So the non-relativistic limit $\varepsilon \ll 1$ that we used
	in QFT is not something we are free to impose in QM. 
	As a result, one should impose only a single projector, $\hat{P}_E$ on the matrix elements in \eqref{eq:sig},
	\[
	\langle b| \,\hat{P}_E \hat{S}\hat{\cal O}\, |0\rangle\,.
	\]
	One should also keep in mind that an anharmonic quantum potential with a non-vanishing 
	$\lambda$ has energy levels 
	$E_n$ that are spaced more densely than the energy states of the harmonic oscillator. Hence $\varepsilon_n$ are negative-valued, and the 
	`decays' $E_n \to n m$ are kinematically forbidden in QM. In comparison, in QFT such decays are only disfavoured by the vanishing phase space 
	and become possible after allowing for arbitrary small particle momenta in the final state leading to a small positive $\varepsilon$. 
	
	\item Importantly, in QM one should distinguish between projecting with $\hat{P}_E$ on the eignestates of the full Hamiltonian with the potential 
	$U(q)$ given by ${\cal V}(\phi)$ in \eqref{eq:Vphical},
	and the projection on the eignestates of the Hamiltonian of the harmonic oscillator. In QFT with $d\ge 2$ spatial dimensions, the field solutions 
	of the Euler-Lagrange equations dissipate in space. Therefore, they become solutions of free equations at early and late times $t \to \pm \infty$.
	This is not the case in QM. Energy eignestates in the full quantum mechanical potential of an anharmonic oscillator are trapped in the potential well 
	and do not linearise.\footnote{One can of course always proceed with the WKB computation of matrix elements between the energy eignestates 
		in a different QM model -- one with a
		potential barrier as in Fig.~\ref{fig:QMpot}. We have analysed this situation in the example of $U(q)=U_0 e^{-\alpha q}$ considered in the previous section. In this case, classical trajectories do indeed become free 
		far away from the barrier at asymptotic times. This however is different from our model \eqref{eq:L2phi}. Free states that are of interest for us 
		are analogues of particle states in QFT, ie those described by the harmonic oscillator potential $m^2 q^2/2$ in \eqref{eq:L2phi} rather than the states in the asymptotically-vanishing potential.} 
	Hence, projecting onto the eigenstates of the harmonic oscillator Hamiltonian $H_0$ as we have done in \eqref{eq:PEdef} becomes problematic in QM.
	
	\item 
	In QM, the integral analogous to the one in \eqref{eq:master}, \eqref{eq:master2} over
	$ \{\xi,  \phi_i,\phi_f, \varphi_i,\varphi_f,b^*, b\}$, is an ordinary non-functional integral. Only the $(0+1)$-dimensional fields 
	$\phi(t)$ and $\varphi(t)$, playing the role of QM coordinates, are functions.
	Hence, proceeding formally in QM, we can write down saddle-point equations resulting from the steepest descent 
	approximation of the integral  \eqref{eq:master}.
	However, we cannot expect that the solutions at asymptotic times linearise and thus we cannot write down boundary conditions
	analogous to \eqref{eq:bc1} and \eqref{eq:bc2} for $q(t)$. Furthermore, the energy $E$ cannot be computed from the late-time asymptotics 
	in analogy to \eqref{eq:al4n}.
	
	\item With no meaningful boundary value problem in the $(0+1)$-dimensional QM model, one cannot proceed to derive 
	the formulae~\eqref{eq:EWq}, \eqref{eq:halfEWqnew}, for
	quantum contributions to the semiclassical rate. As we have already noted earlier, there is no analogue in QM  of the 
	QFT expression \eqref{eq:halfEWqnew2}, which was instrumental in obtaining unsuppressed QFT rates due to a 
	non-trivial stretching of the singularity surface $|\tau_\infty| $ by the force~$nm$.
	
\end{enumerate}
 
\bigskip
\section{Semiclassical rate in (2+1) dimensions}
\label{sec:lowd}
\medskip

Loop contributions to the multiparticle amplitudes at threshold in 1 and 2 spatial dimensions are infrared divergent. 
In the 2+1 dimensional theory at small but non-vanishing $\varepsilon$, 
the terms of order $(\lambda \log \varepsilon)^k$ appear at $k$ loops in perturbation theory. 
These terms were summed up using the renormalization group technique in \cite{Rubakov:1994cz} in the
limit 
where \cite{Libanov:1994ug},
\[
\lambda \to 0\,, \quad n={\rm fixed}\,, \quad \varepsilon \to 0\quad {\rm with} \quad 
 \lambda \log \varepsilon={\rm fixed}\,.
 \]
In this section we will explain that this resummation in fact provides a non-trivial verification of the Higgsploding rate predicted 
in the 2+1 dimensional theory by the
semiclassical approach.

\bigskip

All our 4-dimensional QFT calculations in in sections \ref{sec:son} and \ref{sec:main} can be straightforwardly generalised to any number of 
dimensions $(d+1)$. Consider once more the scalar QFT model \eqref{eq:L} with the VEV $v\neq0$. 

The expression $W(E,N)_d$ in the exponent of the multiparticle rate ${\cal R}_n(E)$ has the same general decomposition
into tree-level and quantum parts as before,
\[
W(E,n)_d \,=\, W(E,n; \lambda)^{\rm tree}_d  \,+\, \Delta W(E,n; \lambda)^{\rm quant}_d\,,
\label{eq:EalWfin2d}
\]
where the tree-level expression in $(d+1)$ dimensions reads ({\it cf.} \eqref{eq:EalWfinfin}),
\[
W(E,n)^{\rm tree}_d \,=\, n \left(\log\frac{\lambda n}{4}-1\right)\,+\, \frac{dn}{2}\left( \log\frac{\varepsilon}{d\pi}+1\right)
\,,
\label{eq:EalWfinfind}
\]
and the quantum contribution is given by,
\[
\Delta W^{\rm quant}_d \,=\ 2nm\,|\tau_\infty|\,+\, 2 \int  d^d x \bigg[ \int^{+\infty}_{\tau_0(\bec{x})} d\tau \, {\cal L}_{E}(h_1) \,-\,
\int_{\tau_0(\bec{x})}^0d\tau \, {\cal L}_{E}(h_2) \bigg]
\label{eq:EWqd233},
\]
extremised over the singularity surfaces $\tau_0(\bec{x})$
in complete analogy with \eqref{eq:EWq}.

\medskip

For the rest of this section we we will consider the case of $d=2$ spatial dimensions and will concentrate on the contribution
of the stationary surface to the quantity $\frac{1}{2} \Delta W^{\rm quant}_{d=2}$, which we write as,
\[
\frac{1}{2}\,\Delta W^{\rm quant}_d \,=\ E\,|\tau_\infty|\,-\, 2\pi\mu \left(
\int_{r_0}^R r \sqrt{1+\dot{r}}\, dr - \int_0^R r\, dr\right)\,.
\label{eq:EWqd}
\]
The surface tension is $\mu =m^3/\lambda$ as before and the critical radius in $d=2$ is given by
$r_0 = E/(2\pi m)$. 
Proceeding with the evaluation of \eqref{eq:EWqd} on the classical trajectory $r(\tau)$, analogously to the calculation in the previous section, we get,
\[
\frac{1}{2} \Delta W^{\rm quant} \,=\, 
-\, \int^{R}_{r_0}  
2\pi \, \mu \, \sqrt{r^2-r_0^2} \,dr
\,+\, {2\pi} \,\mu R^2
\,.\label{eq:pdrNd}
\]
In the $Rm \to \infty$ limit this becomes,
\[
\simeq\, \frac{n^2 \lambda}{m}\,\frac{3}{4\pi} \left(\log (Rm)\,+\, \frac{1}{2} \,+\, \log\left(\frac{2\pi}{3}\frac{m}{\lambda n}\right)
\,+\, {\cal O}\left(\frac{1}{Rm}\frac{\lambda n}{m}\right)\right)\,.
\]
Adopting the infinite volume limit, in which the limit $Rm\to \infty$ is taken first while the quantity $\frac{n\lambda}{m}$ is held fixed,  
we can drop the $R$-independent and $1/R$-suppressed terms, leaving only the logarithmically divergent contribution,
\[
\frac{1}{2} \Delta W^{\rm quant} \,\simeq\,  \frac{3}{4\pi}\,\frac{n^2 \lambda}{m} \log (Rm).
\label{eq:semicld2}
\]
We see that all power-like divergent terms in $mR$ have cancelled in the expressions \eqref{eq:pdrNd} and \eqref{eq:semicld2}, but the 
logarithmic divergence remains. This result is not surprising in $d<3$ dimensions and is the consequence of the infrared divergencies 
in the amplitudes at thresholds due to the rescattering effects of final particles. In fact, the appropriate coupling constant in the lower-dimensional theory
is not the bare coupling $\lambda$ but the running quantity $\lambda t$, where $t$ is the logarithm of the characteristic momentum scale in the final state.
In our case we can set,
\[
t\,=\, \log(Rm),
\]
and treat $R$ as the reciprocal of the average momentum scale in the final state, i.e. $Rm = 1/\varepsilon^{1/2}$.

The semiclassical result obtained in \eqref{eq:semicld2} encodes the effect of taking into account quantum corrections to the
scattering amplitudes into $n$-particle states near their threshold, and implies,
\[
A_n \,\simeq\, A_n^{\rm tree} \exp \left(\frac{3 n^2 \lambda\, t}{4\pi m} \right)\,.
\label{eq:wkb3d}
\]
It is important to recall that the semiclassical limit is assumed in the derivation of the above expression. As always, this is
the weak-coupling large-multiplicity limit, such that\footnote{Recall that in $(2+1)$ dimensions, $\lambda$ has dimensions of mass.}
\[
{\rm dimensionless\, running\, coupling}:\,\, \frac{\lambda\, t}{m} \to 0 \, \qquad{\rm and\,\, multiplicity}:\, n\to \infty,
\]
with the quantity $ n\frac{ \lambda \,t}{m} $ held fixed (and ultimately large) and $t=-1/2\log \varepsilon \to 0$. This enforces the non-relativistic
limit, which selects the amplitudes close to their multiparticle thresholds.

It is important that it is the {\rm running} coupling $\lambda t$ that is required to be small in the semiclassical exponent\footnote{For example, it is completely analogous to the instanton action $S_{\rm inst} = \frac{8\pi^2}{g^2(t)}$ in Yang-Mills theory, where the inclusion of
quantum corrections from the determinants into the instanton measure in the path integral ensures that $S_{\rm inst}$ in the exponent depends on the correct RG coupling 
$g^2(t)$ and not the unphysical bare coupling $g^2_{\rm bare}$.}.
This implies that the semiclassical expression would in general include unknown corrections,
 \[
A_n \,\simeq\, A_n^{\rm tree} \exp \left(\frac{3 n^2 \lambda\, t}{4\pi m}\left(1+ \sum_{k=1}^\infty c_k\left(\frac{\lambda t}{m} \right)^k\right) \right)\,,
\label{eq:wkb3dsum}
\] 
parameterised by the sum $\sum_{k=1} c_k\left(\frac{\lambda t}{m} \right)^k$. Of course, there is a well-defined regime corresponding to 
the small values of the effective coupling $\lambda t$ where these
corrections are negligible and the leading order semiclassical result in \eqref{eq:wkb3d} is justified.

\medskip

Remarkably, the semiclassical formula \eqref{eq:wkb3d} can be tested against an independent computation of quantum effects
in the $(2+1)$-dimensional theory, obtained in~\cite{Rubakov:1994cz,Libanov:1994ug} using the RG resummation of perturbative diagrams.
The result is,
\[
A_n^{\rm RG} \,=\, A_n^{\rm tree} \,\left(1\,-\, \frac{3 \lambda\, t}{2\pi m} \right)^{-\, \frac{n(n-1)}{2}}\,.
\label{eq:RG2d}
\]
This expression is supposed to be valid for any values of $n$ in the regime where the effective coupling $\lambda t$
is in the interval,
\[
0\, \le \, \frac{\lambda\, t}{m}\, \lesssim\, 1\,.
\]
Now, taking the large-$n$ limit, the RG-technique-based result of~\cite{Rubakov:1994cz,Libanov:1994ug} gives,
\[
A_n^{\rm RG} \,=\, A_n^{\rm tree} \,\exp \left(\frac{3 n^2 \lambda\, t}{4\pi m}\left(1+ \sum_{k=1}^\infty \frac{1}{k+1}\,\left(\frac{3\lambda t}{2\pi m} \right)^k\right) \right)\,.
\label{eq:RG2dT}
\]
It is a nice test of the semiclassical approach that the leading-order terms in the exponent in both expressions, \eqref{eq:wkb3dsum} and \eqref{eq:RG2dT}
are exactly the same and given by $\frac{3 n^2 \lambda\, t}{4\pi m}$.
An equally important observation is that the subleading terms are of the form $\sum_{k=1} c_k\left(\frac{\lambda t}{m} \right)^k$, which is suppressed in the 
semiclassical limit $\lambda t\to 0$. There is no contradiction between the two expressions in the regime where the semiclassical approach is justified.

\medskip

It thus follows that there is a regime in the $(2+1)$-dimensional theory where the multiparticle amplitudes near their thresholds, and consequently 
the probabilistic rates ${\cal R}_n(E)$, become large.
In the case of the RG expression \eqref{eq:RG2d}, this is the consequence of taking a large negative power $-n^2/2$ of the term that is smaller than 1.
This implies that there is room for realising Higgsplosion in this $(2+1)$-dimensional model in the broken phase. 

In the case of a much simpler model -- the quantum mechanical anharmonic oscillator in the unbroken phase -- it was  shown 
in Refs.~\cite{Bachas:1991fd,Jaeckel:2018ipq}
that the rates remain exponentially suppressed in accordance with what would be expected from unitarity in QM.

 \bigskip
\section{Conclusions}
\label{sec:concl}
\medskip

In these notes we have  provided a detailed derivation of the semiclassical exponent for the multi-particle 
production rate in the Higgsplosion limit, along with a review of the semiclassical method used to compute it.
The derivation holds in
the high-particle-number $\lambda n \gg 1$ 
limit in the kinematical regime where the final state particles are produced near their mass thresholds.
This corresponds to the limit,
\[ \lambda \to 0\,, \quad n\to \infty\,, \quad {\rm with}\quad
\lambda n = {\rm fixed} \gg 1 \,, \quad \varepsilon ={\rm fixed} \ll 1 \,.
\label{eq:limit2}
\]
Combining the tree-level \eqref{eq:EalWfinfin} and quantum effects \eqref{eq:pdrfinal}
contributions,
\[
W(E,n) \,=\, W(E,n; \lambda)^{\rm tree}  \,+\, \Delta W(E,n; \lambda)^{\rm quant}\,,
\label{eq:EalWfin2222}
\]
we can write down the full semiclassical rate,
\[
{\cal R}_n(E)\,= \, e^{W(E,n)}\,=\, 
\exp \left[ n\, \left( 
\log \frac{\lambda n}{4} \,+\, 0.85\, \sqrt{\lambda n}\,+\,\frac{3}{2}\log \frac{\varepsilon}{3\pi} \,+\, \frac{1}{2}
\right)\right],
\label{eq:Rnp2}
\]
computed in the high-multiplicity non-relativistic limit \eqref{eq:limit2}.
This expression for the multi-particle rates was first written down in the precursor of this work \cite{Khoze:2017ifq}, and
was used in Refs.~\cite{Khoze:2017tjt,Khoze:2017lft} and subsequent papers to introduce and motivate the Higgsplosion mechanism.

The energy in the initial state and the final state multiplicity are related linearly via,
\[ E/m \,=\, (1 + \varepsilon) \, n\,.
\]
Hence, for any fixed non-vanishing value of $\varepsilon$, one can raise the energy to 
achieve any desired large value of $n$ and consequentially a large $\sqrt{\lambda n}$.
Clearly, at the strictly vanishing value of $\varepsilon$, the phase-space volume is zero and 
the entire rate \eqref{eq:Rnp2} vanishes. Then by increasing $\varepsilon$ to positive but still small values, the rate increases.
The competition is between the negative $\log \varepsilon$ term and the positive $\sqrt{\lambda n}$ term 
in \eqref{eq:Rnp2}, and there is always a range of sufficiently high multiplicities where $\sqrt{\lambda n}$
overtakes the logarithmic term $\log \varepsilon$ for any fixed (however small) value of $\varepsilon$.
This leads to the exponentially growing multi-particle rates above a certain critical energy, which in the case
described by the expression in \eqref{eq:Rnp2} is in the regime of $E_c \sim 200 \,m$.

 \begin{figure*}[t]
\begin{center}
\includegraphics[width=1.0\textwidth]{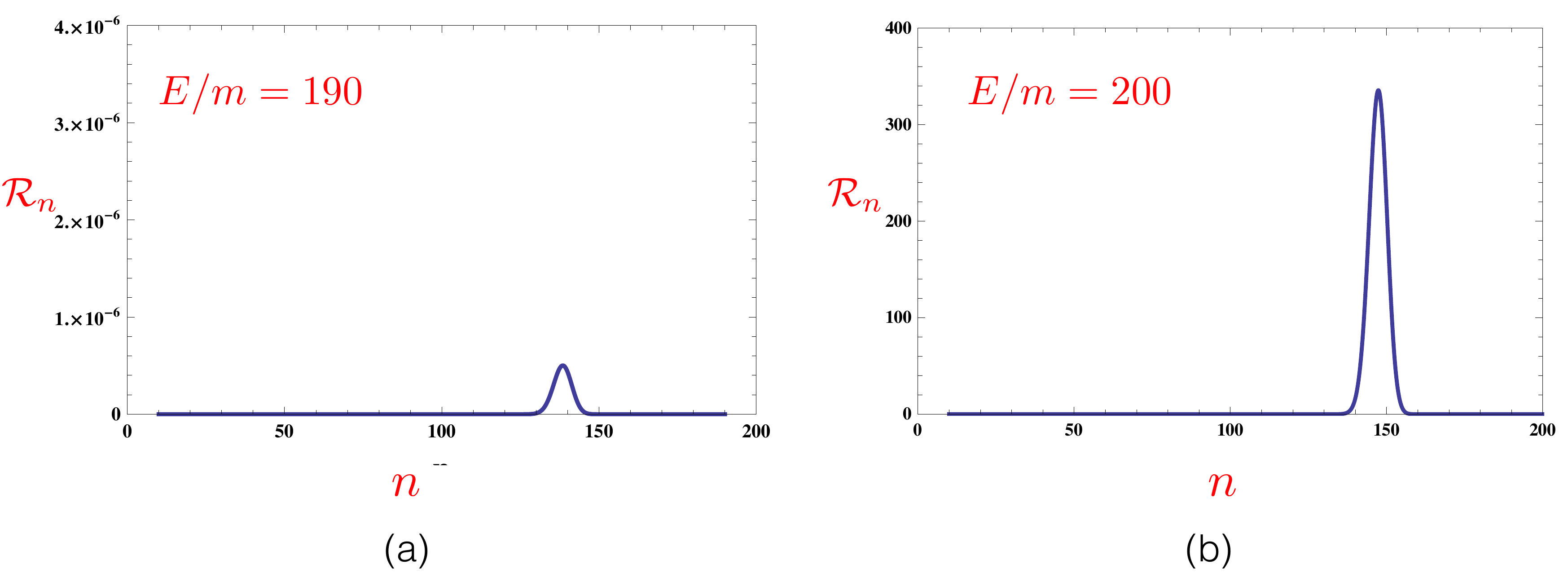}
\end{center}
\vskip-.5cm
\caption{Plots of the semiclassical rate ${\cal R}_n$ in Eq.~\eqref{eq:Rnp2}
 as a function of $n$ for values of the energy/virtuality $E$ fixed at 
190$m$ and at 200$m$. We chose $\lambda=1/8$. There is a sharp exponential dependence of the peak rate
on the energy. The peak multiplicities $n \sim 150$ in these examples are not far below the maximal values $n_{\rm max} = E/m$
allowed by kinematics.
 } 
\label{fig:R2}
\end{figure*}

The expression for the multi-particle rate \eqref{eq:Rnp2} is a leading order semiclassical approximation
and should of course not be taken as anything more than a rough estimate of the Higgsplosion rate. 
We have already emphasised that this result is an
approximation derived in the simplified scalar model \eqref{eq:L} and in the simplifying non-relativistic limit.
Specifically, our result $ \Delta W^{\rm quant}$ for quantum contributions to $W$ in \eqref{eq:pdrfinal}
was derived on the multi-particle threshold, i.e. at $\varepsilon=0$.
Hence the higher-order corrections in $\varepsilon$ will be present in the expression for the rate 
in the $\lambda n$ limit. Let us estimate these corrections following the discussion
in section 5 of Ref.~\cite{Khoze:2017ifq}.

Denote these unknown corrections $f_{\lambda n; \varepsilon} ({\lambda n, \varepsilon} )$, so that
\[ \Delta_{\rm new} W \,=\, \frac{\lambda n}{\lambda}\,\, f_{\lambda n; \varepsilon} ({\lambda n, \varepsilon} )\,,
\]
and the now modified rate becomes,
\[
{\cal R}_n(E)\,\sim \,  \int_0^{\varepsilon_{nr}} d \varepsilon\, \left(\frac{\varepsilon}{3 \pi}\right)^{\frac{3n}{2}} 
\exp \left[ n\, \left( 
0.85\, \sqrt{\lambda n} \,+\, \log \lambda n\,+\,f_{\lambda n; \varepsilon} ({\lambda n, \varepsilon} )
\,+\, c
\right)\right],
\label{eq:RnpN}
\]
where we have included the new correction $ n \sim f_{\lambda n; \varepsilon} ({\lambda n, \varepsilon} )$ 
and have also made explicit the fact that the $3n/2 \log \varepsilon/(3 \pi)$ factor in
the exponent of the rate \eqref{eq:Rnp2} originated from the integration 
over the non-relativistic $n$-particle phase-space with a cut-off at $\varepsilon_{nr} < 1$.
The constant $c$ absorbs various constant factors appearing in the original rate.

The integral above is of course meant to be computed in the large-$n$ limit by finding the
saddle-point value $\varepsilon=\varepsilon_{\star}$. The main point of the exercise is to determine
(1) whether there is a regime where $\varepsilon_{\star} \ll 1$ so that our near-the-threshold approach is justified,
and (2) whether the saddle-point value of the rate itself is large. These requirements should 
tell us something about the function $f_{\lambda n; \varepsilon}$.

Let us assume that the correction to our result has the form,
\[
f_{\lambda n; \varepsilon} ({\lambda n, \varepsilon} ) \,=\, - a\, \varepsilon \, (\lambda n)^p
\,, \label{eq:fnew}
\]
where $a$ and $p$ are constants. This function is supposed to represent the higher-order in 
$\varepsilon$ correction to our result in the small-$\varepsilon$, large-$\lambda n$ limit.
The integral we have to compute is,
\[
{\cal R}_n \,\sim \,
e^{n\, \left( 0.85\, \sqrt{\lambda n} \,+\, \log \lambda n \,+\, \tilde c
\right)}\,\,
\int d \varepsilon\,
e^{n\, \left( \frac{3}{2} \log \varepsilon \,-\, a\, \varepsilon \, (\lambda n)^p\right) } \,.
\label{eq:RnpNsp}
\]
Denoting the $\varepsilon$-dependent function in the exponent $s(\varepsilon)$,
\[ s(\varepsilon) \,=\, \frac{3}{2} \log \varepsilon \,-\, a\, \varepsilon \, (\lambda n)^p\,,
\]
we can compute the saddle-point,
\[ \frac{\partial s(\varepsilon)}{\partial \varepsilon} \,=\, 0  \quad \Rightarrow \quad
\varepsilon_{\star} \,=\, \frac{3}{2}\, \frac{1}{a} \, \frac{1}{(\lambda n)^p}\,,
\]
and the value of the function $s$ at the saddle-point,
\[
s(\varepsilon_{\star}) \,=\, -\frac{3}{2}\, \left(p \log \lambda n \,+\,1\,-\, \log \frac{3}{2a} \right)\,.
\]
Combining this with the function in the exponent in front of the integral in \eqref{eq:RnpNsp}
we find the saddle-point value of the rate,
\[
{\cal R}_n (\varepsilon_\star) \,\sim \,
\exp \left[n\, \left( 0.85\, \sqrt{\lambda n} \,-\, \left(\frac{3p}{2}-1\right) \log \lambda n \,+\, {\rm const} \right)
\right]
 \,.
\label{eq:RnpNsp2}
\]
This is the value of the rate at the local maximum, and since the factor of $\sqrt{\lambda n}$ grows faster
than the $-\log \lambda n $ term, the peak value of the rate is exponentially large
in the limit of $\sqrt{\lambda n}  \to \infty$. It is also easy to verify that this conclusion is
consistent within the validity of the non-relativistic limit.
In fact, the value of $\varepsilon$ at the saddle-point is non-relativistic,
\[
\varepsilon_{\star} \,=\, \frac{3}{2}\, \frac{1}{a} \, \frac{1}{(\lambda n)^p} \, \to\, 0 \,,
\quad {\rm as} \quad 
\lambda n \to\, \infty
\,.
\]
We thus conclude that the appearance of the higher-order $\varepsilon$ corrections to our result 
in the form \eqref{eq:fnew} do not prevent the eventual Higgsplosion in this model at  least in the
formal limit $\sqrt{\lambda n} \to \infty$ where we have found that,
\[
{\cal R}_n (\varepsilon_\star) \, \gg \, 1\,.
\]
The growth persists for any constant values of $a$ and $p$. In fact, if $a$ was negative, the
growth would only be enhanced. In   \eqref{eq:fnew} we have assumed that the function 
goes as $\varepsilon$ to the first power. The higher powers would not change the conclusion, while
the effect of $\sim \varepsilon^0$ is what is already taken into account in \eqref{eq:pdrfinal}.

\medskip

We now also recall from our earlier discussion
that the expression for ${\cal R}_n(E) $ in \eqref{eq:Rnp2} is 
in fact  a distribution-valued function. To obtain the proper $n$-particle production rate one needs to account for the operator-smearing 
effect in the definition of the initial state in \eqref{eq:opdef11}.
The result of this is that the Higgsplosion rate becomes $|\tilde{g} (p)|^2 \, {\cal R}_n(\sqrt{s})$ 
where $\tilde{g}(p)$ is the momentum space Fourier transform
of the spacetime test function $g(x')$. This implies that the Higgsplosion rate can be written in the 
form~\cite{Khoze:2018qhz},
\[
R_g(n,\sqrt{s})\,=\, |\tilde{g} (\sqrt{s})|^2 \,\, {\cal R}_n(\sqrt{s})\, =\, |\tilde{g} (\sqrt{s})|^2 \, e^{ \,n\, F(\lambda n, \varepsilon)}
\,,
\label{eq:ggRnhg}
\]
by dressing the leading order semiclassical result ${\cal R}_n(\sqrt{s})$ with the smearing function $|\tilde{g} (\sqrt{s})|^2$. 
This smearing will also ensure an
acceptable behaviour of the physical production rate at asymptotically high centre-of-mass energies for $2 \to n$ processes,
 in accordance with unitarity.

\medskip

Finally, we note that our discussion concentrated entirely on a simple scalar QFT model.
If more degrees of freedom were included, for example the $W$ and $Z$ vector bosons and
the SM fermions, new coupling parameters (such as the gauge coupling and the Yukawas) 
would appear in the expression for the rate along with the final state particle multiplicities. As there are more parameters,
the simple scaling properties of ${\cal R}_n$ in the pure scalar theory will be modified. Understanding how this would work in practice and 
investigating the appropriate semiclassical limits is one (of the admittedly many) tasks for 
future work on exploring realisations of Higgsplosion in particle physics. 

\bigskip
\bigskip

\section*{Acknowledgements}

We are grateful to Joerg Jaeckel,  Matthew Kirk, Jakub Scholtz and especially to Michael Spannowsky for many useful discussions.

\newpage 
\appendix

\section{Projection of scalar field state onto coherent state}
\label{app:proj}
\medskip

The aim of this appendix is to derive the expression for the overlap $\braket{\phi|\{a\}}$ between the eignstate of the field operator $\phi$ 
and the coherent state in a scalar QFT.
We begin by inserting coherent state definition,
\[
\braket{\phi|\{a\}} \,=\,\braket{\phi|\exp\left[ \int \dk \,a_\bec{k} \ad_\bec{k}\right]|0}\,,
\label{eq:A.1}
\]
and then representing the creation operator $\ad_\bec{k}$ in terms of the original field $\hat\phi$ and conjugate momentum 
$\hat\pi =\partial_t \hat{\phi}$ field operators in Fourier space,
\[
\tilde{\hat{\phi}}(t, \bec{k}):=\int d^3x \, e^{-i\bec{k}\cdot\bec{x}}\, \hat{\phi}(t,\bec{x})\,, \qquad
\tilde{\hat{\pi}}(t, \bec{k}):=\int d^3x \, e^{-i\bec{k}\cdot\bec{x}}\, \partial_t \hat{\phi}(t,\bec{x})\,.
\]
The states in the overlap formula \eqref{eq:A.1} are the eigenstates of the corresponding operators at $t=0$, 
\[
\hat{\tilde{\phi}}(t=0, \bec{k}) \ket{\phi} \,=\, \tilde{\phi}(0, \bec{k}) \ket{\phi}\,, \quad
\an_\bec{k}(t=0) \ket{\{a\}} \,=\, a_\bec{k} \ket{\{a\}}\,.
\]
Hence we should compute the creation operators $\ad_\bec{k}$ in the definition
of the coherent state in 
\eqref{eq:A.1} at $t=0$.
For the annihilation and creation operators we have the standard expressions,
\[
\begin{split}
\hat{a}_\bec{k} &=\, \sqrt{\frac{\omega_\bec{k}}{2}}
\left( \hat{\tilde{\phi}}(0, \bec{k}) + \frac{i \hat{\tilde{\pi}}(0, \bec{k})}{\omega_\bec{k}}\right)\,,
\\
\ad_\bec{k} &=\, (\hat{a}_\bec{k})^\dagger\,=\, \sqrt{\frac{\omega_\bec{k}}{2}}
\left( \hat{\tilde{\phi}}(0, -\bec{k}) - \frac{i \hat{\tilde{\pi}}(0, -\bec{k})}{\omega_\bec{k}}\right)\,.
\label{eq:A.3}
\end{split}
\]
To simplify the expression for the creation operator in \eqref{eq:A.3} we introduce a short-hand notation,
\begin{equation}
\hat{\phi}_\bec{k}\,:=\, \hat{\tilde{\phi}}(0,-\bec{k})\,=\,\int \dx \, e^{i\bec{k}\cdot\bec{x}}\, \hat\phi(0,\bec{x})\,, \quad {\rm and} \quad
\hat{\uppi}_\bec{k}\,:=\,\hat{\tilde{\pi}}(0, -\bec{k})\,.
\end{equation}

For the overlap in \eqref{eq:A.1} we have,
\begin{equation}
\begin{split}
\braket{\phi|\{a\}}&=\braket{\phi|\exp\left[ \int \dk\, a_\bec{k} \ad_\bec{k}\right]|0}\\
&=\braket{\phi|\exp\left[\int \dk\, a_\bec{k} \sqrt{\omk/2}\( \hat\phi_\bec{k}-i\omk^{-1}\hat\uppi_\bec{k}\)\right]|0}.
\end{split}
\end{equation}
In analogy with QM, ${\phi}_\bec{k}$ represents the generalised coordinate, and 
\[
\hat{\uppi}_\bec{k}\,=\, -i\dfrac{\partial}{\partial \phi_\bec{k}} \,,
\]
is the conjugate momentum operator.
Note that since $\phi(x)\in\mathbb{R}$, $\phi^*_\bec{k} = \phi_{-\bec{k}}$ so $|\phi_\bec{k}|^2=\phi_\bec{k}\phi_\bec{k}^*=\phi_\bec{k}\phi_{-\bec{k}}$. Hence,
\begin{equation}
\begin{split}
\braket{\phi|\{a\}}&=\exp\left[\int \dk\, a_\bec{k} \sqrt{\omk/2}\( \phi_\bec{k}-\omk^{-1}\dfrac{\partial}{\partial\phi_\bec{k}}\)\right]\braket{\phi|0}\\
&=N\exp\left[\int \dk\, a_\bec{k} \sqrt{\omk/2}\( \phi_\bec{k}-\omk^{-1}\dfrac{\partial}{\partial\phi_\bec{k}}\)\right]e^{-\int d{\bf p}\,\omega_\bec{p}\phi_\bec{p}\phi_{-\bec{p}}/2},
\end{split}
\end{equation}
where $N$ is some normalisation and $\braket{\phi|0}$ is analogous to $\braket{q|0}$ in QM. We now make use of the following relation,
\begin{equation}
\label{eq:appbch}
e^{\hat{A}+\hat{B}}=e^{-[\hat{A},\hat{B}]/2}e^{\hat{A}}e^{\hat{B}}\quad \mathrm{if}\quad [\hat{A},\hat{B}]\in \mathbb{C},
\end{equation}
with $\hat{A}$ and $\hat{B}$ choices,
\begin{equation}
\begin{split}
\hat{A}&=\int \dk\,\sqrt{\omk/2}\, \an_\bec{k}\phi_\bec{k}\\
\hat{B}&=-\int \dep \, \omega_\bec{p}^{-1}\sqrt{\omega_\bec{p}/2}\, \an_\bec{p}\dfrac{\partial}{\partial\phi_\bec{p}}.
\end{split}
\end{equation}
Recalling that $[\phi_\bec{k},\partial_{\phi_\bec{p}}]=-\delta^3(\bec{k}-\bec{p})$ and $\omk=\omega_{-\bec{k}}$, we find that,
\begin{equation}
\begin{split}
[\hat{A},\hat{B}] &=-\dfrac{1}{2}\int \dk\,\dep\, a_\bec{k} a_\bec{p}\sqrt{\omk/\omega_\bec{p}}[\phi_\bec{k},\partial_{\phi_\bec{p}}]\\
&=\dfrac{1}{2}\int \dk\,\dep\, a_\bec{k} a_\bec{p}\sqrt{\omega_\bec{k}/\omega_\bec{p}}\delta(\bec{k}-\bec{p})\\
&=\dfrac{1}{2}\int \dk\,a_\bec{k} a_{-\bec{k}}.
\end{split}
\end{equation}
Therefore, application of \eqref{eq:appbch} yields,
\begin{equation} 
\begin{split}
e^{\hat{A}+\hat{B}}&=\exp\left[\int \dk\, a_\bec{k} \sqrt{\omk/2}\( \phi_\bec{k}-\omk^{-1}\dfrac{\partial}{\partial\phi_\bec{k}}\)\right]\\
&= e^{-\tfrac{1}{4}\int \dk\, a_\bec{k} a_{-\bec{k}}}\exp\( \int \dk\, a_\bec{k}\sqrt{\omk/2}\phi_\bec{k}\)\exp\( -\int \dk\, a_\bec{k}(2\omk)^{-1/2}\dfrac{\partial}{\partial\phi_\bec{k}}\).
\end{split}
\end{equation}
We note that though slightly abstract, the last term is infact the operator for a translation in $\phi_\bec{k}$-space by $a_\bec{k}(2\omk)^{-1/2}$. Therefore, $\braket{\phi|0}$, has its argument shifted: $\phi_\bec{k}\to\phi_\bec{k}-a_\bec{k}(2\omk)^{-1/2}$, giving,
\begin{equation} 
\begin{split}
\braket{\phi|\{a\}}\propto &e^{-\tfrac{1}{4}\int \dk\, a_\bec{k} a_{-\bec{k}}}\exp\( \int \dk\, a_\bec{k}\sqrt{\omk/2}\phi_\bec{k}\)\times\\
&\exp\( -\dfrac{1}{2}\int \dk\, \omk (\phi_\bec{k}-a_\bec{k}(2\omk)^{-1/2}) (\phi_{-\bec{k}}-a_{-\bec{k}}(2\omk)^{-1/2})\)\\
\propto& e^{-\tfrac{1}{4}\int \dk\, a_\bec{k} a_{-\bec{k}}}\exp\( \int \dk\, a_\bec{k}\sqrt{\omk/2}\phi_\bec{k}\)\times\\
&\exp\( -\dfrac{1}{2}\int \dk\,\omk\phi_\bec{k}\phi_{-\bec{k}} -\dfrac{1}{4}\int \dk\,a_\bec{k}a_{-\bec{k}}+\int \dk\sqrt{\dfrac{\omk}{8}}\left[  \phi_\bec{k} a_{-\bec{k}}+\phi_{-\bec{k}}a_\bec{k} \right]  \)\\
\propto& \exp\left(  -\dfrac{1}{2}\int \dk\, a_\bec{k} a_{-\bec{k}} -\dfrac{1}{2}\int \dk\,\omk\phi_\bec{k}\phi_{-\bec{k}} +\int \dk \sqrt{2\omk}a_\bec{k}\phi_{-\bec{k}}  \right).
\end{split}
\end{equation}
Recall that $\phi_\bec{k}$ is the shorthand for $\tilde{\phi}(-\bec{k})$, and equivalently, 
$\phi_{-\bec{k}}\,=\, \tilde{\phi}(\bec{k})$;
 with this in mind we recover the expression \eqref{eq:projqft}
for the overlap $\braket{\phi|\{a\}}$
used in the rest of the paper.

\bigskip
\section{Contributions of $S[\phi]$ to saddle-point equations of $\tilde{\phi}_{i,f}$}
\label{app:btsp}
\medskip

Recall the action, $S[\phi(x)]$, in $d+1$ dimensions,
\begin{equation}
\begin{split}
S[\phi(x)]=&\int d^{d+1}x\dfrac{1}{2}\left[ (\partial_t\phi)^2-|\nabla\phi|^2 -2V(\phi) \right]\\
=& \dfrac{1}{2}\int^{t_f}_{t_i}dt\int d^dx[\partial_t(\phi\partial_t\phi)-\phi\partial^2_t\phi-|\nabla\phi|^2 -2V(\phi)].
\end{split}
\label{eq:B.1}
\end{equation}
Only the total derivative part of this integral,  
\begin{equation}
S_{\rm Boundary}[\phi_i,\phi_f]=\dfrac{1}{2}\int^{t_f}_{t_i}dt\int d^dx\, \partial_t(\phi\,\partial_t\phi)=\dfrac{1}{2}\int d^dx \,(\phi_f\partial_t\phi_f-\phi_i\partial_t\phi_i),
\label{eq:B.2}
\end{equation}
 will contribute to the saddle-point equations for $\tilde{\phi}_i$ and $\tilde{\phi}_f$. Other terms on the right hand side of 
 \eqref{eq:B.1} 
 contribute instead to the $\phi$ equation. Focusing on the total derivative, we have,  
 \begin{equation}
\int \dx \,\phi_f\partial_t\phi_f \,=\, 
 \int \dep \,\tilde{\phi}_f(\bec{p})\partial_t\tilde{\phi}_f(-\bec{p}).
\end{equation}
where $\partial_t\phi_f$ is $\partial_t\phi$ evaluated at time $t_f$ (and similarly for $\partial_t\phi_i$). 

Hence, the complete saddle-point equation for say $\tilde{\phi}_f$ is,  
\begin{equation}
i\, \dfrac{\delta S_{\rm Boundary}[\phi_i,\phi_f]}{\delta\tilde{\phi}_f(\bec{k})}+
\dfrac{\delta [B(\phi_f;b^*)]^*}{\delta\tilde{\phi}_f(\bec{k})}=0.
\end{equation}
Using the last equation in \eqref{eq:masterbt} for $ [B(\phi_f;b^*)]^*$, we find,
\begin{equation}
i\partial_t\tilde{\phi}_f(\bec{k})-\omk\tilde{\phi}_f(\bec{k})+\sqrt{2\omk}b^*_{-\bec{k}}e^{i\omk t_f} =0,
\end{equation}
which is equivalent to the saddle-point equation \eqref{eq:sp3}. The same logic can be used to recover \eqref{eq:sp2}.

\bigskip
\section{Contribution to $S_E^{(1)}$ from changes to integration range}
\label{app:intrange}
\medskip

Before variation, we have,
\begin{equation}
\label{eq:D1}
S_E^{(1)}\,= \, - \int  d^dx \int_{\infty}^{\tau_0(\bec{x})}d\tau\, \mathcal{L}_E(\phi_1)\,= \, - \int  d^dx \int_{\infty}^{\tau_0(\bec{x})}d\tau\,
 \left( \dfrac{1}{2}(\partial_\mu\phi_1)^2+V(\phi_1)  \right).
\end{equation}
We now consider the difference between this and the case where the singularity surface $A$ is destorted to $A'$, so that the spacetime coordinates describing it, $x^\mu(s_i)\to x^\mu(s_i)+n^\mu\delta x(s_i)$, where $s_i$ are coordinates on the surface. 

It is easily shown for the 1-dimensional integral, that the variation under a small shift in integration range, $\Delta$, gives a boundary term,
\begin{equation}
 \int^b_a[f(x+\Delta)-f(x)]dx \,=\, \left[f(x)\Delta \right]^b_a,
\end{equation}
in the limit of small $\Delta$. Applied to Eq.~\eqref{eq:D1}, with $\tau_0(\bec{x})\to\tau_0(\bec{x})+\delta\tau_0(\bec{x})$, we find,
\begin{equation}
-\left[\int d^dx\mathcal{L}_E(\phi_1)\delta\tau_0(\bec{x})\right]_{\infty}^{\tau_0(\bec{x})}.
\end{equation}
Of course, $\delta\tau_0(\bec{x})$ is only non-zero on surface $A$ and so the above term can be written as an integral over the singularity surface $A$,
\begin{equation}
-\int_A ds \mathcal{L}_E(\phi_1)\delta x(s)\,=\, -\int_A ds  \left[\left( \dfrac{1}{2}(\partial_\mu\phi_1)^2+V(\phi_1)  \right)\delta x(s)\right],
\end{equation}
as stated in Eq.~\eqref{eq:varSE} in \refsec\ref{sec:5.3.2}. Note that $ds$ is shorthand for the appropriate $d$-dimensional integration measure on the $d$-dimensional singularity surface, for surface coordinates $s_i$.

\bibliographystyle{JHEP}
\bibliography{references}

\providecommand{\href}[2]{#2}\begingroup\raggedright\begin{thebibliography}{10}

\bibitem{Khoze:2017tjt}
V.~V. Khoze and M.~Spannowsky, \emph{{Higgsplosion: Solving the Hierarchy
  Problem via rapid decays of heavy states into multiple Higgs bosons}},
  \href{http://dx.doi.org/10.1016/j.nuclphysb.2017.11.002}{\emph{Nucl. Phys.}
  {\bf B926} (2018) 95--111}, [\href{https://arxiv.org/abs/1704.03447}{{\tt
  1704.03447}}].

\bibitem{Son:1995wz}
D.~T. Son, \emph{{Semiclassical approach for multiparticle production in scalar
  theories}}, \href{http://dx.doi.org/10.1016/0550-3213(96)00386-0}{\emph{Nucl.
  Phys.} {\bf B477} (1996) 378--406},
  [\href{https://arxiv.org/abs/hep-ph/9505338}{{\tt hep-ph/9505338}}].

\bibitem{Khoze:2018kkz}
V.~V. Khoze, \emph{{Semiclassical computation of quantum effects in
  multiparticle production at large $\lambda n$}},
  \href{https://arxiv.org/abs/1806.05648}{{\tt 1806.05648}}.

\bibitem{Khoze:2017ifq}
V.~V. Khoze, \emph{{Multiparticle production in the large $\lambda n$ limit:
  realising Higgsplosion in a scalar QFT}},
  \href{http://dx.doi.org/10.1007/JHEP06(2017)148}{\emph{JHEP} {\bf 06} (2017)
  148}, [\href{https://arxiv.org/abs/1705.04365}{{\tt 1705.04365}}].

\bibitem{Cornwall:1990hh}
J.~M. Cornwall, \emph{{On the High-energy Behavior of Weakly Coupled Gauge
  Theories}}, \href{http://dx.doi.org/10.1016/0370-2693(90)90850-6}{\emph{Phys.
  Lett.} {\bf B243} (1990) 271--278}.

\bibitem{Goldberg:1990qk}
H.~Goldberg, \emph{{Breakdown of perturbation theory at tree level in theories
  with scalars}},
  \href{http://dx.doi.org/10.1016/0370-2693(90)90628-J}{\emph{Phys. Lett.} {\bf
  B246} (1990) 445--450}.

\bibitem{Brown:1992ay}
L.~S. Brown, \emph{{Summing tree graphs at threshold}},
  \href{http://dx.doi.org/10.1103/PhysRevD.46.R4125}{\emph{Phys. Rev.} {\bf
  D46} (1992) R4125--R4127}, [\href{https://arxiv.org/abs/hep-ph/9209203}{{\tt
  hep-ph/9209203}}].

\bibitem{Argyres:1992np}
E.~N. Argyres, R.~H.~P. Kleiss and C.~G. Papadopoulos, \emph{{Amplitude
  estimates for multi - Higgs production at high-energies}},
  \href{http://dx.doi.org/10.1016/0550-3213(93)90140-K}{\emph{Nucl. Phys.} {\bf
  B391} (1993) 42--56}.

\bibitem{Voloshin:1992rr}
M.~B. Voloshin, \emph{{Estimate of the onset of nonperturbative particle
  production at high-energy in a scalar theory}},
  \href{http://dx.doi.org/10.1016/0370-2693(92)90901-F}{\emph{Phys. Lett.} {\bf
  B293} (1992) 389--394}.

\bibitem{Voloshin:1992nu}
M.~B. Voloshin, \emph{{Summing one loop graphs at multiparticle threshold}},
  \href{http://dx.doi.org/10.1103/PhysRevD.47.R357}{\emph{Phys. Rev.} {\bf D47}
  (1993) R357--R361}, [\href{https://arxiv.org/abs/hep-ph/9209240}{{\tt
  hep-ph/9209240}}].

\bibitem{Smith:1992rq}
B.~H. Smith, \emph{{Summing one loop graphs in a theory with broken symmetry}},
  \href{http://dx.doi.org/10.1103/PhysRevD.47.3518}{\emph{Phys. Rev.} {\bf D47}
  (1993) 3518--3520}, [\href{https://arxiv.org/abs/hep-ph/9209287}{{\tt
  hep-ph/9209287}}].

\bibitem{Argyres:1993wz}
E.~N. Argyres, R.~H.~P. Kleiss and C.~G. Papadopoulos, \emph{{Multiscalar
  amplitudes to all orders in perturbation theory}},
  \href{http://dx.doi.org/10.1016/0370-2693(93)91287-W}{\emph{Phys. Lett.} {\bf
  B308} (1993) 292--296}, [\href{https://arxiv.org/abs/hep-ph/9303321}{{\tt
  hep-ph/9303321}}].

\bibitem{Libanov:1994ug}
M.~V. Libanov, V.~A. Rubakov, D.~T. Son and S.~V. Troitsky,
  \emph{{Exponentiation of multiparticle amplitudes in scalar theories}},
  \href{http://dx.doi.org/10.1103/PhysRevD.50.7553}{\emph{Phys. Rev.} {\bf D50}
  (1994) 7553--7569}, [\href{https://arxiv.org/abs/hep-ph/9407381}{{\tt
  hep-ph/9407381}}].

\bibitem{Libanov:1995gh}
M.~V. Libanov, D.~T. Son and S.~V. Troitsky, \emph{{Exponentiation of
  multiparticle amplitudes in scalar theories. 2. Universality of the
  exponent}}, \href{http://dx.doi.org/10.1103/PhysRevD.52.3679}{\emph{Phys.
  Rev.} {\bf D52} (1995) 3679--3687},
  [\href{https://arxiv.org/abs/hep-ph/9503412}{{\tt hep-ph/9503412}}].

\bibitem{Voloshin:1994yp}
M.~B. Voloshin, \emph{{Nonperturbative methods}},  in \emph{{27th International
  Conference on High-energy Physics (ICHEP 94) Glasgow, Scotland, July 20-27,
  1994}}, pp.~0121--134, 1994.
\newblock \href{https://arxiv.org/abs/hep-ph/9409344}{{\tt hep-ph/9409344}}.

\bibitem{Khoze:2014zha}
V.~V. Khoze, \emph{{Multiparticle Higgs and Vector Boson Amplitudes at
  Threshold}}, \href{http://dx.doi.org/10.1007/JHEP07(2014)008}{\emph{JHEP}
  {\bf 07} (2014) 008}, [\href{https://arxiv.org/abs/1404.4876}{{\tt
  1404.4876}}].

\bibitem{Khoze:2014kka}
V.~V. Khoze, \emph{{Perturbative growth of high-multiplicity W, Z and Higgs
  production processes at high energies}},
  \href{http://dx.doi.org/10.1007/JHEP03(2015)038}{\emph{JHEP} {\bf 03} (2015)
  038}, [\href{https://arxiv.org/abs/1411.2925}{{\tt 1411.2925}}].

\bibitem{Degrande:2016oan}
C.~Degrande, V.~V. Khoze and O.~Mattelaer, \emph{{Multi-Higgs production in
  gluon fusion at 100 TeV}},
  \href{http://dx.doi.org/10.1103/PhysRevD.94.085031}{\emph{Phys. Rev.} {\bf
  D94} (2016) 085031}, [\href{https://arxiv.org/abs/1605.06372}{{\tt
  1605.06372}}].

\bibitem{Gorsky:1993ix}
A.~S. Gorsky and M.~B. Voloshin, \emph{{Nonperturbative production of
  multiboson states and quantum bubbles}},
  \href{http://dx.doi.org/10.1103/PhysRevD.48.3843}{\emph{Phys. Rev.} {\bf D48}
  (1993) 3843--3851}, [\href{https://arxiv.org/abs/hep-ph/9305219}{{\tt
  hep-ph/9305219}}].

\bibitem{Libanov:1997nt}
M.~V. Libanov, V.~A. Rubakov and S.~V. Troitsky, \emph{{Multiparticle processes
  and semiclassical analysis in bosonic field theories}},
  \href{http://dx.doi.org/10.1134/1.953038}{\emph{Phys. Part. Nucl.} {\bf 28}
  (1997) 217--240}.

\bibitem{Khoze:2017lft}
V.~V. Khoze and M.~Spannowsky, \emph{{Higgsploding universe}},
  \href{http://dx.doi.org/10.1103/PhysRevD.96.075042}{\emph{Phys. Rev.} {\bf
  D96} (2017) 075042}, [\href{https://arxiv.org/abs/1707.01531}{{\tt
  1707.01531}}].

\bibitem{Jaeckel:2014lya}
J.~Jaeckel and V.~V. Khoze, \emph{{Upper limit on the scale of new physics
  phenomena from rising cross sections in high multiplicity Higgs and vector
  boson events}},
  \href{http://dx.doi.org/10.1103/PhysRevD.91.093007}{\emph{Phys. Rev.} {\bf
  D91} (2015) 093007}, [\href{https://arxiv.org/abs/1411.5633}{{\tt
  1411.5633}}].

\bibitem{Gainer:2017jkp}
J.~S. Gainer, \emph{{Measuring the Higgsplosion Yield: Counting Large Higgs
  Multiplicities at Colliders}},  \href{https://arxiv.org/abs/1705.00737}{{\tt
  1705.00737}}.

\bibitem{Khoze:2017uga}
V.~V. Khoze, J.~Reiness, M.~Spannowsky and P.~Waite, \emph{{Precision
  measurements for the Higgsploding Standard Model}},
  \href{https://arxiv.org/abs/1709.08655}{{\tt 1709.08655}}.

\bibitem{Khoze:2018bwa}
V.~V. Khoze, J.~Reiness, J.~Scholtz and M.~Spannowsky, \emph{{A Higgsploding
  Theory of Dark Matter}},  \href{https://arxiv.org/abs/1803.05441}{{\tt
  1803.05441}}.

\bibitem{Faddeev:1980be}
L.~D. Faddeev and A.~A. Slavnov, \emph{{Gauge fields: Introduction to quantum
  theory}}, {\emph{Front. Phys.} {\bf 50} (1980) 1--232}.

\bibitem{Tinyakov:1992dr}
P.~G. Tinyakov, \emph{{Instanton like transitions in high-energy collisions}},
  \href{http://dx.doi.org/10.1142/S0217751X93000771}{\emph{Int. J. Mod. Phys.}
  {\bf A8} (1993) 1823--1886}.

\bibitem{Khlebnikov:1990ue}
S.~{\relax Yu}. Khlebnikov, V.~A. Rubakov and P.~G. Tinyakov, \emph{{Instanton
  induced cross-sections below the sphaleron}}, {\emph{Nucl. Phys.} {\bf B350}
  (1991) 441}.

\bibitem{Landau2}
L.~D. Landau, \emph{{On the theory of transfer of energy at collisions I}},
  {\emph{Phys. Zs. Sowiet.} {\bf 1} (1932) 88}.

\bibitem{Glauber:1963fi}
R.~J. Glauber, \emph{{The Quantum theory of optical coherence}},
  \href{http://dx.doi.org/10.1103/PhysRev.130.2529}{\emph{Phys. Rev.} {\bf 130}
  (1963) 2529--2539}.

\bibitem{Glauber:1963tx}
R.~J. Glauber, \emph{{Coherent and incoherent states of the radiation field}},
  \href{http://dx.doi.org/10.1103/PhysRev.131.2766}{\emph{Phys. Rev.} {\bf 131}
  (1963) 2766--2788}.

\bibitem{Zhang:1990fy}
W.-M. Zhang, D.~H. Feng and R.~Gilmore, \emph{{Coherent states: Theory and some
  Applications}}, \href{http://dx.doi.org/10.1103/RevModPhys.62.867}{\emph{Rev.
  Mod. Phys.} {\bf 62} (1990) 867--927}.

\bibitem{Rubakov:1991fb}
V.~A. Rubakov and P.~G. Tinyakov, \emph{{Towards the semiclassical
  calculability of high-energy instanton cross-sections}},
  \href{http://dx.doi.org/10.1016/0370-2693(92)91859-8}{\emph{Phys. Lett.} {\bf
  B279} (1992) 165--168}.

\bibitem{Rubakov:1992az}
V.~A. Rubakov, D.~T. Son and P.~G. Tinyakov, \emph{{An Example of semiclassical
  instanton like scattering: (1+1)-dimensional sigma model}},
  \href{http://dx.doi.org/10.1016/0550-3213(93)90473-3}{\emph{Nucl. Phys.} {\bf
  B404} (1993) 65--90}, [\href{https://arxiv.org/abs/hep-ph/9212309}{{\tt
  hep-ph/9212309}}].

\bibitem{Jaffe:1967nb}
A.~M. Jaffe, \emph{{High-energy Behavior In Quantum Field Theory. I. Strictly
  Localizable Fields}},
  \href{http://dx.doi.org/10.1103/PhysRev.158.1454}{\emph{Phys. Rev.} {\bf 158}
  (1967) 1454--1461}.

\bibitem{Khoze:2018qhz}
V.~V. Khoze and M.~Spannowsky, \emph{{Consistency of Higgsplosion in
  Localizable QFT}},
  \href{http://dx.doi.org/10.1016/j.physletb.2019.01.052}{\emph{Phys. Lett.}
  {\bf B790} (2019) 466--474}, [\href{https://arxiv.org/abs/1809.11141}{{\tt
  1809.11141}}].

\bibitem{Rajaraman}
R.~Rajaraman, \emph{{Solitons and instantons. An introduction to solitons and
  instantons in quantum field theory}}, {\emph{Amsterdam, Netherlands:
  North-Holland.} (1982) 409}.

\bibitem{Dorey:2002ik}
N.~Dorey, T.~J. Hollowood, V.~V. Khoze and M.~P. Mattis, \emph{{The Calculus of
  many instantons}},
  \href{http://dx.doi.org/10.1016/S0370-1573(02)00301-0}{\emph{Phys. Rept.}
  {\bf 371} (2002) 231--459}, [\href{https://arxiv.org/abs/hep-th/0206063}{{\tt
  hep-th/0206063}}].

\bibitem{Rubakov:1994cz}
V.~A. Rubakov and D.~T. Son, \emph{{Renormalization group for multiparticle
  production in (2+1)-dimensions around the threshold}},  in \emph{{8th
  International Seminar on High-energy Physics (Quarks 94) Vladimir, Russia,
  May 11-18, 1994}}, pp.~233--240, 1994.
\newblock \href{https://arxiv.org/abs/hep-ph/9406362}{{\tt hep-ph/9406362}}.

\bibitem{Farhi:1992pc}
E.~Farhi, V.~V. Khoze and R.~L. Singleton, Jr, \emph{{Minkowski space
  nonAbelian classical solutions with noninteger winding number change}},
  \href{http://dx.doi.org/10.1103/PhysRevD.47.5551}{\emph{Phys. Rev.} {\bf D47}
  (1993) 5551--5564}, [\href{https://arxiv.org/abs/hep-ph/9212239}{{\tt
  hep-ph/9212239}}].

\bibitem{landau1981quantum}
L.~Landau and E.~Lifshitz, \emph{Quantum Mechanics: Non-Relativistic Theory}.
\newblock Course of Theoretical Physics. Elsevier Science, 1981.

\bibitem{Coleman}
S.~R. Coleman, \emph{{The fate of false vacuum. 1. Semiclassical theory}},
  {\emph{Phys. Rev.} {\bf D15} (1977) 2929--2936}.

\bibitem{Shifman}
M.~Shifman, \emph{{Advanced topics in quantum field theory}}, {\emph{Cambridge
  University press} (2012) 622}.

\bibitem{IP}
S.~V. Iordanskii and L.~P. Pitaevskii, \emph{{Multiphoton boundary of the
  excitation spectrum in He II}}, {\emph{Sov. Phys. JETP} {\bf 49} (1979) 386}.

\bibitem{Voloshin:1990mz}
M.~B. Voloshin, \emph{{On strong high-energy scattering in theories with weak
  coupling}}, \href{http://dx.doi.org/10.1103/PhysRevD.43.1726}{\emph{Phys.
  Rev.} {\bf D43} (1991) 1726--1734}.

\bibitem{Khlebnikov:1992af}
S.~{\relax Yu}. Khlebnikov, \emph{{Semiclassical approach to multiparticle
  production}},
  \href{http://dx.doi.org/10.1016/0370-2693(92)90669-U}{\emph{Phys. Lett.} {\bf
  B282} (1992) 459--465}.

\bibitem{Diakonov:1993ha}
D.~Diakonov and V.~Petrov, \emph{{Nonperturbative isotropic multiparticle
  production in Yang-Mills theory}},
  \href{http://dx.doi.org/10.1103/PhysRevD.50.266}{\emph{Phys. Rev.} {\bf D50}
  (1994) 266--282}, [\href{https://arxiv.org/abs/hep-ph/9307356}{{\tt
  hep-ph/9307356}}].

\bibitem{Bachas:1991fd}
C.~Bachas, \emph{{A Proof of exponential suppression of high-energy transitions
  in the anharmonic oscillator}},
  \href{http://dx.doi.org/10.1016/0550-3213(92)90304-T}{\emph{Nucl. Phys.} {\bf
  B377} (1992) 622--648}.

\bibitem{Jaeckel:2018ipq}
J.~Jaeckel and S.~Schenk, \emph{{Exploring High Multiplicity Amplitudes in
  Quantum Mechanics}},
  \href{http://dx.doi.org/10.1103/PhysRevD.98.096007}{\emph{Phys. Rev.} {\bf
  D98} (2018) 096007}, [\href{https://arxiv.org/abs/1806.01857}{{\tt
  1806.01857}}].

\end{thebibliography}\endgroup

\end{document}